\documentclass[twocolumn]{aastex61}

\usepackage{amssymb}
\usepackage{amsmath}
\usepackage{latexsym}
\usepackage{fnpos}
\usepackage{scrextend}
\usepackage{scalerel}
\usepackage{hyperref}
\usepackage{tabularx}
\usepackage{xspace}
\usepackage{enumerate}
\usepackage{graphicx}
\usepackage{url}
\usepackage{longtable}
\usepackage{rotating}
\newcommand{\eal}[2]{\ifmmode{\mathrm{#1\,#2}}\else{#1\textsc{$\,$\lowercase{#2}}}\fi\xspace}
\newcommand{\feal}[2]{\ifmmode{\mathrm{#1\,#2}}\else{[#1\textsc{$\,$\lowercase{#2}}]}\fi\xspace}
\newcommand{\hfeal}[2]{\ifmmode{\mathrm{#1\,#2}}\else{#1\textsc{$\,$\lowercase{#2}}]}\fi\xspace}

\shorttitle{Spectroscopy of novae near optical peak}
\shortauthors{A\MakeLowercase{ydi et al.}}
\begin{document}

%\title{A universal ejection scenario for novae I: high-resolution spectroscopy near optical peak}
%\title{Early spectral evolution of classical novae towards a universal ejection scenario}
\title{Early spectral evolution of classical novae: \\
Consistent evidence for multiple distinct outflows}

\correspondingauthor{Elias Aydi} 
\email{aydielia@msu.edu}
\author[0000-0001-8525-3442]{E.~Aydi}
\affiliation{Center for Data Intensive and Time Domain Astronomy, Department of Physics and Astronomy, Michigan State University, East Lansing, MI 48824, USA \\}

\author[0000-0002-8400-3705]{L.~Chomiuk}
\affiliation{Center for Data Intensive and Time Domain Astronomy, Department of Physics and Astronomy, Michigan State University, East Lansing, MI 48824, USA \\}

\author[0000-0001-9695-8472]{L.~Izzo} 
\affiliation{DARK, Niels Bohr Institute, University of Copenhagen, Jagtvej 128, 2200 Copenhagen {\O}, Denmark \\}

\author[0000-0002-3014-3665]{E.~J.~Harvey} 
\affiliation{Astrophysics Research Institute, Liverpool John Moores University, Liverpool, L3 5RF, UK \\}

\author{J.~Leahy-McGregor}
\affiliation{Center for Data Intensive and Time Domain Astronomy, Department of Physics and Astronomy, Michigan State University, East Lansing, MI 48824, USA \\}

\author{J.~Strader}
\affiliation{Center for Data Intensive and Time Domain Astronomy, Department of Physics and Astronomy, Michigan State University, East Lansing, MI 48824, USA \\}

\author{D.~A.~H.~Buckley}
\affiliation{South African Astronomical Observatory, P.O.\ Box 9, 7935 Observatory, South Africa \\}

%\author[0000-0003-0156-3377]{M.~J.~Darnley}
%\affiliation{Astrophysics Research Institute, Liverpool John Moores University, Liverpool, L3 5RF, UK \\}

\author[0000-0001-5991-6863]{K.~V.~Sokolovsky}
\affiliation{Center for Data Intensive and Time Domain Astronomy, Department of Physics and Astronomy, Michigan State University, East Lansing, MI 48824, USA \\}
\affiliation{Sternberg Astronomical Institute, Moscow State University, Universitetskii~pr.~13, 119992~Moscow, Russia\\}

\author{A.~Kawash}
\affiliation{Center for Data Intensive and Time Domain Astronomy, Department of Physics and Astronomy, Michigan State University, East Lansing, MI 48824, USA \\}

\author{C.~S.~Kochanek}
\affiliation{Department of Astronomy, The Ohio State University, 140 West 18th Avenue, Columbus, OH 43210, USA \\}

\author{J.~D.\ Linford}
\affiliation{Department of Physics and Astronomy, West Virginia University, P.O. Box 6315, Morgantown, WV 26506, USA \\}
\affiliation{Center for Gravitational Waves and Cosmology, West Virginia University, Chestnut Ridge Research Building, Morgantown, WV 26505, USA \\}
\affiliation{National Radio Astronomy Observatory, P.O.\ Box O, Socorro, NM 87801, USA \\}

\author{B.~D.~Metzger}
\affiliation{Columbia Astrophysics Laboratory and Department of Physics, Columbia University, New York, NY 10027, USA\\}

\author{K.~Mukai}
\affiliation{CRESST and X-ray Astrophysics Laboratory, NASA/GSFC, Greenbelt, MD 20771, USA \\}
\affiliation{Department of Physics, University of Maryland, Baltimore County, 1000 Hilltop Circle, Baltimore, MD 21250, USA \\}

\author{M.~Orio}
\affiliation{INAF--Osservatorio di Padova, vicolo dell'Osservatorio 5, I-35122 Padova, Italy \\}
\affiliation{Department of Astronomy, University of Wisconsin, 475 N.\ Charter St., Madison, WI 53704, USA \\}

\author[0000-0003-4631-1149]{B.~J.~Shappee}
\affiliation{Institute for Astronomy, University of Hawai'i, 2680 Woodlawn Drive, Honolulu, HI 96822, USA \\}

\author[0000-0003-0286-7858]{L.~Shishkovsky}
\affiliation{Center for Data Intensive and Time Domain Astronomy, Department of Physics and Astronomy, Michigan State University, East Lansing, MI 48824, USA \\}

%\author{K.~Z.~Stanek}
%\affiliation{Department of Astronomy, The Ohio State University, 140 West 18th Avenue, Columbus, OH 43210, USA \\}

\author{E.~Steinberg}
\affiliation{Columbia Astrophysics Laboratory and Department of Physics, Columbia University, New York, NY 10027, USA\\}
\affiliation{Racah Institute of Physics, Hebrew University, Jerusalem 91904, Israel \\}

\author{S.~J.~Swihart}
\affiliation{Center for Data Intensive and Time Domain Astronomy, Department of Physics and Astronomy, Michigan State University, East Lansing, MI 48824, USA \\}

%\author{S.~C.~Williams}
%\affiliation{Physics Department, Lancaster University, Lancaster, LA1 4YB, UK\\}
%\affiliation{Astrophysics Research Institute, Liverpool John Moores University, Liverpool, L3 5RF, UK \\}

\author{J.~L.\ Sokoloski}
\affiliation{Columbia Astrophysics Laboratory and Department of Physics, Columbia University, New York, NY 10027, USA\\}

\author{F.~M.\ Walter}
\affiliation{Department of Physics \& Astronomy, Stony Brook University, Stony Brook, NY, 11794-3800, USA\\}

%\author{R.~E.\ Williams}
%\affiliation{Department of Astronomy \& Astrophysics, University of California, Santa Cruz, 1156 High Street, Santa Cruz, CA 95064, USA\\}
%\affiliation{Space Telescope Science Institute, 3700 San Martin Drive, Baltimore, MD 21218, USA\\}

\author[0000-0002-6896-1655]{P.~A.\ Woudt}
\affiliation{Department of Astronomy, University of Cape Town, Private Bag X3, Rondebosch 7701, South Africa\\}



\begin{abstract}

The physical mechanism driving mass ejection during a nova eruption is still poorly understood. Possibilities include ejection in a single ballistic event, a common envelope interaction, a continuous wind, or some combination of these processes. Here we present a study of 12 Galactic novae, for which we have pre-maximum high-resolution spectroscopy. All 12 novae show the same spectral evolution. Before optical peak, they show a slow P Cygni component. After peak a fast component quickly arises, while the slow absorption remains superimposed on top of it, implying the presence of at least two physically distinct flows. For novae with high-cadence monitoring, a third, intermediate-velocity component is also observed.

These observations are consistent with a scenario where the slow component is associated with the initial ejection of the accreted material and the fast component with a radiation-driven wind from the white dwarf. When these flows interact, the slow flow is swept up by the fast flow, producing the intermediate component. These colliding flows may produce the $\gamma$-ray emission observed in some novae. Our spectra also show that the transient heavy element absorption lines seen in some novae have the same velocity structure and evolution as the other lines in the spectrum, implying an association with the nova ejecta rather than a pre-existing circumbinary reservoir of gas or material ablated from the secondary. While this basic scenario appears to qualitatively reproduce multi-wavelength observations of classical novae, substantial theoretical and observational work is still needed to untangle the rich diversity of nova properties.

\end{abstract}




\section{Introduction}

A classical nova (CNe) is a transient event powered by a thermonuclear runaway on the surface of an accreting white dwarf in an interacting binary system (e.g., \citealt{Starrfield_etal_2008,2016PASP..128e1001S}). The thermonuclear runaway leads to the ejection of at least part of the accreted envelope ($10^{-7}-10^{-3}$~M$_{\odot}$) with velocities $\sim$ 200\,--\,5000\,km\,s$^{-1}$ and an increase in the optical brightness of the system by 8 to 15 magnitudes \citep{Payne-Gaposchkin_1957,Gallaher_etal_1978}.

There is agreement that the thermonuclear runaway leads to expansion of the accreted envelope, but the mechanism(s) powering the ejection of this envelope is still highly debated and poorly understood. 
The energy output of the nuclear reactions may lead to the prompt ejection of part of the envelope (e.g., \citealt{Starrfield_etal_2008,Shore2014,Mason_etal_2018,Mason_etal_2020}). The envelope will engulf the binary (for main-sequence donor stars), and 
the binary orbital motion energy may help eject the envelope \citep{Livio_etal_1990,Lloyd_etal_1997}. Some accreted material remains on the surface of the white dwarf, and undergoes sustained nuclear burning in a steady-state, near-Eddington luminosity phase \citep{Wolf_etal_2013}; the resultant radiation pressure can drive a wind that lasts for days to months \citep{Kato_Hachisu_1994, Friedjung_2011}. Unfortunately, there are essentially no theoretical studies that consider all of these potential mass loss mechanisms and model them self-consistently, so it remains difficult to predict which mechanism will dominate and under which conditions. Meanwhile, observations suggest that multiple mechanisms may be relevant, even within an individual system (e.g., \citealt{Friedjung_1987,Strope_etal_2010,Chomiuk_etal_2014}), and multiple ejections may occur over a single eruption \citep[e.g.,][]{Pejcha_2009, Aydi_etal_2019_I, Aydi_etal_2020}.

Images of old nova shells show that ejecta geometries are often far from spherical. The ejecta show diverse structures including elliptical morphologies, rings, clumps, and polar caps (e.g., \citealt{Shara_1995,Downes_etal_2000,Obrien_Bode_2008}). These static images imply that mass loss from novae is complex, but unfortunately these observations years-to-decades after eruption are not sufficient to reveal the physics that shaped the ejecta nor its early-time evolution. Observations obtained during the eruption itself---which can track changes in mass ejection as they occur---are needed.
%to understand the complexity of mass loss in nova.
High-resolution optical spectroscopy during the first days of the eruption can track the velocity (and to some extent, the morphology) of the ejecta, and can therefore be used to constrain ejection scenarios in novae (e.g., \citealt{Mclaughlin_1945,Mclaughlin_1947,Payne-Gaposchkin_1957,Friedjung_1987,Williams_Mason_2010,Arai_etal_2016,Aydi_etal_2019_I}).

\subsection{Revisiting the classics: McLaughlin, Payne-Gaposchkin, \& friends}
Novae have been studied with optical spectroscopy for over a century \citep[e.g.,][]{Clerke_1892}. \citet{McLaughlin_1944} and \citet{Payne-Gaposchkin_1957} tracked the evolution of nova spectra from light-curve maximum to eventual quiescence, and noticed the appearance of multiple absorption and emission systems.
These pioneering studies divided the observed systems of spectral lines into five classes based on their chronological appearance throughout the eruption, calling them the ``pre-maximum", ``principal", ``diffuse-enhanced", ``Orion", and ``nebular" spectra. These historic classifications linked these systems to distinct ejecta components or shells, but did not offer extensive speculations about their origin. 

Based on the classification of \citet{McLaughlin_1944}, the pre-maximum spectrum appears before optical peak and is characterized by P Cygni profiles with absorption troughs at velocities of a few hundred km\,s$^{-1}$. The principal spectrum appears at or several days after optical peak, 
%and replaces the pre-maximum spectrum 
with its absorption troughs characterized by higher blueshifted velocities. The two co-exist as distinct systems for a few days before the pre-maximum system disappears. The difference in velocity between the pre-maximum and principal spectrum ranges between $\sim$100--700\,km\,s$^{-1}$ and correlates with the speed class of the nova\footnote{\citet{Payne-Gaposchkin_1957} introduced the ``speed class'' classification of novae, which is based on the time it takes for the light curve to fade by two magnitudes from optical peak.}. 
%Also, soon after optical peak, 
According to \citet{McLaughlin_1944}, once the nova has faded by $\sim$two magnitudes from optical peak, the diffuse-enhanced absorption system appears and is characterized by velocities around twice that of the pre-maximum components. It also shows broad emission components whose width is again correlated with the nova speed class \citep{Payne-Gaposchkin_1957}. 
%\citet{McLaughlin_1944} suggests that around two magnitudes fainter than the optical peak, the diffuse-enhanced spectrum appears. 
Less than three magnitudes below optical peak, the Orion system sometimes appears, 
%named by McLaughlin as the Orion spectrum. This system of absorption has 
with more extreme velocities than the diffuse-enhanced spectrum and from more highly ionized species.
%mainly shows a diffuse set of lines of \eal{He}{II}, \eal{O}{II}, and \eal{N}{II}. 
The last distinct system of the nova eruption is the nebular spectrum, characterized by emission features of nebular and auroral forbidden lines, such as the \feal{O}{III} lines and high ionization forbidden lines of iron. %The full width at zero intensity (FWZI) of nebular lines is generally $\sim$2 times the terminal (late) velocity observed in the ``principal spectrum''. 
%While the \citet{McLaughlin_1942,McLaughlin_1944} classification system is in chronological and velocity order, we revisit the chronological appearance of these systems, particularly in comparison to the optical light curve evolution.
%These early papers by McLaughlin and Payne-Gaposchkin have noted that these absorption systems show changes in velocity, either deceleration or acceleration depending on the evolutionary stage of the eruption.  

Although \citet{McLaughlin_1944,Mclaughlin_1947,McLaughlin_1964} and \citet{Payne-Gaposchkin_1957} provide eloquent descriptions of the evolution of the different spectral systems, the spectra themselves are not clearly illustrated in these works.
This is mainly due to the use of different tools $\sim$70 years ago, which makes it difficult to compare with recent nova data sets. For example, early spectra were recorded with photographic plates, and when published, were represented in grey scale; the spectra have not reproduced well electronically \citep[e.g.,][]{McLaughlin_1944}, and are difficult to compare with modern one-dimensional spectra extracted from CCDs.
Therefore, a primary aim of this paper is to revisit this pioneering work with a more modern data set, and clearly illustrate the spectral evolution of a large sample of novae near optical peak.

\subsection{Proposed explanations for multiple spectral components}\label{sec:proposedex}

\citet{Mclaughlin_1947} unequivocally concluded from his spectroscopic observations that there must be multiple %at least two
ejecta components or shells in a given nova eruption, with the pre-maximum component external to, and expanding relatively slowly, compared to ejecta associated with the diffuse-enhanced component (see \S \ref{two_flows} for modern examples). He speculates that the intermediate-velocity principal component may represent material from the pre-maximum ejection, swept up and accelerated by radiation pressure from the hot white dwarf \citep{McLaughlin_1943}.
While \citet{Russell36} agreed that there must be multiple ejecta components, he proposed that the principal system forms due to shock interaction rather than radiation pressure. 

This scenario was revisited in a series of studies by \citet{Friedjung_1966_I,Friedjung_1966_II,Friedjung_1966_III,Friedjung_1987}, who agreed that the light curves and spectral evolution of novae could not be explained by a single ballistic ejection, and instead conclude that multiple outflows are present.
Friedjung explained McLaughlin's observations by suggesting that the principal spectrum originates from a shell that is formed by the collision of two flows---a fast wind, associated with the diffuse enhanced and Orion spectra, slamming into a slow flow, which is associated with the pre-maximum spectrum. \citet{Friedjung_1992,Friedjung_2011} also interpreted the observed acceleration of the spectral components after optical peak in the context of a continuous radiation-driven wind from the white dwarf.
%rather than a single ballistic ejection.
%In a series of studies, Friedjung elaborated on the origin of the different spectral systems and the scenario of mass ejection governing the "nova phenomenon" \citep{Friedjung_1966_I,Friedjung_1966_II,Friedjung_1966_III,Friedjung_1987}. They argued that the light curves and spectral evolution of novae could not be explained by a single ballistic ejection, and instead conclude that multiple outflows are present during the early stages of the nova eruption, which interact and give rise to the observed spectral features. They also suggest that a continuous, radiation driven wind is necessary to reproduce the observations, particularly the observed Doppler broadening in the emission line profiles after optical peak. 

In contrast to multiple ejections, %\citet{Mason_etal_2018, Mason_etal_2020} 
\citet{Shore_etal_2011, Shore_etal_2013, Shore_etal_2016} and \citet{Mason_etal_2018,Mason_etal_2020}
suggested that nova ejecta are expelled in a single impulse and expand ballistically, structured as a clumpy medium with a biconical geometry. Changes in the profile of a particular line may be attributed to changes in the optical depth and ionization state of the expanding ejecta \citep[e.g.,][]{Shore_etal_2011}.
\citet{Mason_etal_2020} studied the spectra of nova ASASSN-17hx, and argued that some spectral features are observed at the same velocities in different species at different stages of the eruption, implying that they originate in clumps frozen in from the very start of eruption.

%LC-- i don't think we really need this paragraph. we never really talk about it again in this paper.
%In a different context, \citet{Williams_2012} and \citet{Shore_etal_2014} discussed the origin of the post-maximum spectra of novae, particularly focusing on the differences between ``\eal{Fe}{II}'' and ``He/N'' spectroscopic classes. As defined in \citet{Williams_1992}, ``\eal{Fe}{II}'' novae have spectra dominated by low-ionization \eal{Fe}{II} and CNO emission lines, in addition to H Balmer lines. The emission lines are usually characterized by a FWHM lower than 2500\,km\,s$^{-1}$ and have rounded profiles. ``He/N'' novae have spectra wherein the strongest lines after the Balmer sequence are high excitation lines of He and N with broad, rectangular profiles.  \citet{Williams_1992, Williams_2012} suggested that the ``\eal{Fe}{II}'' spectral features originate from a prolonged wind, while the ``He/N'' spectral features originate from ejections dominated by a single shell. However, \citet{Shore_etal_2014} argued that both features are present in the spectrum of every nova, appearing in turn based on the time-dependent opacity of the ejecta. The ``\eal{Fe}{II}'' spectral features appear when the ejecta are optically thick, while the ``He/N'' features appear when the ejecta become optically thin. 

Recent work by R.~E.\ Williams et al.\ has highlighted the possibility that not all spectral features originate in nova ejecta expelled from the white dwarf surface.
\citet{Williams_etal_2008} pointed out the presence of absorption lines from heavy elements, such as Ti, Ba, Sc, and Y, with relatively low velocities ($\sim$400 to 1000\,km\,s$^{-1}$). The lines are present from early on in the eruption, and last for a few days/weeks after optical peak before disappearing---they are therefore named \textit{transient heavy element absorption} (THEA) lines.
%several unidentified absorption features in nova spectra, characterized by velocities ranging from 400 to 1000\,km\,s$^{-1}$. By comparing them to  features in the \eal{Na}{I} D lines at equivalent velocities, they associated these absorption lines with heavy elements, such as Ti, Ba, Sc, and Y. \citet{Williams_etal_2008} noticed that these lines last for a few days/weeks after peak before disappearing and thus named them \textit{transient heavy element absorption} (THEA) lines.
\citet{Williams_etal_2008} and \citet{Williams_Mason_2010} associated these lines with a pre-existing circumbinary reservoir of gas, perhaps from non-conservative mass transfer funneled out of the binary's outer Lagrangian points \citep[e.g.,][]{Taam&Spruit01, Sytov+07}.
However, to explain the strengths and kinematics of the THEA lines, the amount of mass and energy in this circumstellar material (CSM) would need to be uncomfortably high, rivaling the nova ejecta themselves.
There is no evidence for such circumbinary reservoirs in observations of cataclysmic variables during quiescence (e.g., \citealt{Dubus_etal_2004,Froning_2005,Hoard_etal_2014}).
\citet{Williams_2012, Williams_2013} revised the hypothesized origin of the THEA lines to be material irradiated or ablated from the secondary star during the nova eruption.
Today, the origin of the THEA lines remains a matter of debate.
%Studying the evolution of these lines can help us understand the mass-transfer prior the nova eruption
%, the common envelop evolution during the early days of the eruption, 
%and identify the different bodies of gas responsible for the multiple spectral features observed in novae near optical peak.

\subsection{Why revisit the basics now?}
The presence of multiple spectral features and a possible link to multiple ejections and shock interaction is not limited to novae, but extends to supernovae (SNe), particularly to those showing evidence for interaction between the ejecta and a dense CSM, such as Type IIn, Type Ia-CSM, and superluminous SNe  (e.g., \citealt{Chugai_Danziger_1994,Chugai_etal_1995,Smith_etal_2008,Fox_etal_2015,Dessart_etal_2016,Gangopadhyay_2020,Jerkstrand_etal_2020}). Several of these studies argue that the diversity of spectral features observed in the optical and infrared spectra of SNe 
%originate in distinct body of gases, 
%are the results of multiple phases of mass-loss or 
originate in shells created by the interaction of the SN ejecta with a complex CSM.

In addition, understanding how mass is ejected in novae has gained new urgency with the detection of GeV $\gamma$-ray emission from
Galactic novae by the Large Area Telescope (LAT) on the \textit{Fermi Gamma-Ray Space Telescope}
\citep{Ackermann_etal_2014,Cheung_etal_2016,Franckowiak_etal_2018}. The $\gamma$-rays imply that shocks are (1) present in novae, (2) energetically important, and (3) can dominate the optical luminosity of eruptions \citep{Metzger_etal_2015,Li_etal_2017_nature,Aydi_etal_2020}. 
The first GeV-detected nova %detected with \textit{Fermi}-LAT 
was V407 Cyg in 2010, a system with a Mira giant secondary \citep{Abdo_etal_2010}. The white dwarf in V407 Cyg is surrounded by a dense CSM enriched by the giant wind, so it was proposed that the $\gamma$-ray producing shocks occur between the nova ejecta and this pre-existing medium  \citep{Nelson_etal_2012, Martin_Dubus_2013}. However, many other \textit{Fermi}-detected novae have dwarf companions and thus are characterized by low-density CSM. In these cases, the shocks are likely to be internal to the nova ejecta---the result of interaction between multiple colliding flows \citep{Chomiuk_etal_2014,Metzger_etal_2015}.

Radio interferometric imaging of the $\gamma$-ray detected nova V959~Mon over the first $\sim$2 years of eruption showed the presence of two flows. \citet{Chomiuk_etal_2014} interpreted these two flows as an initial slow torus directed by the binary motion in the equatorial plane, followed by a fast wind which propagates more freely in the polar directions. At the interface of these two flows there are shocks that produce radio synchrotron and $\gamma$-ray emission (see Figures~2 and~3 of \citealt{Chomiuk_etal_2014}). Imaging of the peculiar Helium nova V445~Pup also shows a bipolar shell, equatorially confined by a dusty disk \citep{Woudt_etal_2009}, and evidence of shocks and particle acceleration in the form of luminous radio synchrotron emission (\citealt{Rupen_etal_2001}). Again, high-resolution radio imaging shows that that the synchrotron emission originates near the dusty disk
\citep{Nyamai_etal_2020}.

%If this mass-loss scenario is indeed common across novae, is it also the universal driver of $\gamma$-ray emission in novae? 
These multi-wavelength observations highlight the complexity and importance of mass loss in novae, and optical spectroscopy remains a critical tool for understanding it.
However, only a few studies presenting modern pre-maximum spectroscopic observations of novae are available in the literature. These consist mainly of very slow novae which rise to optical peak over several weeks, and focus on explaining the peculiarities of individual novae 
%and show multiple secondary maxima 
%lasting for several days/weeks, 
such as HR Del \citep{Friedjung_1992}, V723~Cas \citep{Iijima_etal_1998}, V5558~Sgr \citep{Poggiani_etal_2008,Tanaka_etal_2011_159}, and ASASSN-17pf \citep{Aydi_etal_2019_I}.
%These studies individually interpret the early spectral evolution in context of the multiple peaks observed in the light curve. %, some studies have targeted the spectral evolution of individual novae with the aim of interpreting their mass-loss mechanisms (e.g., \citealt{Cassatella_etal_2004,Arai_etal_2016}). 
Thus, for more than 70 years and since the pioneering work of \citet{McLaughlin_1944,Mclaughlin_1947,McLaughlin_1964} and \citet{Payne-Gaposchkin_1957}, no studies have tackled the early spectral evolution of a large sample of novae, particularly during the pre-maximum phase and early decline.

Given the ongoing debate about the mass ejection scenario in novae, we aim to solidify a unifying picture for how novae eject their accreted envelopes. In the current paper, we present pre-maximum 
%early medium- and high-resolution 
optical spectra for a sample of 12 novae and compare them with later spectra to test whether the mass-loss scenario proposed for 
 V959~Mon can be extrapolated to other novae. 
%While, some studies have targeted the spectral evolution of individual novae with the aim of interpreting their mass-loss mechanisms (e.g., \citealt{Cassatella_etal_2004,Arai_etal_2016}), we present a sample that uniquely combines early time coverage and high spectral resolution for a relatively large and diverse sample of novae. 
Here we present a sample that uniquely combines early time coverage and high spectral resolution for a relatively large and diverse sample of novae,
%, to revisit the pioneering work of \citet{McLaughlin_1944}, \citet{Payne-Gaposchkin_1957}, and \citet{Friedjung_1987} using modern data and to interpret these data in context of the multi-wavelength discoveries and advances made in the field.
%Our sample 
at least doubling the sample of novae with near-optical-peak spectra in the literature. %We also include a more detailed look at two novae with exceptionally good time coverage of their spectroscopic observations, V906~Car and FM~Cir.
Section~\ref{sec_obs} describes our sample and the observations. In Section~\ref{sec_results} we present the spectroscopic results, illustrating the early spectral evolution of our nova sample, and offering more details on two particularly well-observed novae, V906~Car and FM~Cir. In Sections~\ref{sec_disc} we discuss these results in the context of nova mass ejection and interaction between different flows, while in Section~\ref{sec_conc} we present our conclusions.

\begin{table*}[!t]
\centering
\caption{The nova sample.}
%including details for each nova, such as the discovery date ($t_0$), a reference to the discovery, the date of optical peak ($t_{\mathrm{max}}$), the magnitude at optical peak, $t_2$, and whether the nova was detected by \textit{Fermi}-LAT. $t_0$, $t_{\mathrm{max}}$, $V_{\mathrm{max}}$, and $t_2$ were derived using data from ASAS-SN and AAVSO.}
\begin{tabular}{lcccccc}
\hline
Name & $t_0$ & Discovery &$t_{\mathrm{max}}$ & $V_{\mathrm{max}}$ & $t_2$ & \textit{Fermi}-detected?$^{b}$\\
 & (UT date) & Ref.$^{a}$ & (UT date) & (mag) & (days) & \\
\hline
\hline
V1369~Cen & 2013 Dec 02.7 & (1) & 2013 Dec 06.3 & 3.6 & 14 & Y (13) \\
V5855~Sgr & 2016 Oct 20.5 & (2) & 2016 Oct 24.8 & 7.8 & 13 & Y (14)\\
%V612~Sct (ASASSN-17hx) & 2017-06-19.40 & 2017-07-30.00 & $-4$ & 3\\
V549~Vel (ASASSN-17mt) & 2017 Sept 23.4 & (3) & 2017 Oct 17.4 & 9.0 & 100 & Y (15)\\
ASASSN-17pf (LMCN-2017-11a) & 2017 Nov 17.2 & (4) & 2017 Dec 07.2 & 11.8  & 121 & N\\
FM~Cir & 2018 Jan 19.7 & (5) & 2018 Jan 28.7 & 6.4 & 150 & N\\
V906~Car (ASASSN-18fv) & 2018 Mar 16.0 & (6) &  2018 Mar 26.5 & 5.8 & 44 & Y (16)\\
V435~CMa & 2018 Mar 24.5 & (7) &  2018 Mar 29.0 & 10.3 & 60 & N\\
V613~Sct & 2018 June 29.6 & (8) &  2018 July 01.0 & 10.5 & 52 & N\\
%V3666~Oph & 2018-08-08.95 & 2018-08-11.90 & $-1.1$ & 16\\
V1706~Sco (ASASSN-19mo) & 2019 May 13.2 & (9) &  2019 May 22.0 & 12.3 & 108 & N\\
ASASSN-19qv (SMCN-2019-07a) & 2019 July 04.3 & (10) &  2019 July 06.4 & 11.2 & 15 & N\\
LMCN-2019-07a & 2019 July 29.1 & (11) &  2019 July 31.4 & 10.9 & 20 & N\\
V1707~Sco & 2019 Sept 14.1 & (12) & 2019 Sept 16.0 & 11.5 & 6 & Y (17)\\
%V659~Sct (ASASSN-19aad) & 2019-10-29.05 & 2019-10-30.75 & $-0.9$ & 5.25\\
\hline
\end{tabular}

{\raggedright $^{a}$ Discovery references: (1) = \citet{2013IAUC.9265....1G}; (2) = \citet{2016IAUC.9284....5N}; (3) = ASAS-SN \citep{ATel_10772}; (4) = ASAS-SN \citep{ATEL_11132}; 
(5) = 
% CBET 4482
\citet{2018CBET.4482....1S}; 
(6) = ASAS-SN \citep{ATel_11454}; (7) = 
% CBET 4499
\citet{2018CBET.4499....1N}; (8) = 
% CBET 4530
\citet{2018CBET.4530....1S}; 
(9) = ASAS-SN \citep{2019TNSTR.774....1S}; (10) = ASAS-SN \citep{2019TNSTR1142....1S}; (11) = \citet{2019TNSTR1337....1J}; (12) = 
% CBET 4667
\citet{2019CBET.4667....1I}.
\\
\raggedright $^{b}$References for \emph{Fermi}-LAT observations: (13) = \citet{Cheung_etal_2016}; (14) = \citet{Nelson_etal_2019}; (15) = \citet[][2020, in prep]{ATel_10977}; (16) = \citet{Aydi_etal_2020}; (17) = \citet{ATel_13116}. \par}
\label{table:sample}
\end{table*}

\begin{table*}
\centering
\caption{Log of the spectroscopic observations of the novae in our sample.}
%including the times of the spectroscopic observations relative to optical peak ($t_{\mathrm{spec}} - t_{\mathrm{max}})$, the instruments used, and the resolution ($R$) of the spectra.}
\begin{tabular}{l|cccc|cccc}
\hline
Name & $t_{\mathrm{s1}} - t_{\mathrm{max}}$ & Instrument & $R$ & $\lambda$ Range & $t_{\mathrm{s2}} - t_{\mathrm{max}}$ & Instrument & $R$ & $\lambda$ Range\\
& (days) & & & ($\mathrm{\AA}$)  & (days) & & & ($\mathrm{\AA}$) \\
\hline
\hline
V1369~Cen & $-1.0$ & ARAS & 11,000 & 6400\,--\,6720 & 3.0 & FEROS & 59,000 & 3750\,--\,9000\\
V5855~Sgr & $-1.4$ & ARAS & 1,500 & 3800\,--\,7260 & 7.2 & ARAS & 1,500 & 3800\,--\,7260 \\
V549~Vel  & $-13$ & SOAR-Good. & 5,000 & 4500\,--\,5170 & 41.9 & SOAR-Good. & 1,000 & 4500\,--\,5170\\
ASASSN-17pf & $-4.0$ & Mage-MIKE & 65,000 & 4850--4920 & 3.0 & Du Pont & 40,000& 4840\,--\,4905 \\
FM~Cir & $-0.3$ & SALT-HRS & 67,000 & 3900\,--\,8800 & 4.3 & SALT-HRS & 67,000 & 3900\,--\,8800 \\
V906~Car & $-4.4$ & VLT-UVES & 59,000 & 3050\,--\,9000 & 1.5 & VLT-UVES & 59,000 & 3050\,--\,9000 \\
V435~CMa & $-0.2$ & ARAS & 9,000 & 4000\,--\,7500 & 5.8 & ARAS & 9,000 & 4000\,--\,7500 \\
V613~Sct & $-0.2$ & ARAS & 11,000 & 4250\,--\,7550 & 2.8 & SALT-HRS & 14,000 & 3900\,--\,8800\\
V1706~Sco & $-0.0$ & SALT-HRS & 14,000 & 3900\,--\,8800 & 7.9 & SALT-HRS & 14,000 & 3900\,--\,8800 \\
ASASSN-19qv & $-0.0$ & SOAR-Good. & 1,000 & 4050\,--\,8000 & 3.8 & SALT-HRS & 14,000 & 3900\,--\,8800\\
LMCN-2019-07a & $-0.0$ & SOAR-Good. &  5,000 & 4500\,--\,5170 &  1.8 & SOAR-Good. & 5,000 & 4500\,--\,5170 \\
V1707~Sco & $-0.0$ & SOAR-Good. & 5,000 & 4500\,--\,5170 & 1.8 & SALT-HRS & 14,000 & 3900\,--\,8800\\
%V659~Sco (ASASSN-19aad)$^{a}$ & -- & -- & -- & 5.3 & SOAR-Goodman & 5,000 \\
%V407 Lup (ASASSN-16kt)$^{a}$ & -- & -- & -- & 5.0 & VLT-XSHOOTER & 59,000 \\
\hline
\end{tabular}
\label{table:spec}
\end{table*}

\section{Observations}
\label{sec_obs}

\subsection{Sample Selection}
Since novae rise to optical peak in a short period of time (from a few hours to a few days, to weeks in some extreme cases), observing novae spectroscopically before they reach optical peak is a challenging task. However, with new all-sky surveys such as the All-Sky Automated Survey for Supernovae (ASAS-SN;  \citealt{Shappee_etal_2014,Kochanek_etal_2017}) and the advanced capabilities of citizen scientists, it is becoming more feasible to discover and report on novae before they reach their optical peak. In addition, with large telescopes capable of rapid follow-up, such as SALT, SOAR, and VLT, it is also becoming more feasible to obtain high-resolution spectra for novae near optical peak. Some citizen scientists have telescopes equipped with spectrographs, and can obtain time series of high-resolution spectroscopy for bright sources (e.g., \citealt{Teyssier_2019}).

Therefore, we selected all recent (2013--2019) southern novae for which we obtained at least one pre-maximum spectrum (using one or more of SOAR, SALT, Magellan, or VLT). 
In addition we include four novae which were bright enough to be observed by citizen scientists during the rise to optical peak.
Our nova sample sums up to 12 novae, all of which were observed before and after optical peak with
%low, medium, or high-resolution ($R \approx$ 1,000\,--\,70,000) 
optical spectroscopy. Each of the selected novae also had a light curve of sufficient cadence to constrain the date of the optical peak. In Table~\ref{table:sample} we present the nova sample, listing the date of first detection in eruption ($t_0$) and a reference to the announcement of discovery. We also list the time of optical peak ($t_{\mathrm{max}}$; or first optical peak in case of a nova with multiple peaks) %the times of the spectroscopic observations relative to optical peak ($t_{\mathrm{spec}} - t_{\mathrm{max}})$, and 
and the $V$-band  magnitude at optical peak ($V_{\mathrm{max}}$). Also cataloged are the time for the light curve to decline by two magnitudes from optical peak ($t_2$), and whether the nova was detected by \textit{Fermi}-LAT. The quantities $t_0$, $t_{\mathrm{max}}$, $V_{\mathrm{max}}$, and $t_2$ were derived using data from  ASAS-SN and the American Association of Variable Star Observers (AAVSO; \citealt{Kafka_2020}; see Section~\ref{sec_lc}). $t_2$ is measured as the duration between the first peak and the last time the nova reaches two magnitudes fainter than this peak.
\\

\begin{figure*}
\begin{center}
  \includegraphics[width=\textwidth]{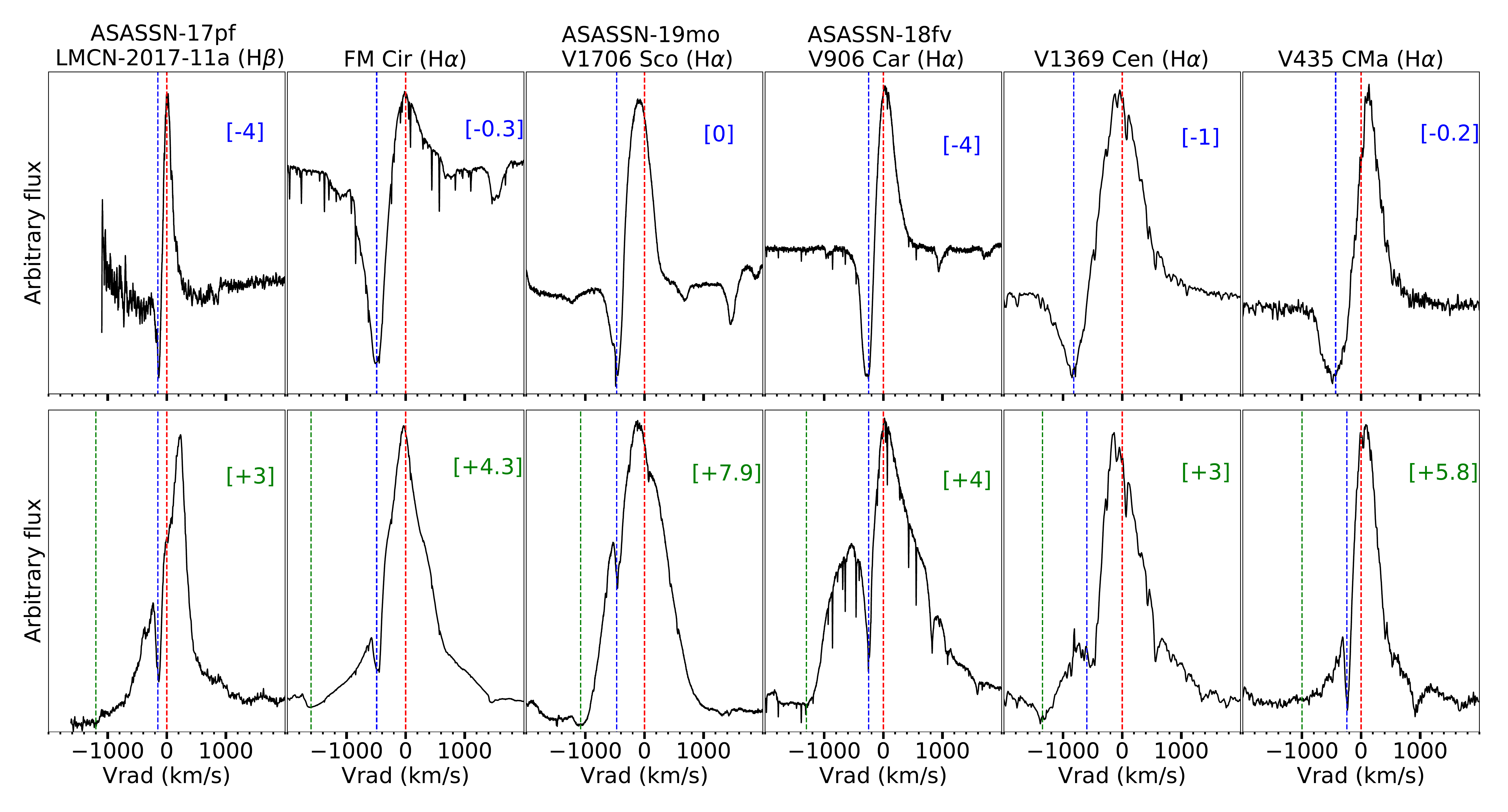}
\caption{The line profiles of H$\alpha$ or H$\beta$ before (\textit{top}) and after (\textit{bottom}) optical peak for novae LMCN-2017-11a (ASASSN-17pf), FM~Cir, V1706~Sco (ASASSN-19mo), V906~Car (ASASSN-18fv), V1369~Cen, and V435~CMa. The red dashed lines represent rest velocity ($v_{\mathrm{rad}}$ = 0\,km\,s$^{-1}$). The blue and green dashed lines represent the velocities of the slow and fast components, respectively; they are centered at the minima of the absorption features or the edge of the broad emission. The numbers in brackets are the day of observation relative to the optical peak ($t_{\mathrm{s}} - t_{\mathrm{max}})$. Heliocentric correction is applied to the radial velocities in all the plots across the paper.}% For nova ASASSN-17pf, which is in the Large Magellanic Cloud (LMC), a correction of 250\,km\,s$^{-1}$ for the radial velocity of the LMC is applied.}
\label{Fig:line_profiles_slow}
\end{center}
\end{figure*}

\begin{figure*}
\begin{center}
  \includegraphics[width=\textwidth]{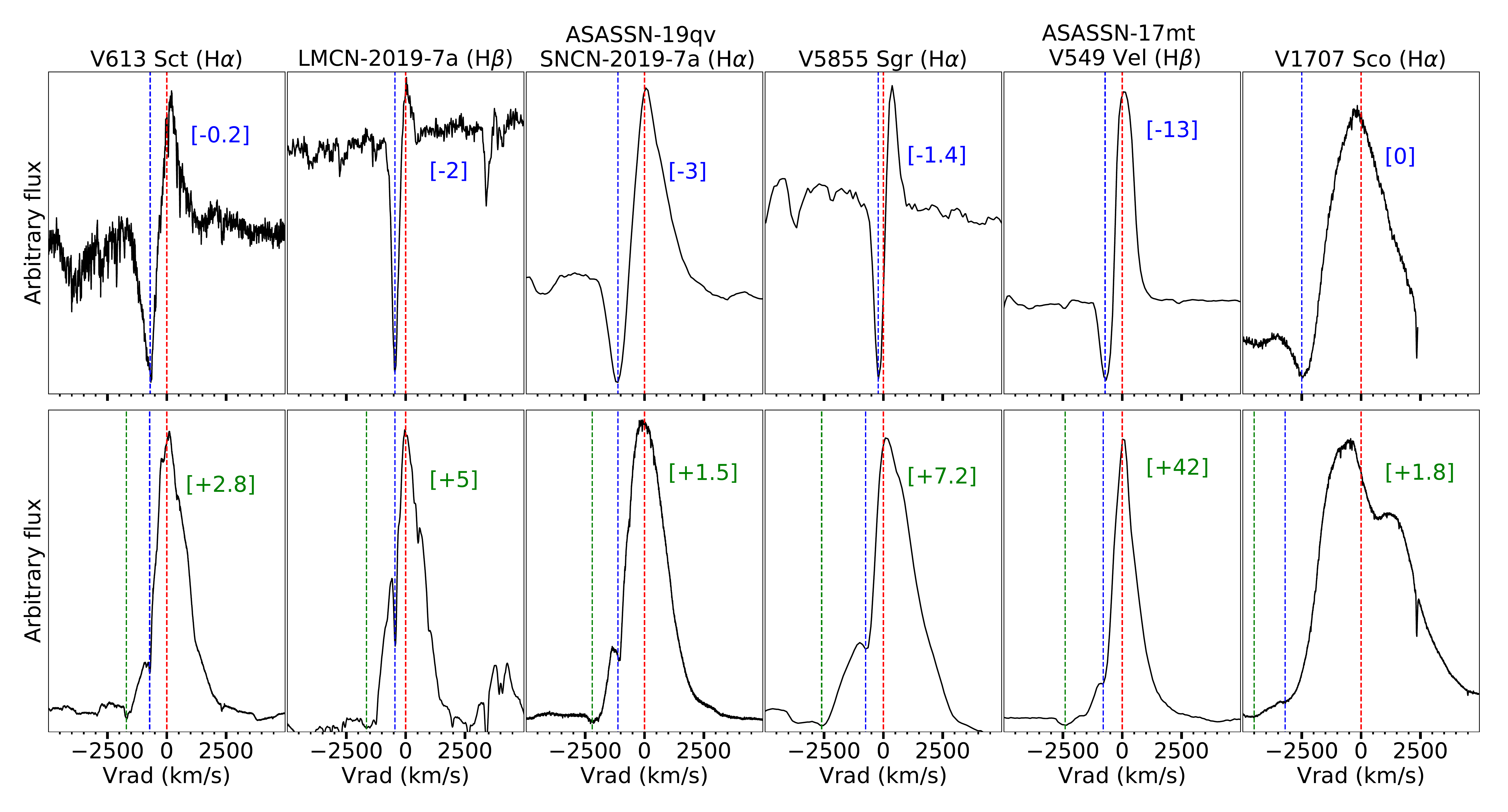}
\caption{Same as Figure~\ref{Fig:line_profiles_slow}, but for the novae V613~Sct, LMCN-2019-7a, ASASSN-19qv (SMCN-2019-7a), V5855~Sgr, V549 Vel (ASASSN-17mt), and V1707~Sco. Note the broader velocity scale here, as compared with Figure~\ref{Fig:line_profiles_slow}; the novae plotted here show relatively fast outflows.}% For nova LMCN-2019-7a, which is in the LMC, a correction of 250\,km\,s$^{-1}$ for the radial velocity of the LMC is applied. For nova ASASSN-19qv which is in the Small Magellanic Cloud (SMC), a correction of 160\,km\,s$^{-1}$ for the radial velocity of the SMC is applied.}
\label{Fig:line_profiles_mod}
\end{center}
\end{figure*}

\iffalse
\begin{figure*}
\begin{center}
  \includegraphics[width=0.95\textwidth]{line_profiles_fast.pdf}
\caption{Same as Figure~\ref{Fig:line_profiles_slow}, but for the novae ASASSN-19qv (SMCN-2019-7a), V5855~Sgr, V549 Vel (ASASSN-17mt), and V1707~Sco.}% For nova ASASSN-19qv which is a member of the Small Magellanic Cloud, a correction of 160\,km\,s$^{-1}$ for the radial velocity of the SMC is applied.}
\label{Fig:line_profiles_fast}
\end{center}
\end{figure*}
\fi

\subsection{Spectroscopic observations}

\subsubsection{The spectra of the 12 nova sample}

We carried out medium-resolution optical spectroscopy for several novae using the Goodman spectrograph \citep{Clemens_etal_2004} on the 4.1\,m Southern Astrophysical Research (SOAR) telescope located on Cerro Pach\'on, Chile. Most observations used a setup with a 2100~l\,mm$^{-1}$ grating and a 0.95\arcsec\ slit, yielding a resolution $R \approx$ 5000 in a region centered on H$\alpha$ or H$\beta$ that is 570\,\AA\ wide. The spectra were reduced and optimally extracted using the \textsc{apall} package in the Image Reduction and Analysis Facility (IRAF; \citealt{Tody_1986}). A few low-resolution spectra were obtained using the 400~l\,mm$^{-1}$ grating and a 0.95\arcsec\ slit, yielding a resolution $R \approx$ 1000 over the wavelength range 3820--7850\,\AA.

We also obtained spectra using the High Resolution Spectrograph (HRS; \citealt{Barnes_etal_2008,Bramall_etal_2010,Bramall_etal_2012,Crause_etal_2014}) mounted on the Southern African Large Telescope (SALT; \citealt{Buckley_etal_2006,Odonoghue_etal_2006}) in Sutherland, South Africa. HRS was used in three distinct modes with a range of moderate to high resolutions, yielding $R \approx$ 14,000, 40,000, and 65,000, over the range 4000\,--\,9000\,\AA. The primary reduction of the HRS spectroscopy was conducted using the SALT science pipeline \citep{Crawford_etal_2010}, which includes over-scan correction, bias subtraction, and gain correction. The rest of the reduction was done using the MIDAS FEROS \citep{Stahl_etal_1999} and $echelle$ \citep{Ballester_1992} packages. The reduction procedure is described by \citet{kniazev_etal_2016}.

We obtained high--resolution spectra from the ESO/MPG 2.2m telescope with FEROS \citep{Kaufer_etal_1999} %for V1369 Cen 
and from the ESO-VLT 8m telescope with UVES \cite{Dekker_etal_2000}. %for V407 Lup (ASASSN-16kt). 
The UVES data were reduced using the reflex environment \citep{Freudling2013} while %for the FEROS data we used a dedicated script based on the standard ESO Recipe (esorex). 
the FEROS data %for V1369 Cen 
were reduced using the ESO-MIDAS data analysis software \citep{MIDAS}.

We also make use of publicly available medium- and high-resolution spectra from the Astronomical Ring for Access to Spectroscopy (ARAS\footnote{\url{http://www.astrosurf.com/aras/Aras_DataBase/Novae.htm}}; \citealt{Teyssier_2019}). For nova ASASSN-17pf, we use data from \citet{Aydi_etal_2019_I}. These data were obtained using the Magellan Inamori Kyocera Echelle (MIKE) spectrograph mounted on the Magellan Clay telescope \citep{Bernstein_etal_2003,Shectman_Johns_2003} and the Echelle Spectrograph mounted on the 2.5\,m Ir$\acute{\mathrm{e}}$n$\acute{\mathrm{e}}$e Du Pont Telescope\footnote{\url{http://www.lco.cl/telescopes-information/lco/telescopes-information/irenee-du-pont/instruments}}.

In Table~\ref{table:spec} we present the details of all the spectroscopic observations, such as the times of the spectroscopic observations relative to optical peak ($t_{\mathrm{s}} - t_{\mathrm{max}})$, the instruments used, the resolution ($R$) of the spectra, and their spectral ranges.

\subsubsection{Additional observations of specific novae}

Nova V906~Car and FM~Cir were two bright, naked-eye, novae for which we have dedicated, high-resolution spectra during the rise to maximum and early decline phases. Therefore, in Sections~\ref{V906_Car_sec} and~\ref{FM_Cir_sec} we discuss in detail the early spectral evolution of these two novae, respectively. The spectra obtained over the first 30 days of the eruption of nova V906 Car are presented in \citealt{Aydi_etal_2020}, and a log of these observations can be found in Table~\ref{table:spec_V906_Car} in the Appendix. 

The pre-maximum spectra of nova FM~Cir, during the first week of the eruption, were obtained using the CHIRON echelle spectrograph \citep{Tokovinin_etal_2013} mounted on the CTIO 1.5m telescope.
All observations were made in queue mode by professional observers. Integration times were typically 10 minutes in a single integration, but
multiple integrations were obtained and summed on some nights.  Spectra were taken in both ``fiber mode'', with $4\times4$ on-chip binning yielding a resolution $R \approx$27,800, and with the image slicer (``slicer mode''), with $3\times1$ on-chip binning yielding a resolution $R \approx$78,000 over the range of 4100\,--\,8900\,$\mathrm{\AA}$. A log of these observations is presented in Table~\ref{table:spec_FM_Cir}. 

%which is operated by the SMARTS\footnote{The archive of optical spectra is available at \url{http://www.astro.sunysb.edu/fwalter/SMARTS/NovaAtlas/}} Consortium. CHIRON was used in the fibre mode to provide a resolution $R \sim$ 25,000 over the range of 4080\,--\,8900\,$\mathrm{\AA}$. The spectra were reduced using a pipeline coded in IDL\footnote{\url{http://www.astro.sunysb.edu/fwalter/SMARTS/CHIRON/ch\_reduce.pdf}}.

%The images were flat-fielded. Cosmic rays are removed using the L.A.Cosmic algorithm \citep{vD01}. The 74 echelle orders were extracted, and instrumental background was subtracted. As Chiron is fiber-fed, there is no simple method to subtract the sky. Wavelength calibration is accomplished using ThAr calibration lamp exposures at the start and end of the night, and occasionally throughout the night. Chiron in fiber mode is stable to better than 200 m/s over the course of many nights. The instrumental response was removed from the individual orders by dividing by the spectra of a flux-standard star, $\mu$~Col. This provides flux-calibrated orders with a systemic uncertainty due to sky conditions. The individual orders are then spliced together, resulting in a calibrated spectrum from 4080-8900\AA.

\subsubsection{Sample bias and very fast novae}

Our sample is  biased towards slower evolving novae,  since faster novae rise to maximum in $\lesssim$1 day and therefore are more challenging to observe before peak. We have obtained post-maximum spectra of two very fast novae (V407~Lup and V659~Sct). We did not have pre-maximum spectra for these objects, so they are not included in our main sample. Still, these post-maximum spectra, obtained using VLT-UVES and SOAR during the early decline, are useful for extending conclusions about spectral evolution to the fastest-evolving ejections. The log of these observations is listed in Table~\ref{table:spec_two_novae}.

\begin{table}
\centering
\caption{Log of the spectroscopic observations of the very fast novae V659~Sct (ASASSN-19aad) and V407~Lup (ASASSN-16kt).}
\begin{tabular}{lccc}
\hline
Name & $t_{\mathrm{s2}} - t_{\mathrm{max}}$ & Instrument & $R$\\
 (days) & & \\
\hline
\hline
V659~Sct & 5.3 & SOAR-Goodman & 5,000 \\
V407 Lup & 5.0 & VLT-UVES & 59,000 \\
\hline
\end{tabular}
\label{table:spec_two_novae}
\end{table}

\subsection{Light curves} 
\label{sec_lc}
We construct an optical light curve for each nova using $V$-, and/or $g$-band data from ASAS-SN and the AAVSO. For some novae, we augment $V$ and $g$ light curves with visual and $CV$ (unfiltered observations with $V$ magnitude zero-point). We present the light curves for all the novae in the sample in Appendix~\ref{appB}. These light curves are plotted around optical peak to highlight the early evolution and show the dates of the spectroscopic observations on the plot. The peak is measured from the $V$ or $g$-band light curves as the date when the light curve reached its highest brightness. For novae with multiple peaks/flares, we consider the peak as the date when the light curve reaches its first peak.

%For some novae we augment $V$ and $g$ light curves with visual and $CV$ (unfiltered observations with $V$ magnitude zero-point). %The typical uncertainty of visual observations is $\sim0.1$\,mag, both for an individual observer and when comparing multiple observers. Visual observations of very red objects have larger and harder to quantify uncertainties. $V$ magnitudes from the APASS survey \citep{2014CoSka..43..518H} and other literature are typically used for comparison stars, so visual magnitude estimates have the same zero-point as $V$-band CCD observations. The unfiltered CCD observations with $V$ zero-point ($CV$-band) typically have uncertainty $<0.02$\,mag for each individual observer. However, $CV$ photometry may have large zero-point offsets from visual and $V$ light curves, that is different between observers. The common practice is to shift the $CV$ zero-point of each observer to match $V$-band observations taken close in time.

\section{Results and analysis}
\label{sec_results}

In this section, we first present the early spectral evolution of the Balmer lines for all the novae in our sample. Then we present a more detailed view on the spectral evolution of novae V906~Car and FM~Cir, considering lines beyond the H Balmer series and including THEA lines during the rise to peak and early decline. At the end we highlight some general trends that we noticed in these two novae and other well-observed novae in our sample. 

\subsection{Consistent evidence for two flows in all the novae of our sample}
\label{two_flows}
In Figures~\ref{Fig:line_profiles_slow} and~\ref{Fig:line_profiles_mod} we present the H$\alpha$ or H$\beta$ line profiles, shortly before and after optical peak, for each nova in our sample. The novae are grouped in increasing order of the full width at zero intensity (FWZI) of the post-peak emission lines, for a better illustration. The reason we choose to highlight the evolution of the Balmer lines for all novae is that 
several of our medium- and high-resolution spectra cover limited spectral ranges centered on H$\alpha$ or H$\beta$. Additional lines are studied for some well-observed novae in our sample and will be discussed in following sections. 

All the novae show a similar spectral evolution: before optical peak, the emission lines show P Cygni profiles %characterized by low
with absorption troughs at velocities ranging between $-200$ and $-1000$\,km\,s$^{-1}$, correlated to the speed class of the nova. Shortly after optical peak, a broad emission component emerges with the base of the emission extending to velocities $>$1000\,km\,s$^{-1}$ (more than twice the velocity of the pre-maximum component), while the pre-existing P Cygni profiles are superimposed on top of the broad emission line profiles. The broad emission is sometimes accompanied by a blue-shifted absorption feature that is relatively weak compared to the emission. For some of the fast novae in our sample, such as V1707~Sco (Figure~\ref{Fig:line_profiles_mod}), the narrow component is difficult to see after optical peak without a zoom-in on the absorption of the P Cygni profile. This is mainly due to the nova being very fast and thus the absorption feature weakening rapidly. 

Hereafter we call the pre-maximum P Cygni profile the ``slow component'' and the post-maximum broad emission + higher-velocity P Cygni absorption the ``fast component''. These are probably the same as the ``pre-maximum'' and ``diffuse enhanced'' systems of \citet{McLaughlin_1944}, though quantitative comparison between modern one-dimensional spectra extracted from CCD images and reproduced older two-dimensional photographic plate spectra is challenging.

%Our findings here echo the results of \citet{Mclaughlin_1947}, who states:
%\begin{quote}
%\emph{The broad emissions of the ``diffuse-enhanced" system extend across the emission and absorption of the same lines from the principal shell. Nevertheless, the principal absorption remains strong and well defined, without the filling in that would surely occur if the atoms that produce the ``diffuse-enhanced" emission were outermost.} 
%\end{quote}
%McLaughlin concludes that ``\emph{The `diffuse-enhanced' emission and absorption must be assigned to a P Cygni-like expanding atmosphere close to the central star and wholly inside the principal shell.}"
%Decades worth of progress in spectroscopic observations leave these basic conclusions unchanged. 
The co-existence of the slow and fast components
%---and the abrupt transition of the spectral profiles over roughly a day (see Section~\ref{V906_Car_sec})---
indicates the presence of at least two physically distinct flows, a slow and a fast one. The superimposition of the slow P Cygni profile on top of the broad emission indicates that a significant portion of the fast flow is inside the slow flow. We return to these points in Section \ref{sec:universal}.

%In some cases, such as for novae V1707~Sco and V5855~Sgr, the P Cygni profiles show a sudden large blueshift in velocities after optical peak, compared to their pre-maximum velocities. The likely explanation is that the post-optical-peak P Cygni profiles are associated with the so-called ``principal spectrum'' which replaces the ``pre-maximum'' spectrum several days after optical peak, with higher blueshifted velocities. In the discussion we will elaborate on the origin of each of these components.  %Our results also show that this same evolution is observed in all novae, even the very fast evolving ones. %The presence of multiple ejections/outflows/winds have been suggested earlier due to the co-existence of different spectral spectral systems, (REF), and several suggestions have been presented in the literature to explain their origin. In Section~\ref{sec_disc} we discuss the implications of the presence of multiple ejections in the context of a universal ejection scenario, $\gamma$-ray emission, and shock interactions. 

\subsection{Very fast novae}
\label{sec_very_fast_novae}

The light curves of very fast novae reach optical peak in roughly a day, which makes it challenging to obtain spectroscopic observations before peak brightness \citep{Warner_2008}. For the fast novae in our sample. which we manage to observe before or at optical peak (V1707~Sco and ASASSN-19qv in Figure~\ref{Fig:line_profiles_mod}), we observe the same spectral evolution as seen in slower novae.

\begin{figure}
\begin{center}
  \includegraphics[width=1.05\columnwidth]{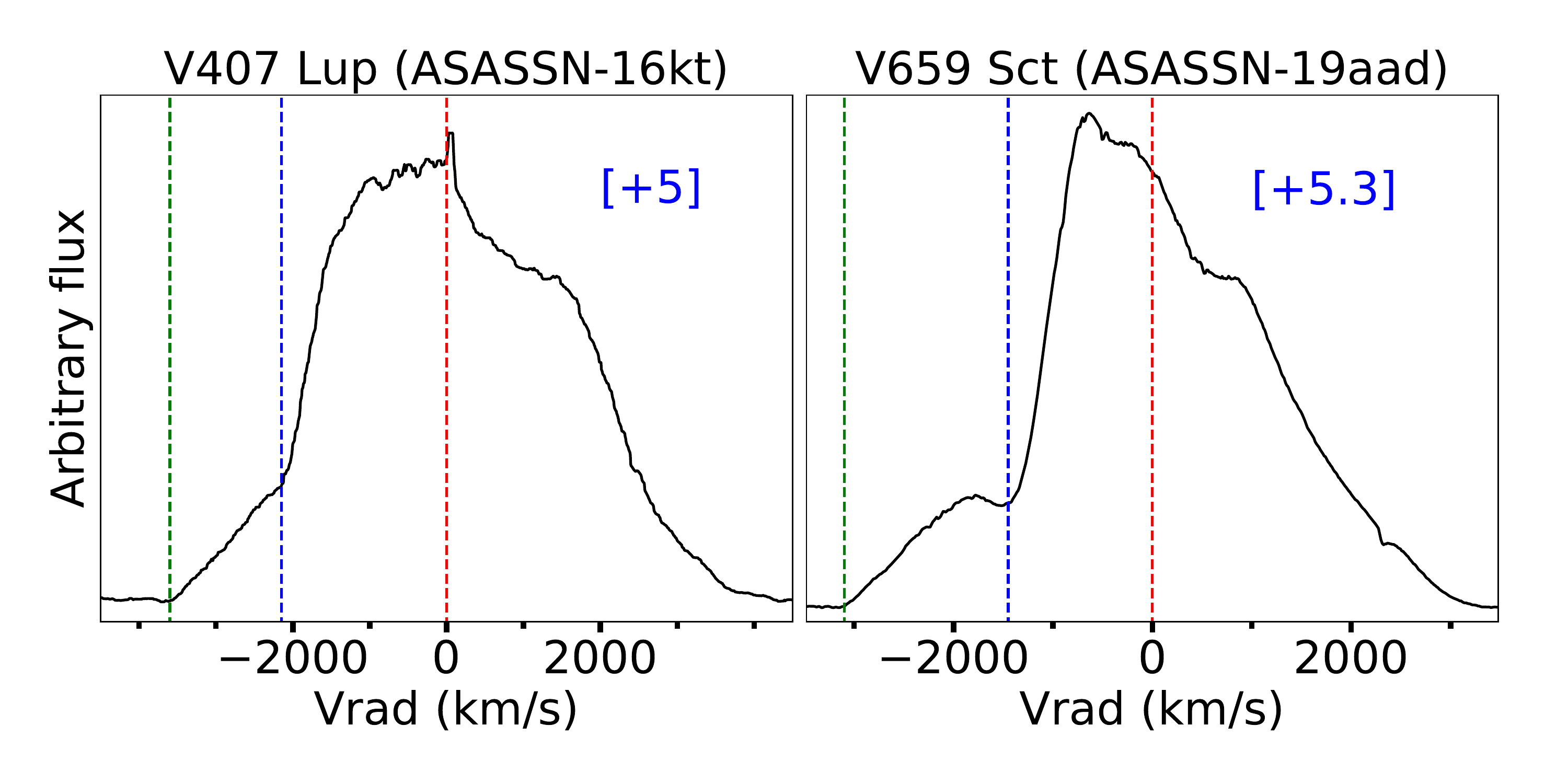}
\caption{The H$\alpha$ line profiles of the very fast novae V407~Lup (\textit{left}) and V659~Sct (\textit{right}), taken around 5 days after their respective optical peaks. Despite the very rapidly evolving light curves of these novae, we still observe the slow (blue dashed line) and fast (green-dashed line) components co-existing in their line profiles. The red-dashed line represents $v_{\mathrm{rad}}$ = 0\,km\,s$^{-1}$.}% Heliocentric correction is applied to the radial velocities.}
\label{Fig:fast_novae}
\end{center}
\end{figure}

For other very fast novae, observations before optical peak were not feasible, but the post-maximum spectra still show evidence of the slow and fast components. In Figure~\ref{Fig:fast_novae} we present two additional examples of very fast novae, V407~Lup (ASASSN-16kt) and V659~Sct (ASASSN-19aad). These novae have $t_2$ $\approx$ 3 days \citep{Aydi_etal_2018_2} and 6 days (based on AAVSO data), respectively. The H$\alpha$ line profiles of both novae show a slower P Cygni profile superimposed on top of a broader emission base. This is additional evidence that the two-flows scenario is common to all novae, regardless of their speed class.

\begin{figure*}
\begin{center}
  \includegraphics[width=0.65\textwidth]{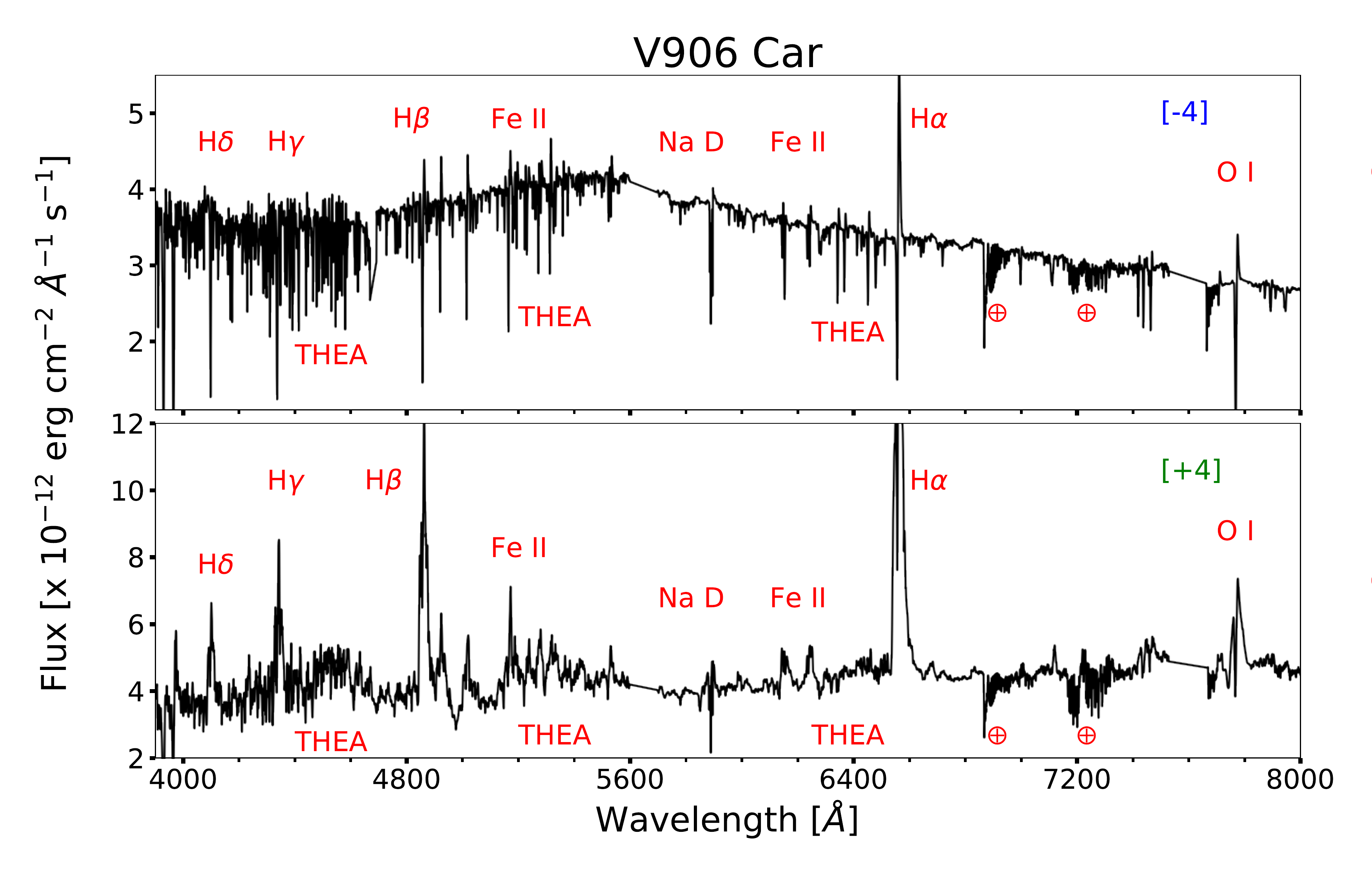}
  \vspace{-0.1in}
\caption{ UVES high-resolution spectra of \textbf{nova V906~Car}. The numbers in brackets are days relative to optical peak. Prominent lines are labelled, and telluric features are marked with an Earth symbol.}
\label{Fig:V906_Car_spec}
\end{center}
\end{figure*}

\subsection{The spectral evolution of nova V906 Car}
\label{V906_Car_sec}
Nova V906 Car (ASASSN-18fv) was discovered in eruption in March 2018 by ASAS-SN \citep{ATel_11454} and peaked at $\sim$5 mag in the optical. The nova was observed extensively from radio to $\gamma$-ray and has already been the topic of multiple studies \citep{Molaro_etal_2020,McLoughlin_etal_2020,Aydi_etal_2020,Pavana_etal_2020,Wee_etal_2020,Sokolovsky_etal_2020}. \citet{Aydi_etal_2020} detected a correlation between flares in its optical and $\gamma$-ray light curves, which led them to conclude that shocks are the source of a substantial fraction of the nova optical luminosity (in contrast to the standard picture of a nuclear-burning white dwarf powering the bolometric luminosity). High resolution optical spectra were obtained for this moderately fast nova during the rise and early decline, with a cadence of around 1 day, making it an ideal case to study the spectral evolution of novae near optical peak.

Figure~\ref{Fig:V906_Car_spec} shows two full spectra of V906~Car, obtained before and after optical peak. In addition to the typical nova emission lines of \eal{H}{I},  \eal{Fe}{II}, and \eal{O}{I}, the spectra are characterized by large numbers of absorption features, which we associate with THEA lines 
%(see e.g., \citealt{Williams_etal_2008,Williams_Mason_2010})
and which we discuss in \S~\ref{sec:v906_thea}. These lines appears less prominent in the post-maximum spectrum because at this stage the emission lines become very strong. 
%Using the \citet{Williams_2012} classification criteria, the spectrum is that of an \eal{Fe}{II}-type nova.

\subsubsection{The Balmer lines}
\begin{figure}
\begin{center}
  \includegraphics[width=\columnwidth]{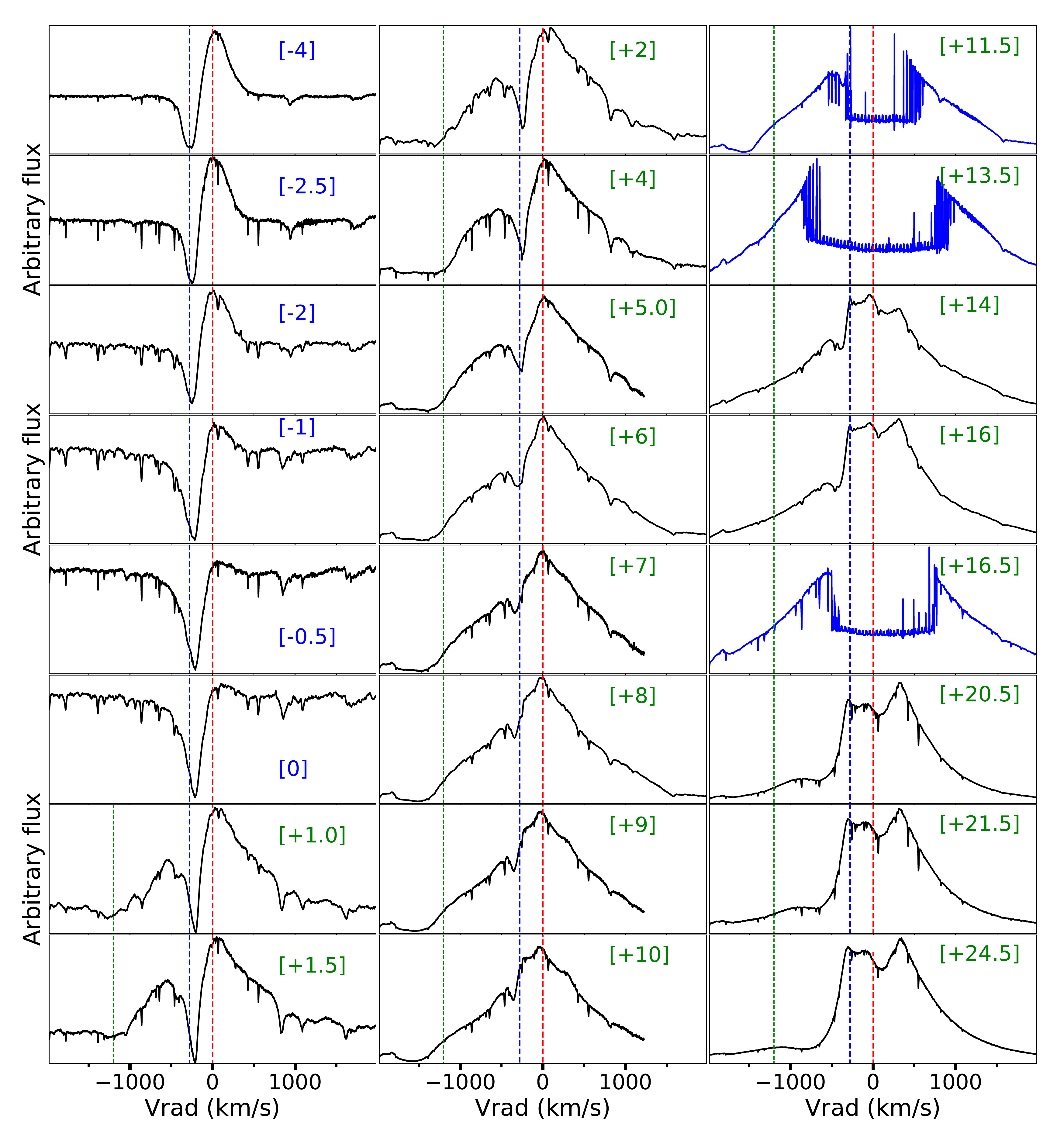}
\caption{The evolution of the H$\alpha$ line profile of \textbf{nova V906~Car}. The numbers in brackets are days relative to optical peak. The red vertical dashed line represents $v_{\mathrm{rad}}$ = 0\,km\,s$^{-1}$ (rest wavelength). The blue vertical dashed line marks a velocity of $-250$\,km\,s$^{-1}$ to highlight the velocity variation of the slow component. The green dashed line marks a velocity of $-1200$\,km\,s$^{-1}$ to highlight the velocity variation of the fast component. The profiles shown in blue with a centred ``notch'' are saturated and are shown to illustrate the changing width of the fast component (see \citealt{Aydi_etal_2020}).}
\label{Fig:V906_Car_profiles}
\end{center}
\end{figure}

\begin{figure}
\begin{center}
  \includegraphics[width=\columnwidth]{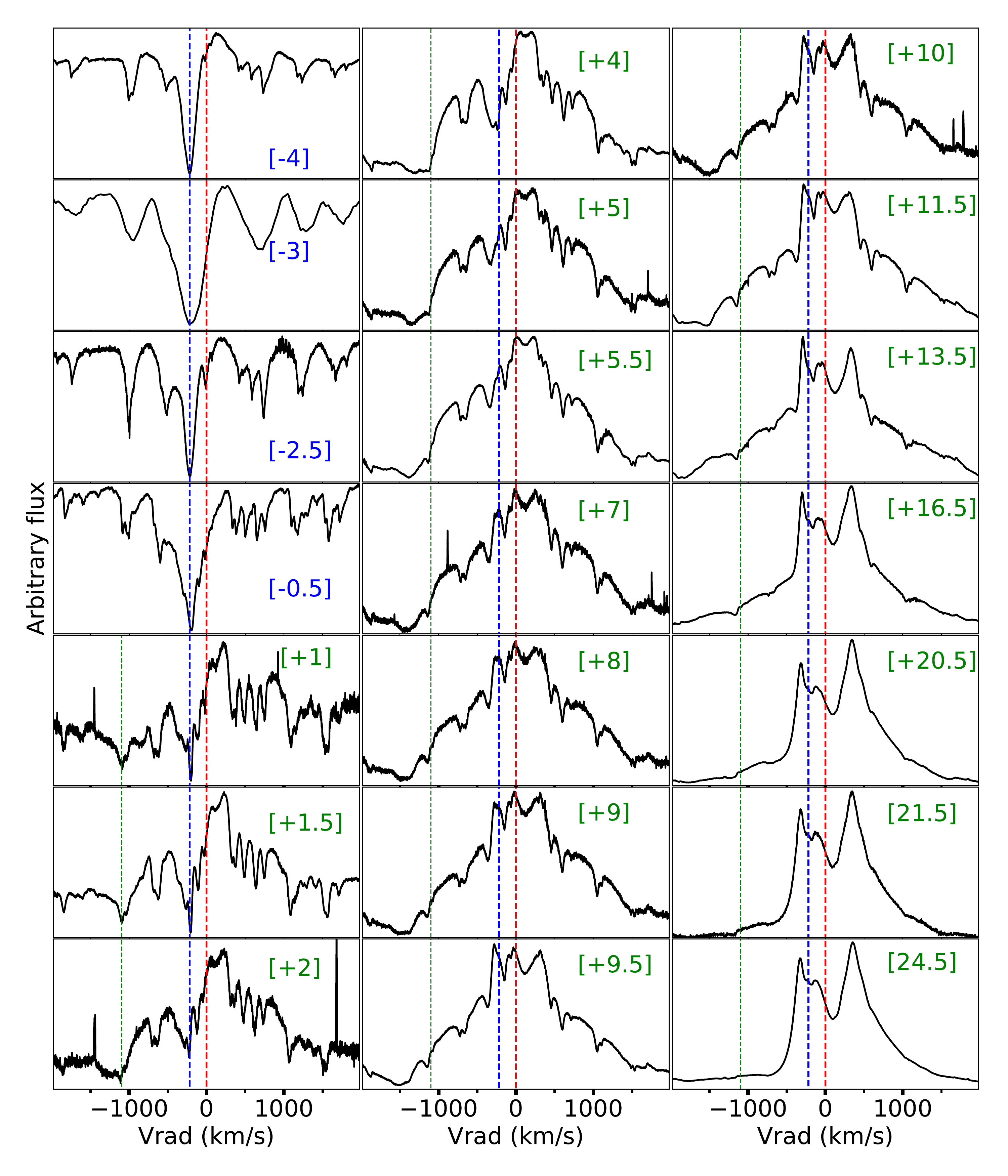}
\caption{The evolution of the H$\beta$ line profile of \textbf{nova V906~Car}. The red vertical dashed line represents $v_{\mathrm{rad}}$ = 0\,km\,s$^{-1}$ (rest wavelength). The blue vertical dashed line marks a velocity of $-200$\,km\,s$^{-1}$ to highlight the velocity variation of the slow component. The green dashed line marks a velocity of $-1100$\,km\,s$^{-1}$ to highlight the velocity variation of the fast component. The numbers in brackets are days relative to optical peak.}
\label{Fig:V906_Car_Hbeta}
\end{center}
\end{figure}

In Figures~\ref{Fig:V906_Car_profiles} and~\ref{Fig:V906_Car_Hbeta}, we plot the line profile evolution of H$\alpha$ and H$\beta$ for V906~Car. During the rise to maximum, 
%which was reached 10 days after the start of the eruption, 
the lines show P Cygni profiles with absorption troughs centered at blueshifted velocities of $\sim$ 200\,--\,250\,km\,s$^{-1}$ (the slow component). The emission components of the P Cygni profiles fade relative to the continuum during the rise to maximum, as the equivalent width (EW) of the emission components weakens from EW = $-5\,\mathrm{\AA}$ on day $-4$ to EW = $-0.3\,\mathrm{\AA}$ around peak.%We measure the equivalent widths (EW) of the emission components of the P Cygni profiles before optical peak for H$\alpha$ and we plot them in Figure~\ref{Fig:V906_Car_EW}.

Around a day after peak, a broad emission line component emerges with the P Cygni profile superimposed on top of it. The broad emission is accompanied by a broad and relatively shallow P Cygni absorption component at $\sim$ $-1200$\,km\,s$^{-1}$ (the fast component). Note how suddenly this fast component appears, between day 0 and 1.
The fast emission line component gradually broadens after optical peak and its P Cygni absorption moves blueward. %In Figure x we plot the evolution of the velocity of the fast component relative to the light curve. 
%Around 13 days after peak, the fast component shows a sudden dramatic increase in velocity. We discuss this in the following sections.
After the light curve peaks, the pre-maximum P Cygni profile changes gradually into a double peaked emission profile, which becomes prominent around 10 days after optical peak. %This could indicate that the expanding slow flow is becoming less optically thick while it expands and that it is moving preferentially in a specific direction (e.g., the orbital plane). %This means that the inclination of the system could affect the line profile evolution of the slow flow after optical peak. 

%\begin{figure}
%\begin{center}
%  \includegraphics[width=\columnwidth]{V906Car_LC_EW.pdf}
%  \vspace{-0.3in}
%\caption{The evolution of the Equivalent Width of the H$\alpha$ emission during the rise to optical peak of\textbf{nova V906~Car}.}
%\label{Fig:V906_Car_EW}
%\end{center}
%\end{figure}

\subsubsection{The \eal{Fe}{II} (42) and \eal{O}{I} lines}

\begin{figure}
\begin{center}
  \includegraphics[width=\columnwidth]{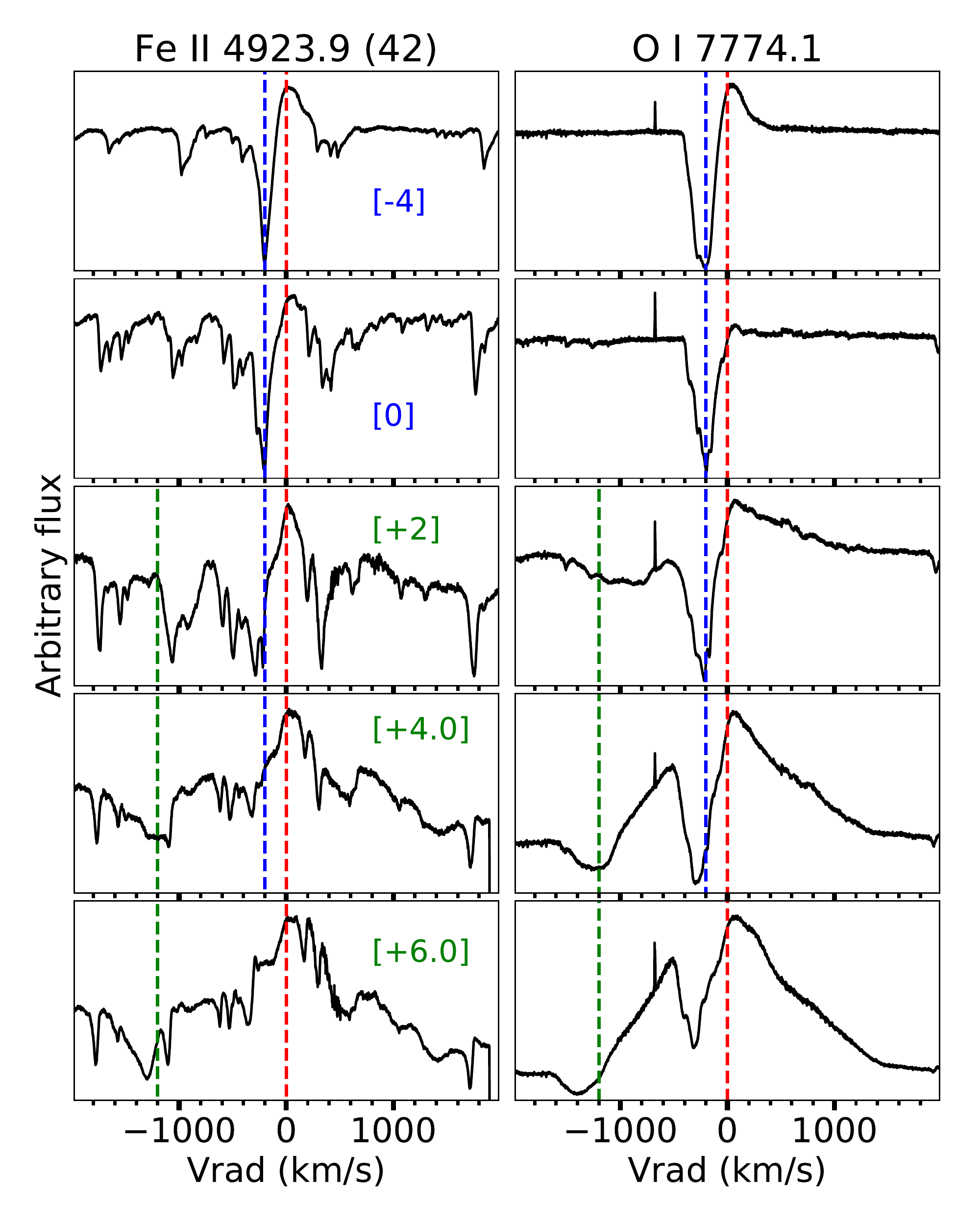}
\caption{The line profile evolution of one of the \eal{Fe}{II} (42) and \eal{O}{I} lines for \textbf{nova V906~Car}. The red vertical dashed line represents $v_{\mathrm{rad}}$ = 0\,km\,s$^{-1}$ (rest wavelength). The blue vertical dashed line marks a velocity of $-200$\,km\,s$^{-1}$ highlighting the slow component. The black dashed line marks a velocity of $-1200$\,km\,s$^{-1}$ highlighting the fast component.}
\label{Fig:FeII_OI_broad}
\end{center}
\end{figure}

The other prominent lines in the spectrum such as the \eal{Fe}{II} (42) multiplet and \eal{O}{I} lines show the same spectral evolution as the Balmer lines (Figure~\ref{Fig:FeII_OI_broad}). They start with pre-maximum slow P Cygni profiles, and later develop another broader
P Cygni component, while the original slow absorption remains superimposed on the fast component, showing that this evolution is consistent across the different prominent lines. 

\subsubsection{The intermediate component}

A closer look at the low velocities of the prominent lines shows that at optical peak a new distinct component appears at a velocity of around $-300$\,km\,s$^{-1}$, co-existing with the slower pre-maximum component at $-200$\,km\,s$^{-1}$ (see Figure~\ref{Fig:THEA_Balmer} where we plot the \eal{Na}{I} D, \eal{H}{I}, \eal{Fe}{II}, and \eal{O}{I} lines, focusing on the lower velocities). This intermediate velocity system is what McLaughlin calls the ``principal system'' \citep{McLaughlin_1944,Payne-Gaposchkin_1957} and we call it the ``intermediate component'', hereafter. While initially blended with the slower pre-maximum component, a few days after peak the intermediate component starts replacing the slower, pre-maximum component. Moreover, both components move blueward, showing a gradual acceleration after peak. In Section~\ref{sec_disc} we interpret the origin of the intermediate component.

For the \eal{Na}{I} D line the slow component disappears after a couple of days from peak, while for the Balmer lines the slow component lasts a bit longer and stays dominant till around 4 days
%6 days 
after peak, when the intermediate component starts replacing it. This means that comparing the Balmer line evolution near optical peak for other novae and assuming that the same slow component is still superimposed on the fast component a few days after optical peak is reasonable.

\subsubsection{The THEA lines}\label{sec:v906_thea}

The spectra of nova V906~Car show a large number of THEA lines present from the first epoch obtained 4 days before optical peak (see Figure~\ref{Fig:V906_Car_spec}). We show some of the THEA lines in comparison with the \eal{Na}{I} D2 line at 5889.9\,$\mathrm{\AA}$ in Figure~\ref{Fig:THEA_1}. More THEA lines are plotted in Figures~\ref{Fig:THEA_2_3}\,--\,\ref{Fig:THEA_8_9}.

\begin{figure*}
\begin{center}
  \includegraphics[width=0.5\textwidth]{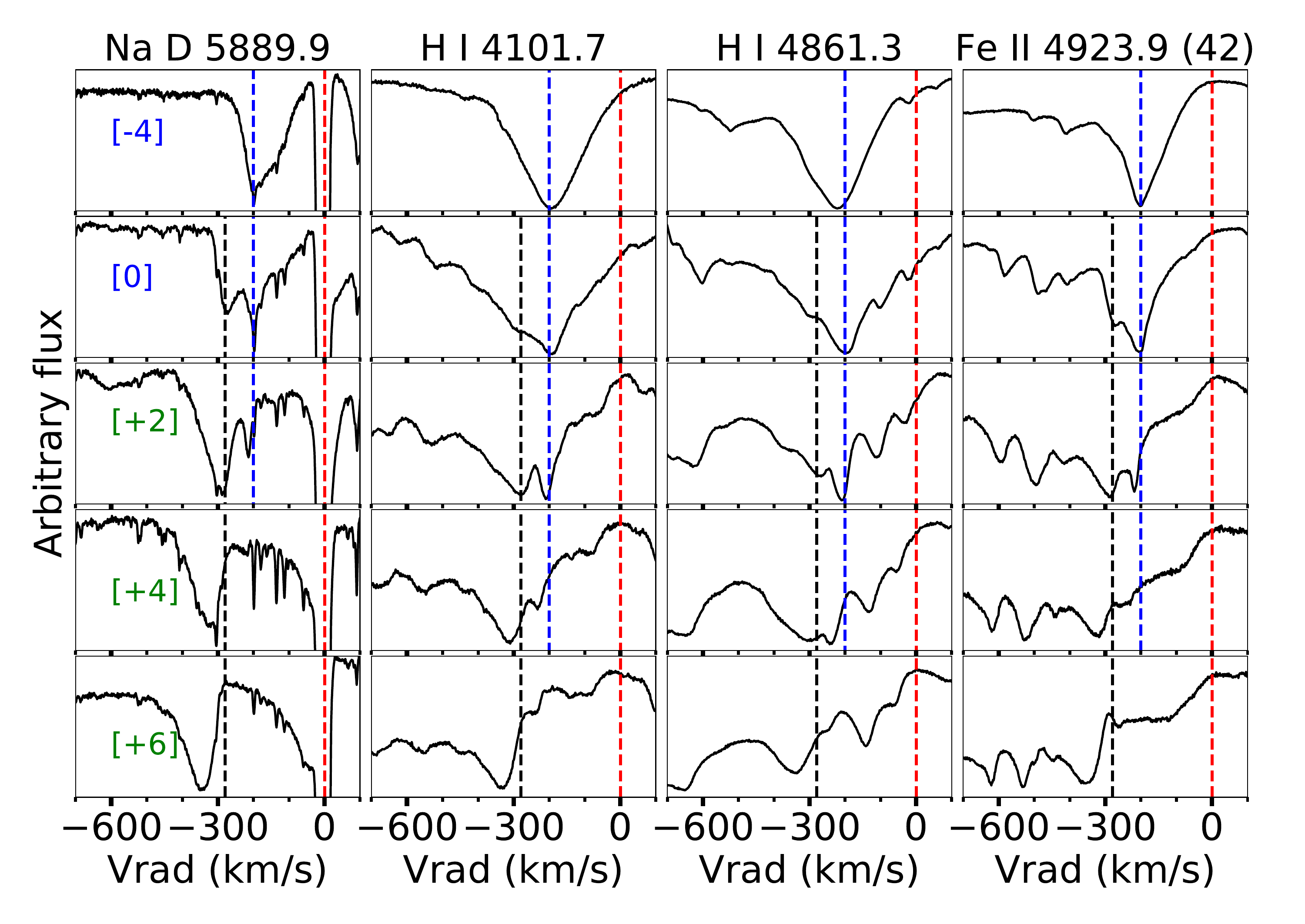}
  \hspace{-0.2in}
  \includegraphics[width=0.5\textwidth]{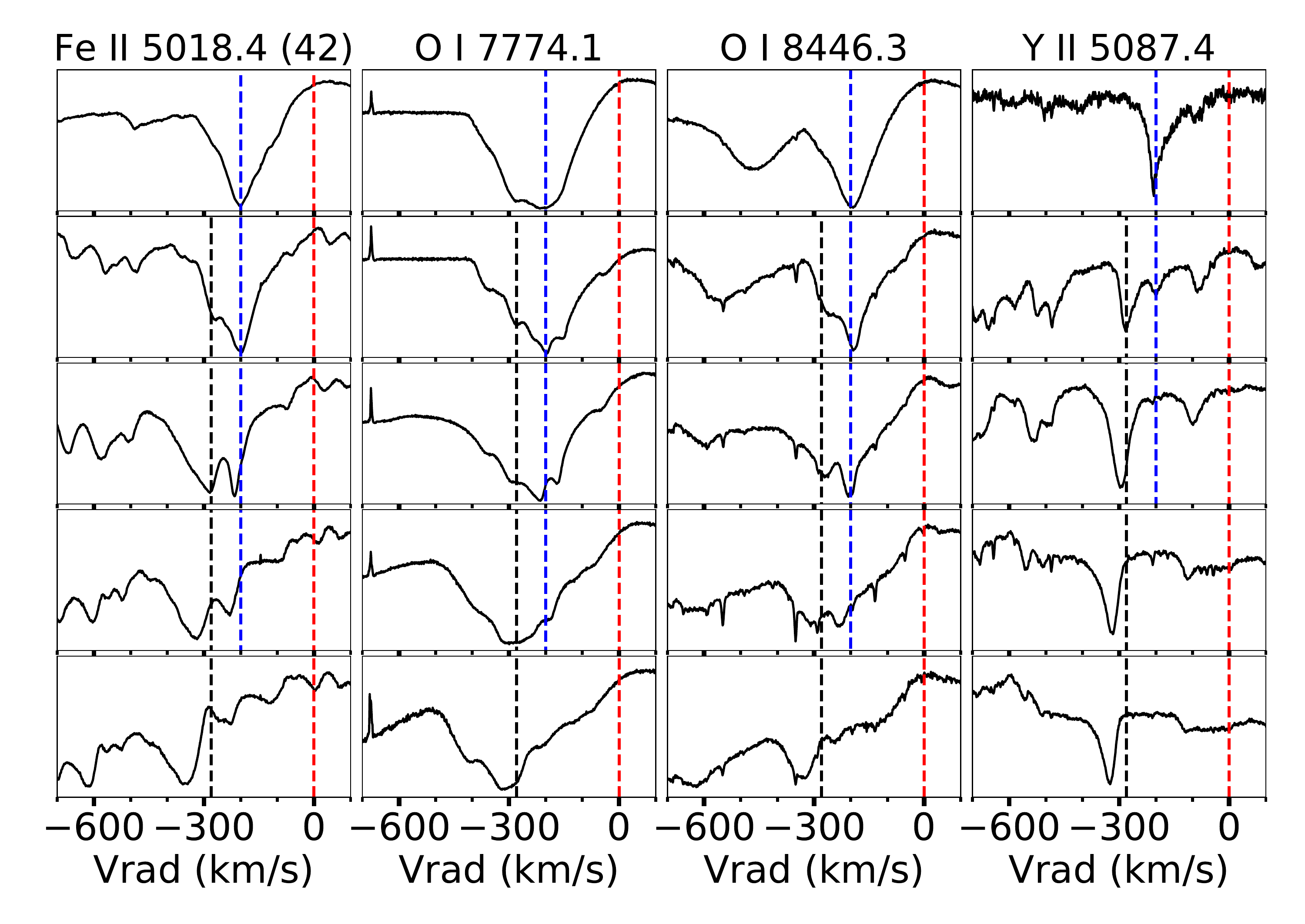}
\caption{The line profile evolution of some of the Balmer, \eal{Fe}{II} (42), and \eal{O}{I} lines in comparison with one of the THEA lines and \eal{Na}{I} D2 for \textbf{nova V906~Car}. The red vertical dashed line represents $v_{\mathrm{rad}}$ = 0\,km\,s$^{-1}$ (rest wavelength). The blue vertical dashed line marks a velocity of $-200$\,km\,s$^{-1}$ highlighting the slow component. The black dashed line marks a velocity of $-300$\,km\,s$^{-1}$ highlighting the intermediate component.}
\label{Fig:THEA_Balmer}
\end{center}
\end{figure*}

\begin{figure*}[ht]
\begin{center}
  \includegraphics[width=0.63\textwidth]{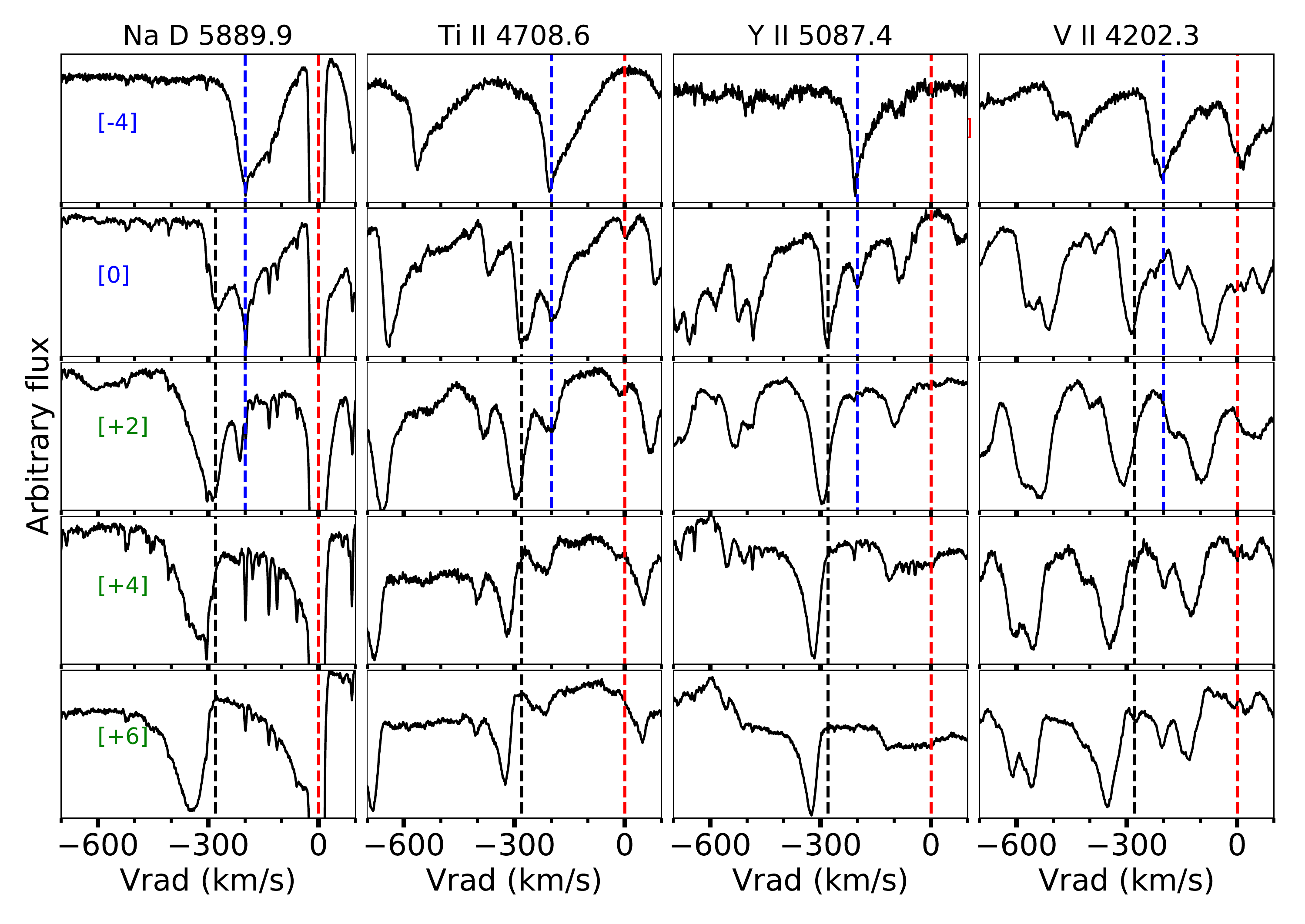}
\caption{The line profile evolution of a sample of THEA lines plotted in comparison to \eal{Na}{I} D2 at 5889.9\,$\mathrm{\AA}$ for \textbf{nova V906~Car}. The red vertical dashed line represents $v_{\mathrm{rad}}$ = 0\,km\,s$^{-1}$ (rest wavelength). The blue vertical dashed line marks a velocity of $-200$\,km\,s$^{-1}$ highlighting the slow component. The black dashed line marks a velocity of $-300$\,km\,s$^{-1}$ highlighting the intermediate component.}
\label{Fig:THEA_1}
\end{center}
\end{figure*}

Four days before peak, the lines show an absorption feature at around $-200$\,km\,s$^{-1}$ (consistent with the slow component seen in other lines). Around optical peak a new system appears at a velocity of around $-300$\,km\,s$^{-1}$, co-existing with the slower feature (the intermediate component). %This new system is what McLaughlin call the ``principal system'' \citep{McLaughlin_1944,Payne-Gaposchkin_1957} and we call it the ``intermediate component'', hereafter.
A couple of days after peak, the intermediate component replaces the slower, pre-maximum system and shows a gradual acceleration.

Unlike the \eal{Na}{I} D, \eal{H}{I}, \eal{Fe}{II}, and \eal{O}{I} lines, we do not detect the ``fast component'' of $\sim -$1200\,km\,s$^{-1}$ in the THEA lines. In Figure~\ref{Fig:THEA_long_range} we show some THEA line profiles plotted with an extended velocity scale, to $-$1500\,km\,s$^{-1}$ to show the absence of a fast component. It is possible that the THEA lines do not develop a fast component, or 
%It could be 
that the fast component in the THEA lines is too weak to be detected. We compare the EW of the pre-maximum components of the \eal{Fe}{II}~(42) lines with the EW of the same components in THEA lines. The THEA EW are 0.1--0.6$\times$ the EW of the \eal{Fe}{II} lines. Therefore, if the line ratio between the slow and fast components in a specific transition is the same for all transitions (e.g., the EW ratio between the slow and fast components of \eal{Fe}{II} 5018 is equal to that of the THEA lines), we should expect to detect the fast component in some THEA lines, particularly the stronger ones. However, the optical depth of some transitions of \eal{Fe}{II} (42) is larger than for the THEA lines and therefore might affect this ratio. Also, the difference in densities and possibly abundances between the fast and slow flows might be another reason for why we do not observe a fast component in certain transitions, particularly the THEA lines.

Except for the fast component, the THEA lines show the same evolution as the \eal{Na}{I} D, \eal{Fe}{II} (42), and \eal{O}{I} lines (Figure~\ref{Fig:THEA_Balmer}). That is, they develop the same slow and intermediate components at approximately the same velocities at approximately the same time.

\subsection{The pre-maximum spectral evolution of THEA lines in nova FM~Cir}
\label{FM_Cir_sec}
\begin{figure*}[ht]
\begin{center}
  \includegraphics[width=0.7\textwidth]{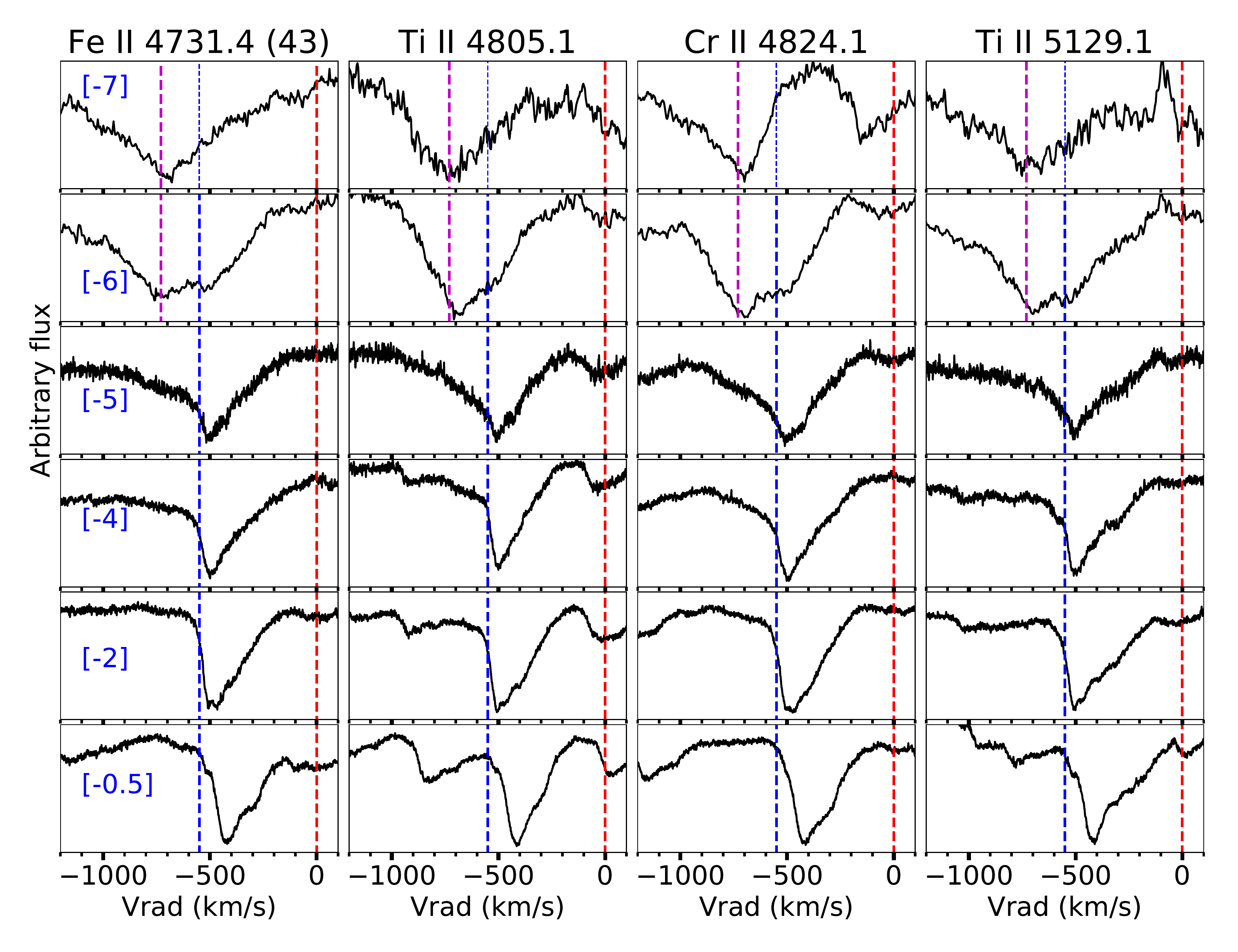}
  \vspace{-0.1in}
\caption{The pre-maximum line profile evolution of a sample of THEA lines for \textbf{nova FM~Cir}. The red vertical dashed line represents $v_{\mathrm{rad}}$ = 0\,km\,s$^{-1}$ (rest wavelength). The blue vertical dashed line marks a velocity of $-550$\,km\,s$^{-1}$ highlighting the slow component. The magenta vertical dashed line marks a velocity of $-730$\,km\,s$^{-1}$. The numbers between brackets are days relative to optical peak.}
\label{Fig:THEA_FM_Cir}
\end{center}
\end{figure*}

Nova FM~Cir was another bright nova (peaking at $\sim$5 mag), discovered in January 2019 by
\citep{2018CBET.4482....1S}. The nova rose to peak in around 9 days, allowing dedicated pre-maximum spectroscopy on a daily cadence. In Figure~\ref{Fig:THEA_FM_Cir} we present the evolution of some of the THEA lines during the rise to optical peak. %, obtained by the CHIRON spectrograph on the 1.5m SMARTS telescope.
Initially we observe a broad absorption component at a velocity of around $-730$\,km\,s$^{-1}$ (highlighted in magenta), which disappears in $\sim$1 day, as the narrower slow component (at a velocity of $-550$\,km\,s$^{-1}$) becomes dominant. While rising to peak, the absorption troughs of the slow component moves redward, receding to a velocity of around $-400$\,km\,s$^{-1}$ near peak. The implications of this apparent deceleration is discussed in the following sections. A gap in our spectroscopic coverage after peak, followed by secondary maxima, complicates the later spectral evolution and identification of the principal component for this nova.

%In Figure~\ref{Fig:Balmer_FM_Cir} we show the line profile evolution of H$\beta$ and the \eal{Fe}{II} (42) lines before optical peak. The lines show P Cygni profiles with diffuse absorption troughs at a velocity of around $-850$\,km\,s$^{-1}$. These diffuse absorption troughs are merged with the slow component, which has a velocity of around $-550$\,km\,s$^{-1}$. Similarly to the THEA lines, the diffuse absorptions disappear as the slow component  dominates and becomes more narrow with time. While rising to peak, the absorption troughs of the slow component move redward, reaching a velocity of  around $-400$\,km\,s$^{-1}$ near peak. 

\iffalse
\begin{figure*}[ht]
\begin{center}
  \includegraphics[width=0.7\textwidth]{Balmer_FM_Cir.pdf}
  \vspace{-0.1in}
\caption{The pre-maximum line profile evolution of H$\beta$ and the \eal{Fe}{II} (42) lines for \textbf{nova FM~Cir}. The red vertical dashed line represents $v_{\mathrm{rad}}$ = 0\,km\,s$^{-1}$ (rest wavelength). The blue vertical dashed line marks a velocity of $-550$\,km\,s$^{-1}$ highlighting the slow component. The magenta vertical dashed line marks a velocity of $-850$\,km\,s$^{-1}$. The numbers between brackets are days relative to optical peak.}
\label{Fig:Balmer_FM_Cir}
\end{center}
\end{figure*}
\fi

\subsection{Additional conclusions about the early spectral evolution of novae}
\begin{figure*}[!t]
\begin{center}
  \includegraphics[width=0.48\textwidth]{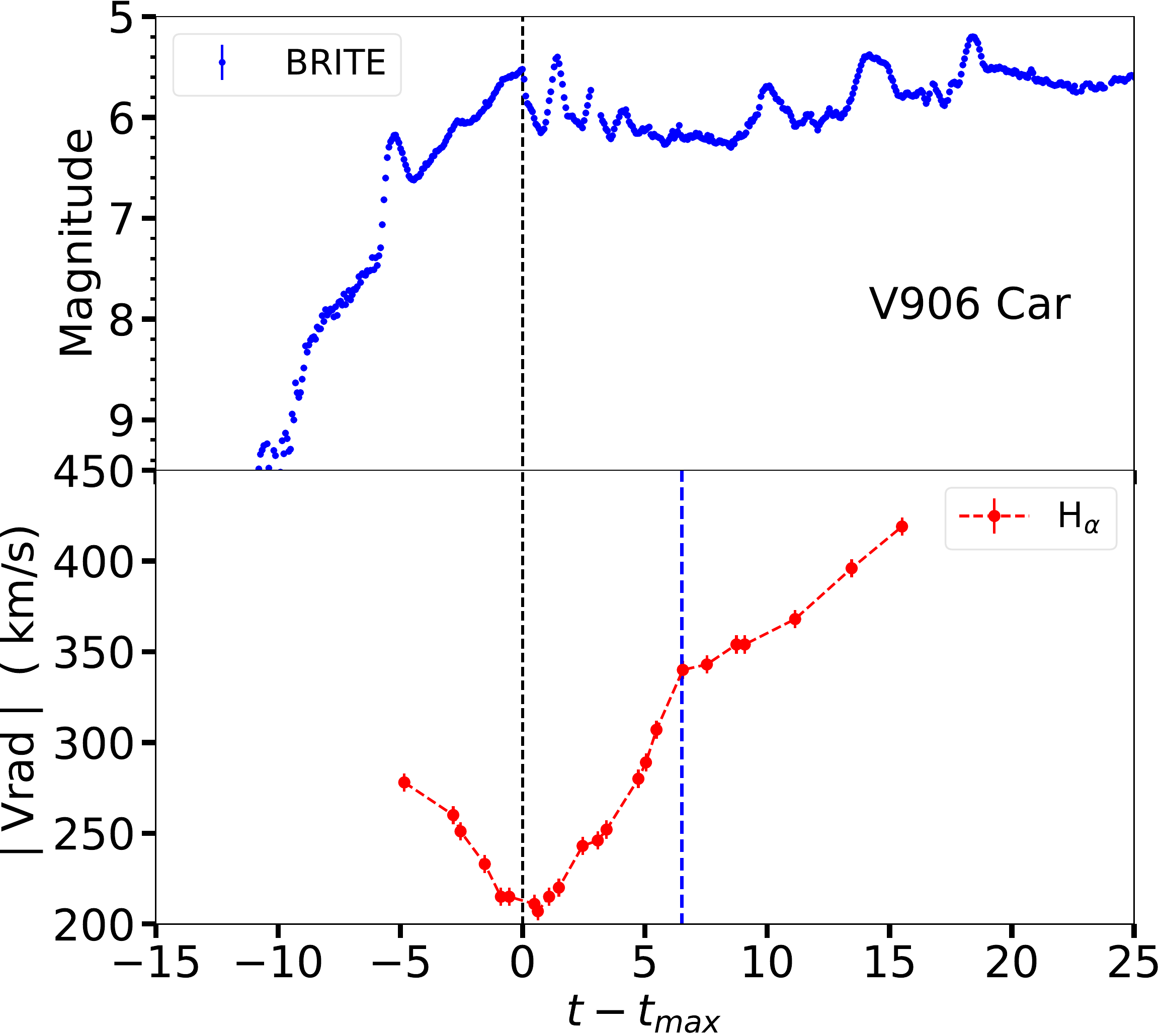}
  \includegraphics[width=0.48\textwidth]{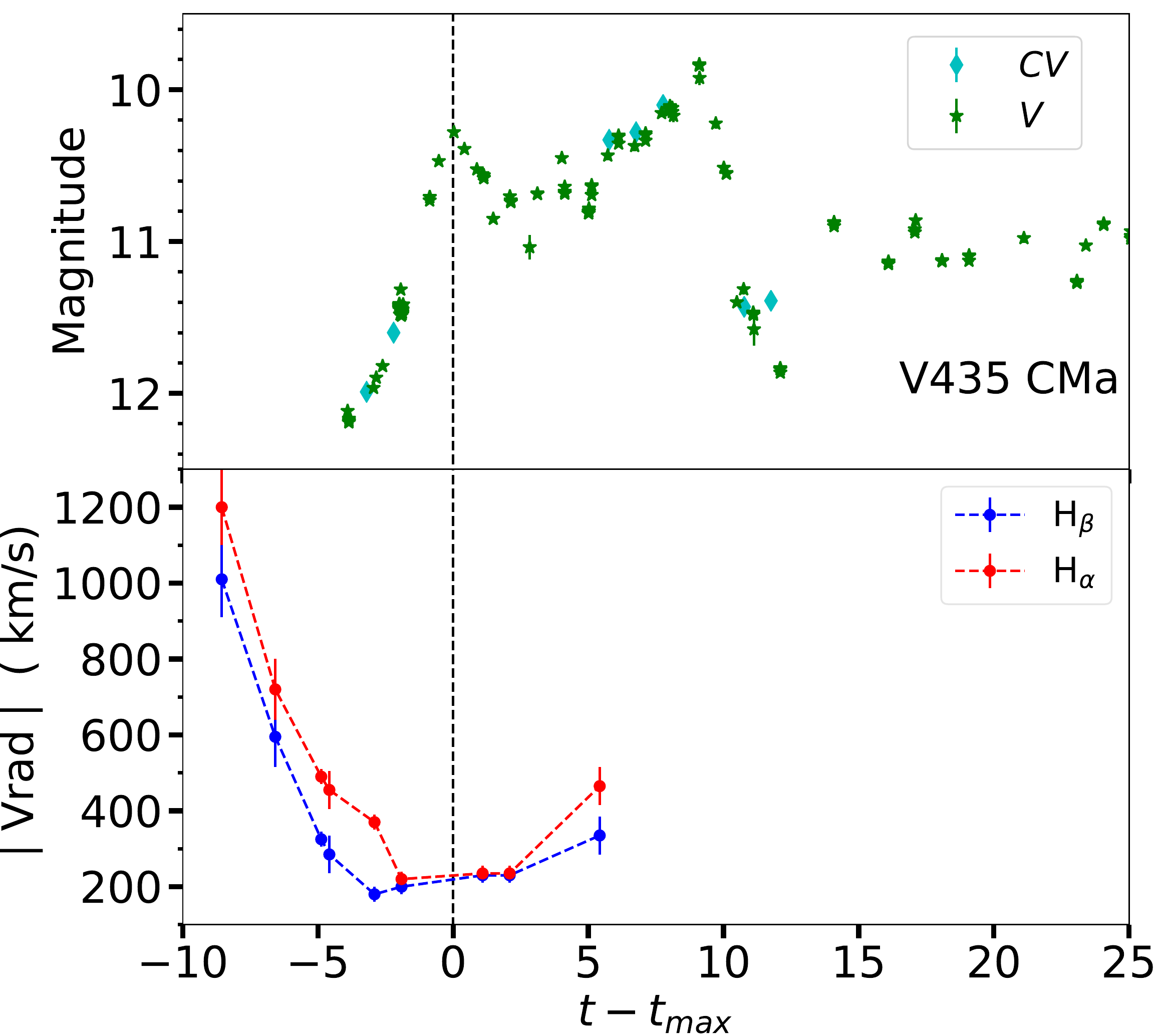}
  \includegraphics[width=0.48\textwidth]{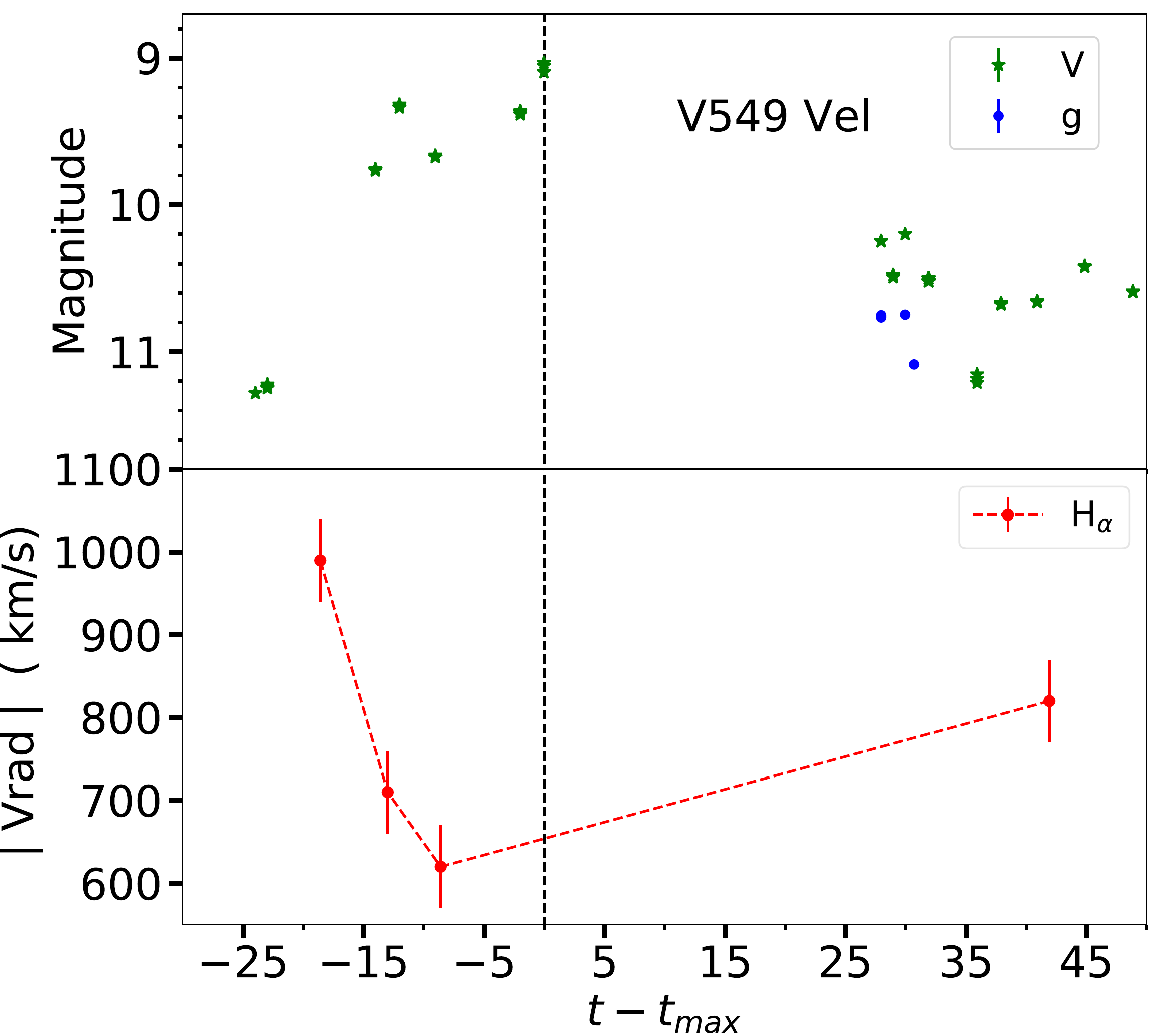}
  \includegraphics[width=0.48\textwidth]{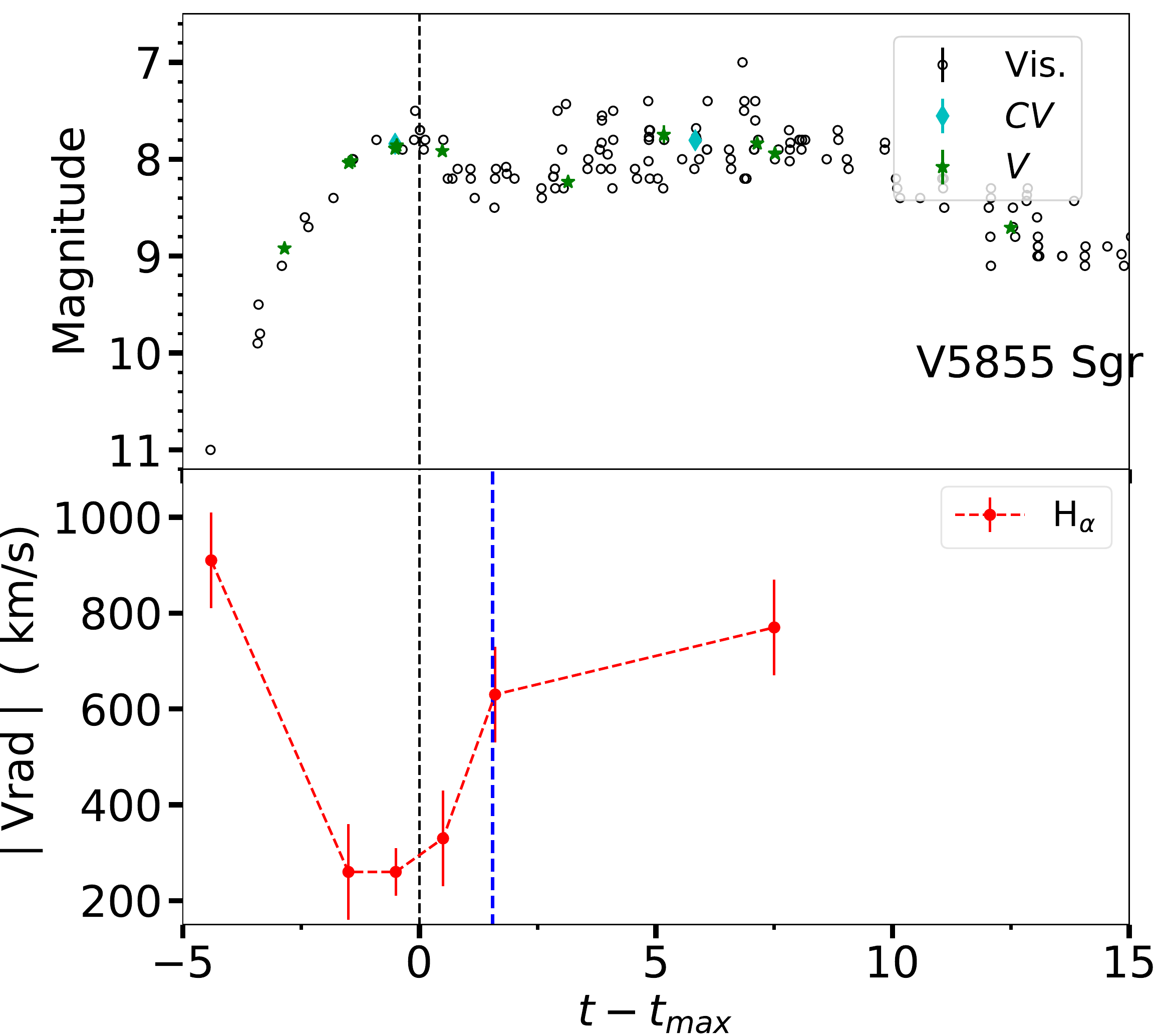}
\caption{For each of four novae, the $V/g$-band light curve \emph{(top panel)} is compared with the evolution of velocity for the slow H$\alpha$ and/or H$\beta$ component \emph{(bottom panel}; measured as the velocity of the absorption trough center). Novae V906~Car, V435~CMa, V549~Vel, an V5855~Sgr are shown. The black dashed lines represent the date of the optical peak. For V906~Car we use the BRITE optical light curve from \citet{Aydi_etal_2020}; all others are from AAVSO/ASAS-SN (\S~\ref{sec_lc}). The blue dashed lines in some panels represent the tentative replacement of the slow component by the intermediate component.}
\label{Fig:LC_abs}
\end{center}
\end{figure*}

In this section we list some of the aspects of the spectral evolution we found in novae V906~Car, FM~Cir, and the other novae in our sample. For some novae, particularly those with slowly evolving light curves, it was feasible to obtain multiple spectra before the light curve reached its optical peak. In Figure~\ref{Fig:LC_abs} we present the evolution of the velocity of the absorption trough of the slow component of H$\alpha$ and/or H$\beta$ in comparison to the optical light curves for novae V906~Car, V453~CMa, V5855~Sgr, and V459~Vel. The full evolution of the line profiles for the last three novae are also presented in Figures~\ref{Fig:V435_CMa_profiles} and~\ref{Fig:V5855_Sgr_profiles}. In addition to the early slow component, followed by---and superimposed upon---a fast component (Section~\ref{two_flows}), we find several other intriguing patterns in their spectral evolution:
\begin{itemize}
    \item A pre-maximum deceleration of the slow component. The trough of the blue-shifted absorption feature moves redward
    %absorption features of the P Cygni profiles show a deceleration 
    while the nova is rising to its peak (Figure~\ref{Fig:LC_abs}).%. This has been observed in several novae with multiple observations before the peak
    
    %\item The emission feature of the slow component P~Cygni profile fades relative to the continuum while the nova is rising to its peak (see Figure~\ref{Fig:V906_Car_EW}, as well as Figures~\ref{Fig:V906_Car_profiles} and~\ref{Fig:V435_CMa_profiles}).    
    
    \item A gradual post-maximum acceleration of the slow component. After the nova reaches optical peak, the troughs of the blue-shifted absorption features of the slow component move blueward. We observed the same acceleration in the intermediate component as well, in nova V906~Car. Nova V5855~Sgr shows a sudden large increase in the velocity of the slow P Cygni absorption component 2 days after peak (see Figure~\ref{Fig:LC_abs}). This is likely due to the slow component being replaced by the intermediate component. A distinction between the slow and intermediate components was not feasible for this nova due to the poor cadence. \citet{McLaughlin_1944} noted that dedicated monitoring is needed to simultaneously detect the slow and intermediate components. 
    
    \item A post-maximum acceleration of the fast component. Similar to the other components, the fast component appears to accelerate (the emission line base gradually broadens and the broad absorption feature moves blueward; see Figures~\ref{Fig:V906_Car_profiles} and~\ref{Fig:V906_Car_Hbeta}). This line broadening has been observed in many novae after peak (see e.g., \citealt{Friedjung_2011} and references therein),
    %as the emission line base gradually broadens (see Figure~\ref{Fig:V906_Car_profiles}).
    % similarly to the slow component, the fast component show acceleration as the emission base exhibits gradual broadening
    
    %\item The earliest spectra of nova FM~Cir show diffuse absorptions merged with the slow component at slightly blueward velocities and disappear in a couple of days (Figure~\ref{Fig:Balmer_FM_Cir}). Something similar is observed in the early spectra of nova V435~CMa (Figure~\ref{Fig:V435_CMa_profiles}). We suggest that such diffuse absorptions are possibly from a small body of gas, impulsively ejected during the thermonuclear runaway and dissipating rapidly during the early days of the eruption. Such absorptions are possibly missed in other novae and can only be detected during the first few days of the eruption. 
    
    %\item Some novae, particularly ones with multiple maxima (e.g., ASASSN-17pf and V1369 Cen), show new systems of absorption features whose appearance somtimes correlates with peaks in the light curve (see, e.g., \citealt{Aydi_etal_2019_I,Walter_2016}). The origin of these multiple absorption systems has been discussed in other works \citep{Tanaka_etal_2011,Mason_etal_2018,Aydi_etal_2019_I} and will not be explored here.

\end{itemize}

\section{Discussion}
\label{sec_disc}

\subsection{A universal ejection scenario}\label{sec:universal}

The co-existence of the pre-maximum P Cygni profile (slow component) on top of the broad emission lines (fast component), and the large velocity difference between these two spectral components, lead us to conclude that there are at least two physically distinct flows. Since all the novae in our sample follow the same spectral evolution, we suggest that this behavior is common and may even be ubiquitous in classical novae. 

Our findings here echo the results of \citet{Mclaughlin_1947}, who states:
\begin{quote}
\emph{The broad emissions of the ``diffuse-enhanced'' system extend across the emission and absorption of the same lines from the principal shell. Nevertheless, the principal absorption remains strong and well defined, without the filling in that would surely occur if the atoms that produce the ``diffuse-enhanced'' emission were outermost.} 
\end{quote}
McLaughlin concludes that ``\emph{The `diffuse-enhanced' emission and absorption must be assigned to a P Cygni-like expanding atmosphere close to the central star and wholly inside the principal shell.}''
Decades worth of progress in spectroscopic observations leave these basic conclusions unchanged. In order for the slow absorption to be superimposed on the broad emission, much of the ejecta associated with the fast or diffuse-enhanced components must be located at smaller radius, compared to the slow or pre-maximum ejecta. 

%Coupling these smaller radii with the higher expansion velocities of the fast component, the implication is that the fast component must be expelled at a later time, relative to the slow component. This implies that the post-maximum appearance of the fast component in spectra may actually coincide with the launching of the fast component.

\citet{McLaughlin_1964} pictured the two ejections as spheres, with the fast component entirely within the slow component. However, given their relative velocities,
%if the morphology of the ejecta is more complex, then portions of 
the fast component should quickly catch up with and grow beyond the confines of the slow component. Let us take, for example, V906~Car. The slow component expands at $\sim$200 km~s$^{-1}$ and we assume that it is expelled at the beginning of optical rise (11 days before optical peak; \citealt{Aydi_etal_2020}). The fast component is observed to begin expanding $\sim$ 12 days later, a couple of days after peak, at $\sim$1200 km~s$^{-1}$. This implies that the fast component would catch up with the slow component just 2.4 days after the launch of the fast component (i.e., 14.4 days after the start of the optical eruption). The fact that the fast component is not clearly seen to decelerate around this time, and instead actually accelerates (Figure \ref{Fig:V906_Car_profiles}), implies that some portions of the fast flow continue to freely expand beyond the radius of the slow flow.

\begin{figure*}[!t]
\begin{center}
  \includegraphics[width=0.8\textwidth]{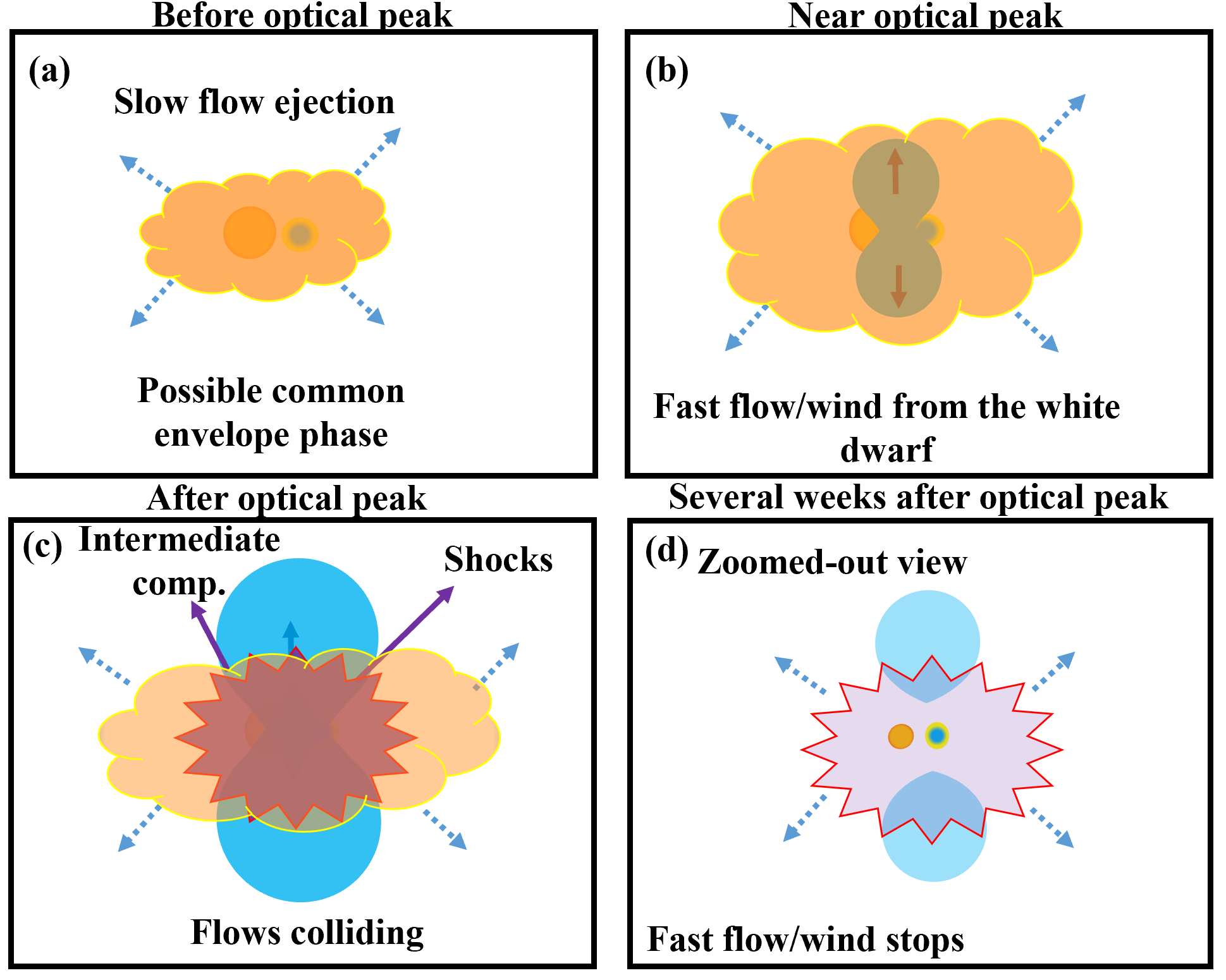}
\caption{Our proposed universal model of nova ejection: (a) before optical peak the accreted envelope puffs up due to energy output from the thermonuclear reactions, engulfing the system in a common envelope phase and becoming concentrated in the equatorial plane (e.g., \citealt{Livio_etal_1990,Chomiuk_etal_2014,Sokoloski_etal_2017}). This is the slow or pre-maximum component.
(b) A continuous fast wind starts, driven by radiation from the ongoing nuclear burning on the surface of the white dwarf (e.g., \citealt{Bath_Shaviv_1976,Kato_Hachisu_1994}). The fast flow could propagate more freely in the polar directions due to the oblate shape of the pre-existing slower ejecta. 
(c) As the fast flow collides with the slow flow, it leads to shock interaction and physically accelerates the slow ejecta. This interaction could form a shell which sweeps through the slow flow and is the origin of the intermediate spectroscopic component \citep{Friedjung_1987}. This same interaction could power the $\gamma$-ray and a substantial fraction of the optical emission of the nova \citep{Aydi_etal_2020}. (d) Several days/months after optical peak, the continuous ejection of the fast outflow stops, and the ejecta dissipate and become optically thin to the remnant nuclear burning on the white dwarf, observable as a supersoft X-ray source.} %Depending on the inclination of the system relative to the observer, either the slow or the fast component could dominate the emission line profiles.}
\label{Fig:nova_ejection}
\end{center}
\end{figure*}

Therefore, the spectroscopic observations imply that the two ejections are aspherical, as illustrated in  Figure~\ref{Fig:nova_ejection}. 
%two-flows scenario is consistent with the spectral evolution of the  nova sample we present here and is illustrated in  Figure~\ref{Fig:nova_ejection}.
Although the fast flow begins after the expansion of the slow component, it is able to expand relatively freely and unimpeded in the polar directions.
Absorption associated with the slow component weakens over time, as the emitting area of the fast component expands beyond the extent of the slow flow.

\citet{Chomiuk_etal_2014}, \citet{Metzger_etal_2015} and \citet{Li_etal_2017_nature} have suggested a similar scenario to explain high-resolution images, $\gamma$-ray emission, and shock formation in novae.
%Similarly, \citet{Metzger_etal_2015} and \citet{Li_etal_2017_nature}, adopted the same two-flows scenario to explain the $\gamma$-ray emission and shock formation in novae. %This two-flows scenario is consistent with what we observe in the early spectral evolution of our nova sample.
%This two-flows scenario is consistent with the spectral evolution of the  nova sample we present here and is illustrated in  Figure~\ref{Fig:nova_ejection}. %we present an illustration of the suggested scenario.%, which consists of at least two phases of mass-loss. 
Initially, parts of the accreted envelope expand due to the energy of the thermonuclear runaway, and engulf the binary system in a common envelope stage.
%Depending on the energetics of the thermonuclear runaway, parts of the envelope could be ballistically ejected (e.g., \citealt{Sparks_etal_1978, Mason_etal_2018}), or
%Parts of the expanding envelope remains gravitationally bound to the binary (e.g., \citealt{Starrfield_etal72}). In this case 
%As parts of the envelope might remain gravitationally bound to the system (e.g., \citealt{Starrfield_etal72}), 
The binary motion might help to expel the envelope due to frictional drag (e.g., \citealt{MacDonald_etal_1985, Shankar_1991}, though see \citealt{kato_Hachisu_1991a,kato_Hachisu_1991b} for another view), or at least direct the ejection,  concentrating it in the orbital plane 
%and shaping it into an oblate morphology or torus 
\citep{Livio_etal_1990, Lloyd_etal_1997}. 
This  manifests as a slow flow with P Cygni spectral line profiles and velocities $\lesssim$1000 km\,s$^{-1}$, apparent from the earliest times in the eruption (i.e., during the light curve rise to its peak). 

%\cite{kato_Hachisu_1991a,kato_Hachisu_1991b} suggest that have found that near the binary, the density of the envelope is too low for frictional drag to have a significant effect and it is possible that radiation-driven winds are enough to expel the envelope \citep{Kato_Hachisu_1994}. 

%Since we observe P Cygni profiles with rounded emission features in all novae, this means that the slow flow is either spherical or oblate. It is unlikely that the slow flow is bipolar, otherwise the inclination of the system would lead to a diversity in the structures of the line profiles between different novae. On the other hand, for an oblate flow, different inclinations would only lead to differences in the structure of the absorption feature. Indeed, we observe a diversity of absorption feature shapes between the novae, suggesting that the slow flow might have an oblate morphology (Figures~\ref{Fig:line_profiles_slow}\,--\,\ref{Fig:line_profiles_fast}).

The slow flow is followed by a faster wind,
%manifesting as a broad base in the emission line profiles, often accompanied by a blueshifted absorption. The fast flow
which propagates more freely in the bipolar direction since the slower flow is concentrated in the equatorial plane. 
The origin of the fast flow is possibly a radiation-driven wind from the continuous nuclear burning on the surface of the white dwarf \citep{Bath_Shaviv_1976,Kato_Hachisu_1994}. As discussed in Section \ref{sec:intermediate}, the interaction of the two flows gives rise to high energy emission and additional spectral features. 
\\
\\

\subsection{The origin of the intermediate component and the link to $\gamma$-ray emitting shocks and optical peak} \label{sec:intermediate}

\citet{McLaughlin_1944} and \citet{Mclaughlin_1947} pointed out that the intermediate component (principal spectrum) appears a $\sim$ few hours to a couple of days after optical peak and has an intermediate velocity between the slow and fast components. 
%\citet{Mclaughlin_1947} ruled out the possibility of the intermediate component being the result of radiation pressure acting on the inner layers that accelerate through the outer layers, particularly due to the absence of intermediate velocities to the slow and intermediate components. 
The observed change is quite abrupt and both components (slow and intermediate) co-exist for a few days. 
%Therefore, \citet{Mclaughlin_1947} concluded that the intermediate component originates from a distinct shell or a body of gas, which retains its identity until it has expanded sufficiently and become so faint that it can no longer be detected.

As stated by \citet{McLaughlin_1943}, the intermediate component originates in a swept-up shell, but it is unclear if it is swept up by shock interaction or radiation pressure:
\begin{quote}
{\it At this point we must attempt to account for the emergence of the principal spectrum and the disappearance of the pre-maximum one. In Nova Herculis, Russell suggested that an inner and swifter shell of gas swept up the outer one. If that were a unique case we might accept such an interpretation, but it taxes one's credulity to suggest that each nova had two such discrete shells and that the inner one has overtaken the outer one just after maximum light in each case.}

{\it One possibility appears to be that the temperature of the inner star has continued to rise, with consequent increase of the radiation pressure which acts upon the thick ``shell" from within. This may become so great that it blows the inner layers right through the outer ones...
%Since this could only happen after the ``shell" has become somewhat detached, it could only occur after maximum light, and perhaps it is the last violent push that expands the shell beyond the critical density below which it is no longer opaque. 
This is admittedly the most conjectural feature of the suggested model.}
\end{quote}

We argue that, in fact, the shock interaction hypothesis does \emph{not} ``tax one's credulity". First, as shown in Section~\ref{sec_results}, the presence of two flows is common to all the novae in our sample. Second, as discussed in Section~\ref{sec:universal}, it is inevitable that the fast component will catch up with the slow component, and given their relative velocities, this should occur soon after the ejection of the fast component. Third, shocks internal to nova ejecta appear to be common. More than 15 novae have been detected with \textit{Femi}-LAT since 2010 as $\gamma$-ray sources (e.g., \citep{Ackermann_etal_2014,Cheung_etal_2016,Franckowiak_etal_2018}, and even more as sources of hard X-ray or radio synchrotron emission \citep[e.g.,][]{Taylor_etal_1987, Mukai_Ishida_2001, Mukai_etal_2008, Weston_etal_2016}. The time of $\gamma$-ray detection of novae coincides with their optical peak and therefore, around the same time as the intermediate component emerges. If the intermediate component is the result of the collision of the fast and slow flow, this could be the same shock interaction responsible for the $\gamma$-ray emission.

\citet{Friedjung_1987} discussed  the formation of the intermediate component in the context of shock interaction between the fast and slow flows.
%, suggesting that it is the results of a collision between the continuously ejected fast material (fast component) and the pre-maximum slow flow. 
He hypothesizes that the shell created by the shocked material sweeps through the slow flow, and when most of the slow flow is swept up, the slow component disappears. \citet{Friedjung_1987} also pointed out that the shell responsible for the intermediate component may lose its internal energy via radiative cooling. Additionally, \citet{Friedjung_1987} derived a shock temperature that would produce X-ray emission, and conjectures that it could be absorbed by the dense medium ahead of the shell, explaining the lack of X-ray detection with the limited X-ray facilities at the time.

\citet{McLaughlin_1943} noted that the emergence of the intermediate component always occurs around optical peak, so this timing cannot be a coincidence. In this case, correlation may indeed imply causality: the interaction (shock) that creates the intermediate component could also produce the optical peak. This would explain the coincidence between the optical peak and emergence of the intermediate component for all novae. \citet{Munari_etal_2017} suggested that the optical light curves of some novae around maximum can be decomposed into one peak from the fireball expansion, followed by a brighter ``$\gamma$-ray'' peak powered by shock interaction. 
%This could explain the coincidence of the light curve peak with the emergence of the intermediate component in novae as they both share the same origin. 
\citet{Li_etal_2017_nature} and \citet{Aydi_etal_2020} demonstrated that $\gamma$-ray emitting shocks can indeed power a significant fraction of the optical luminosity during the early days of nova eruptions, rivalling the radiative luminosity from the white dwarf.

\subsection{The pre-maximum deceleration} \label{sec:deceleration}
The deceleration of the slow flow before optical maximum, which we observe in some novae (Figure~\ref{Fig:LC_abs}; particularly those characterized by slowly rising light curves), could be explained by two effects. First, it could be an optical depth effect. If the slow flow is homologous (composed of ejecta with a range of velocities, where $v \propto r$), as the slow flow expands, the photosphere will retreat to slower-moving interior layers, leading to a shift of the absorption  to lower velocities %If the slow flow is expanding as a Hubble, homologous flow, the photosphere expands in velocity coordinates but recedes in mass coordinate, shifting the origin of the absorption to deeper, slower regions 
\citep{Friedjung_1992,Mason_etal_2018}.

The second possible explanation is that it is due to a real physical deceleration of the slow flow. Unless the slow flow is ejected with velocities substantially larger than the local escape velocity, the material will quickly decelerate in the combined potential of the secondary and the white dwarf, leading to an observed deceleration in the spectral line profiles. \citet{Pejcha_etal_2016} showed that mass ejected through the L2 outer Lagrangian point could stay bound to the system, depending on a number of parameters such as binary mass ratio, temperature, and degree of co-rotation of the envelope. 

Based on the observed velocities of the slow flows in most novae ($\sim$ 500\,--\,1000\,km\,s$^{-1}$), the average mass of the binary system, and the estimated radius of the photosphere, we can test if the second explanation is plausible. For a total system mass of 1\,M$_{\odot}$ and a photospheric radius of 10$^{13}$\,cm at maximum light (see e.g., \citealt{Bath_Shaviv_1976,Bath_1978,Kato_Hachisu_1994}), the escape velocity from the system would be around 50\,km\,s$^{-1}$. This value is small compared to the observed velocity of the slow flow near optical peak. In addition, we see a deceleration of around 150\,km\,s$^{-1}$ in FM~Cir during a week. This is $\gtrsim$10 times larger than the expected deceleration based on the gravitational potential of a $\sim$1\,M$_{\odot}$ binary. All this implies that the optical depth effect must be the main contributor in the apparent velocity deceleration observed before optical peak. Nevertheless, there is a chance that some of the material near the binary system might not reach escape velocities and fall back into the system. 
%In this case, if the slow flow is ejected with low velocities (below the escape velocity from the system), the gravitational pull of the system would lead to a physical deceleration of the bulk of the ejecta, manifesting as a deceleration in the spectral line profiles (see Figure~\ref{Fig:nova_ejection}). A combination of optical depth effect and physical deceleration could also be the reason behind the deceleration.

%While the optical light curve of the nova rises to optical peak, the emission component of the early P Cygni profiles is weakening relative to the continuum. This is mainly due to a rising continuum as the photosphere expands to its maximum radius \citep{Warner_2008}. This weakening of the emission profiles lead to a redward shift of the absorption trough of the P Cygni profiles, mimicking an apparent deceleration in some cases. However, in most cases the bluer edge of the P Cygni profiles moves blueward and therefore the observed deceleration is not a measurement bias (see Figure A.3, A.4, and A.5). 

\subsection{The post-maximum acceleration}
\label{acc}
We also observe a gradual acceleration in the slow and intermediate components after the optical peak (Figure~\ref{Fig:LC_abs}).  Again, this is not a new result; for example, \cite{Hutchings70} pointed out similar movement of absorption components to more extreme blueshifts as the eruption of LV~Vul proceeded.

We suggest that this acceleration is caused by interaction with the fast flow. 
%As the shock sweeps through the slow flow, the observed velocity of the system increases. %If the fast component is a continuous wind originating from the white dwarf, then when it collides with the slow flow, it may sweep it up and accelerate it (see Figure~\ref{Fig:LC_abs}). 
\citet{Steinberg_etal_2020} show that, as the fast outflow adds momentum to the swept-up shell separating the fast and slow outflow components, this can manifest as an acceleration in the spectral lines of the intermediate velocity component (assuming the latter is generated in the shell).

The fast flow also shows evidence for acceleration as the base of the emission lines broaden with time and the accompanying absorptions increase in velocity (see Figure~\ref{Fig:V906_Car_profiles}). The broadening of the fast spectroscopic component can be explained as a wind whose velocity increases with time, as expected for nova radiation-driven winds \citep{Kato_Hachisu_1994}.

%is again evidence for a continuous wind with increasing velocity rather than a single ballistic ejection of a homologous flow (see e.g., \citealt{Seitter_1990,Cassatella_etal_2004,Friedjung_2011}) --- as the photosphere recedes during the optical light curve decline, the emission originate from deeper regions.
%Again, the acceleration of the fast component runs counter to expectations for a single ballistic ejection, where velocities should decelerate as the photosphere recedes into the inner, slower-expanding (for a homologous flow) regions  (see e.g., \citealt{Seitter_1990,Cassatella_etal_2004,Friedjung_2011}).

%\section{the diversity of gamma-ray luminosities in novae}
\subsection{The origin of the THEA lines}
\label{THEA}
In Figure~\ref{Fig:THEA_1} we showed that the THEA lines of nova V906~Car exhibit a slow (pre-maximum) component and an intermediate component, essentially identical to the velocities and evolution demonstrated in the \eal{H}{I}, \eal{Fe}{II}, \eal{O}{I}, and \eal{Na}{I} lines (Figure~\ref{Fig:THEA_Balmer}). Therefore, the THEA absorptions originate from the same body of gas responsible for the P Cygni profiles in prominent lines, which we associate with the slow nova ejecta. 

The THEA lines observed in nova FM~Cir (Figure~\ref{Fig:THEA_FM_Cir}) shows an apparent deceleration by around 150\,km\,s$^{-1}$ during the rise to optical peak. In addition, the lines appear to be broader initially and they become more narrow as the nova rises to peak. Again, this mimics the evolution of more prominent lines associated with the slow ejecta, like \eal{Fe}{II}.

As discussed in \S\ref{sec:v906_thea}, no fast component is associated with the THEA lines. Although this could be an optical depth effect (i.e., the THEA lines are too weak to be detected in the fast flow), it is possible that this observation denotes a real absence.
The THEA lines are s-process elements and therefore it is unlikely that they are synthesized during the thermonuclear runaway. The THEA elements that are accreted onto the white dwarf throughout the years prior to the nova eruption might selectively diffuse into the white dwarf interior \citep{Williams_etal_2008}. It is therefore possible that these elements are freshly accreted onto the white dwarf or that they still reside in the accretion disk prior to the eruption and thus are ejected during the early common envelope phase. %In all cases, we suggest that these lines are associated with the nova ejecta.

\begin{figure*}
\begin{center}
  \includegraphics[width=0.8\textwidth]{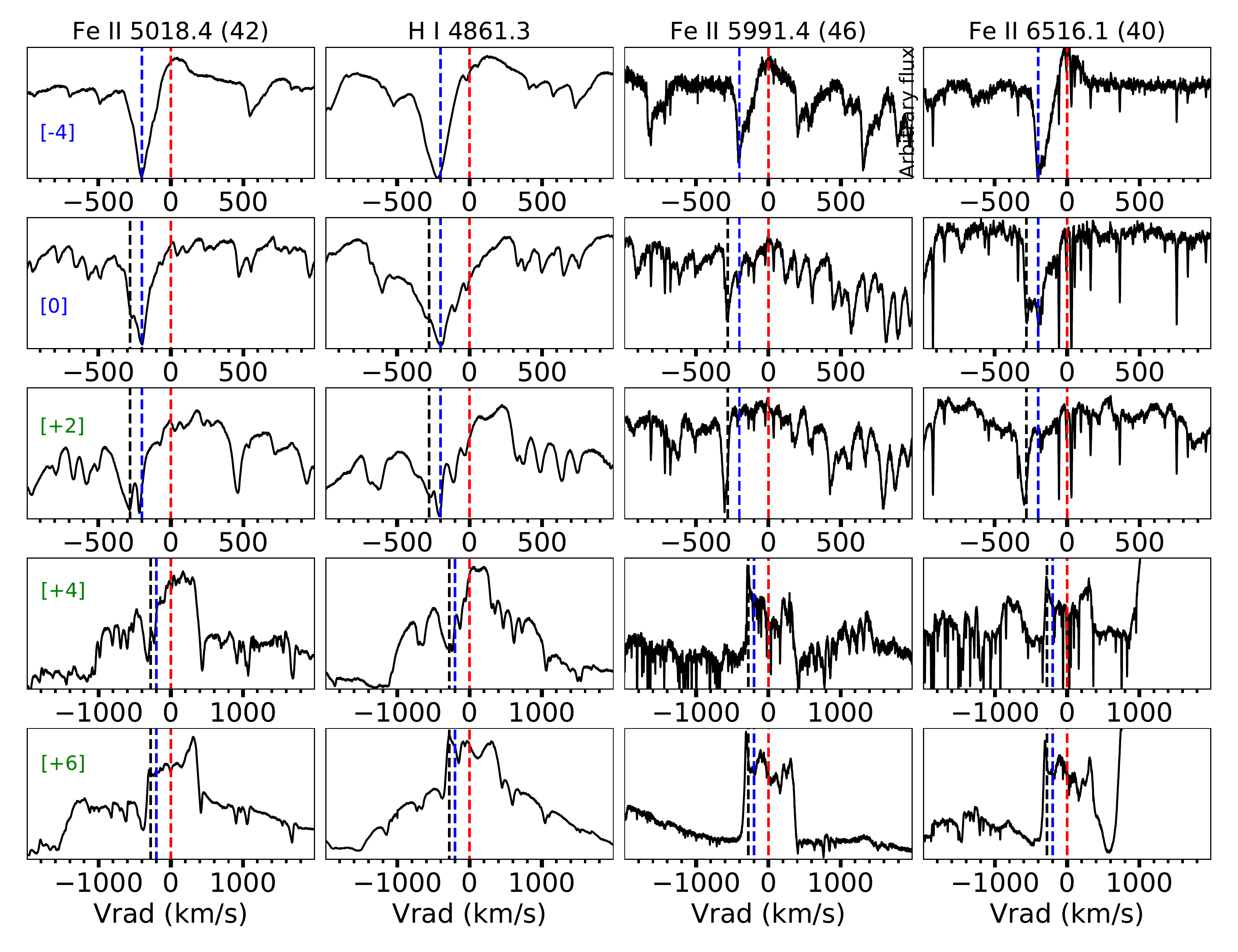}
\caption{The line profile evolution of a sample of \eal{Fe}{II} 5018\,$\mathrm{\AA}$ and H$\beta$ in comparison \eal{Fe}{II} 5991 and 6516\,$\mathrm{\AA}$ for nova V906~Car. The red vertical dashed line represents $v_{\mathrm{rad}}$ = 0\,km\,s$^{-1}$ (rest wavelength). The blue vertical dashed line marks a velocity of $-200$\,km\,s$^{-1}$ highlighting the slow component. The black dashed line marks a velocity of $-300$\,km\,s$^{-1}$ highlighting the intermediate component. The top three panels extend between $-$1000\,km\,s$^{-1}$ and 1000\,km\,s$^{-1}$ while the bottom two panels extend between $-$2000\,km\,s$^{-1}$ and 2000\,km\,s$^{-1}$ to show the fast component.}
\label{Fig:Mcloughlin}
\end{center}
\end{figure*}

\subsection{Alternative scenarios}

In this section we explore alternative explanations for the
%shaping of the nova ejecta and the 
origin of the different absorption systems.
%, and the formation of the $\gamma$-ray emitting shocks. 

\subsubsection{Circumbinary material?}

As mentioned in Section \ref{sec:proposedex}, \citet{Williams_etal_2008} and \citet{Williams_Mason_2010}
%, and \citet{Williams_2013} 
suggested a circumbinary reservoir as the source of the THEA lines.
%and the low velocity ($\sim$ a few hunderds km\,s$^{-1}$) absorption features observed in the nova spectra near optical peak. 
However, these studies missed the presence of similar low-velocity components in the \eal{H}{I}, \eal{Fe}{II}, and \eal{O}{I} lines during the pre-maximum stage. This suggests that the 
THEA lines
%low-velocity components 
are associated with the nova ejecta rather than a circumbinary reservoir. If the THEA lines and the slow, pre-maximum P Cygni profiles of \eal{H}{I}, \eal{Fe}{II}, and \eal{O}{I} all originate in a pre-existing circumbinary reservoir, then there would be no features in the pre-maximum spectra to explain and associate with the rise to peak phase, when the nova brightness increases by 8 to 15 magnitudes.
%In Section~\ref{THEA} we offered further evidence arguing against the presence of a circumbinary reservoir prior to the thermonuclear runaway.

%An association of the THEA lines with a reservoir of circumbinary material is also unlikely for several additional reasons. 
%(1) the THEA lines show the same evolution of the P Cygni profiles (\eal{H}{I}, \eal{Fe}{II}, \eal{O}{I} etc.), which are presumably originating from the nova ejecta; 
Similar to the other lines in the spectra, the THEA lines show deceleration during the rise to optical peak (\S\ref{FM_Cir_sec}), which is consistent with a photosphere receding through expanding ejecta (\S\ref{sec:deceleration}), rather than a pre-existing circumbinary reservoir of gas. Before eruption and in its early stages, we would expect CSM to exhibit constant velocity.
In later stages, \citet{Williams_etal_2008,Williams_Mason_2010} used the \emph{acceleration} of the THEA lines observed post-maximum to argue for a circumbinary origin, with the CSM accelerated outward by the radiation pressure from the white dwarf. %Whether this radiation pressure is acting on a cirumbinary reservoir or the nova ejecta, it cannot be ruled out as a possible explanation for the accleration we observe in the post-maximum spectra (see e.g., \citealt{Kato_Hachisu_1994}).

%We note that \citet{Williams_Mason_2010} distinguished between the broad fast components observed in the nova ejecta lines and the THEA/Na D intermediate components without discussing the slow, pre-maximum component. Therefore they might have missed the early presence of slow P Cygni profiles in the ejecta lines.

\citet{Williams_2012, Williams_2013} argues that the intermediate component (and potentially the slow component), along with the THEA lines, are ablated/irradiated from the secondary star, rather than originating on the white dwarf. He cites as evidence that the velocities of these components are consistent with the escape velocities of the secondary star ($\sim$ a few hundreds km\,s$^{-1}$). While this is true for slow novae, faster novae such as ASASSN-19qv and V1707~Sco show pre-maximum P Cygni profiles with absorption troughs at velocities $\gtrsim$ 1000--2000\,km\,s$^{-1}$, much higher than the escape velocities of the secondary star (see Figure~\ref{Fig:line_profiles_mod}). In addition, we observe a correlation between the velocities of the slow and fast flows with that of the speed class of the nova (see the marked velocities in Figures~\ref{Fig:line_profiles_slow}, and~\ref{Fig:line_profiles_mod}). \citet{McLaughlin_1944} also points out that the fast component always has around twice the velocity of the intermediate component (see also \citealt{Gallaher_etal_1978}). This means that both the slow/intermediate flows and their velocities are associated with the eruption and not with the secondary star.

The first spectra of nova FM~Cir show a component in the THEA lines at around 800\,km\,s$^{-1}$ (marked by magenta lines in Figure~\ref{Fig:THEA_FM_Cir}) co-existing with the slow component (550\,km\,s$^{-1}$) before disappearing in one day. The large velocity and behavior of this component is also not consistent with a pre-existing circubminary reservoir. The origin of this early component could be early ejection of a small body of gas during the TNR. This low density body of gas dissipates rapidly before the bulk of the ejecta expands and engulf the binary. We see a similar behaviour in the early spectra of nova V435 CMa (Figure~\ref{Fig:V435_CMa_profiles}).  

Recently, \citet{McLoughlin_etal_2020} suggested the presence of a circumbinary disk rich with Fe and O in nova V906~Car. Their evidence consists of particular emission lines of \eal{Fe}{II} at 5991\,\AA\ (46), 6432\,\AA\ (40), 6456\,\AA\ (74), and 6516\,\AA\ (40), in addition to \feal{O}{I} 6300\,$\mathrm{\AA}$, which are characterized by double-peaked profiles and modest expansion velocities (FWZI $\approx$ 900\,km\,s$^{-1}$). These lines do not develop a broad emission component (FWZI $\approx$ 2500\,km\,s$^{-1}$) like the other lines in the spectrum, such as \eal{H}{I}, and other transitions of \eal{O}{I} and \eal{Fe}{II} (42). McLoughlin et al.\ associate these double-peaked emission lines with a circumbinary disk and suggest that this disk could also be responsible for the THEA lines. 

In Figure~\ref{Fig:Mcloughlin} we plot two of these lines in comparison with H$\beta$ and \eal{Fe}{II} 5018\,$\mathrm{\AA}$ (42). All four lines show ``slow component'' P Cygni profiles before optical peak, as do other lines in the spectrum. After optical peak, all of the lines also develop the intermediate component.
%at the same velocities as the other lines. 
A few days after optical peak, the \eal{Fe}{II}(40) and (46) lines develop a double-peaked narrow emission, but not a fast component. The narrow double-peaked emission is not unique to these lines and is present in the other lines in the spectrum, such as the Balmer and \eal{Fe}{II} (42) lines, but these other lines are superimposed on top of a broad emission (see also Figures~\ref{Fig:V906_Car_Hbeta}). The spectral evolution of the lines presented in \citet{McLoughlin_etal_2020} is then similar to other prominent lines associated with the ejecta, and therefore it is reasonable to associate them with the slow flow rather than a circumbinary disk. The slow flow could propagate aspherically, leading to the observed double peak in the slow component emission profiles.

One explanation for the absence of a fast component in the \eal{Fe}{II} (40 \& 46) multiplet lines could be the $f$-value (oscillator strength) of these transitions. The $f$-values of the \eal{Fe}{II} (42) multiplet transitions are $\sim$ 10$^{-3}$\,--\,10$^{-2}$, while the $f$-values of the \eal{Fe}{II} (40 \& 46) multiplets are $\sim$ 10$^{-5}$. The 2\,--\,3 orders of magnitude smaller $f$-values for the (40 \& 46) multiplets could result in relatively weak, and possibly undetectable, broad emission lines produced by the lower-density fast wind. %In addition the opacity of the \eal{Fe}{II} (42) lines   

\subsubsection{A single ballistic ejection?}

\citet{Shore_etal_2011, Shore_etal_2013, Shore_etal_2016} and \citet{Mason_etal_2018,Mason_etal_2020} model nova line profiles as a single biconical and clumpy ejection. When the (pseudo)photosphere recedes through the homologously expanding ejecta, lines can appear to decelerate. On the other hand, lines can appear to accelerate if a recombination wave sweeps outward through the ejecta \citep{Shore_etal_2011}. 
\citet{Mason_etal_2018} point out that some absorption components are present at the same velocities in earlier spectra (in low ionization lines) and later spectra (in high excitation lines). This led them to conclude that the ejecta are stationary in velocity, characterized by a single ballistic explosion. In this scenario, shock signatures can be produced at early times if the ejecta are expelled 
%over a very short duration, but 
with a large range in velocities. Clumps will crash into one another as the flow relaxes to homologous expansion \citep{Shore_etal_2013}.

A single ballistic ejection cannot explain the early spectral evolution of novae: specifically, a relatively slow absorption component superimposed on fast emission. As pointed out decades ago by McLaughlin, the fact that the slow component is seen in absorption necessitates that it is external to the fast component---and this, in turn, necessitates that the fast component must have been ejected after the slow component (\S\ref{sec:universal}).
\citet{Mason_etal_2018,Mason_etal_2020} did not tackle the early spectral evolution of the novae they consider; instead, their observations are obtained at later stages (more than one hundred days after eruption). During this late stage, the ejecta probably are expanding ballistically, after the mass loss has ceased. On the other hand, \citet{Hauschildt_etal_1994, Hauschildt_etal_1995, Hauschildt_etal_1996, Hauschildt_etal_1997} were able to explain many aspects of the early spectra of novae with a single ballistic ejection, but their models always predict lines composed of a single P~Cygni profile; they provide no explanation for the slow/intermediate component superimposed on top of the fast component.

In addition, in a scenario where impulsively ejected clumps of different velocities collide to produce the shocks and $\gamma$-rays, these collisions should occur immediately, and last no longer than the impulsive ejection itself. Instead, we see that gamma-rays do not appear until optical maximum %($\sim$11 days after the start of eruption and expansion, in the well observed case of V906~Car; \citealt{Aydi_etal_2020}), 
and can remain detectable for weeks (e.g., \citealt{Cheung_etal_2016}). In order to explain the prolonged periods of gamma-ray detection and other shock signatures, mass ejection itself must be prolonged in time.

While a photosphere receding through homologously expanding ejecta can reveal slower spectral components with time, and while a recombination wave moving outward through the ejecta might reveal faster components, neither phenomenon can explain the sudden appearance of an intermediate-velocity component. Take, for example, V906 Car. Its intermediate component appears at optical peak (day $\sim$ 0; Figure \ref{Fig:THEA_Balmer}), around the same time as the fast component (Figure \ref{Fig:V906_Car_profiles}). The fact that the fast component is visible would imply, in the ballistic outflow scenario, that the recombination wave has reached the outer extent of the ejecta. Meanwhile, the slow component is also present (Figure \ref{Fig:THEA_Balmer}), which would imply that inner ejecta are contributing to the line profile. The appearance of the intermediate component cannot be explained by changes in opacity or ionization state of the ejecta, and therefore implies an actual change in the ejecta configuration (i.e., a new swept-up shell). In addition, the acceleration of spectral components after optical peak (e.g., Figure \ref{Fig:LC_abs}) is difficult to explain as a recombination front, given that the fast component tracing the outer ejecta is present in the line profile.

%The emergence of spectral features with increasing velocities, such as the intermediate and fast components, also implies multiple flows. An outward moving recombination wave could possibly account for the acceleration observed in some lines,  but it cannot explain the appearance of new distinct absorption at greater velocities for the same transition. 

%The observed shift from a deceleration to an acceleration of the slow flow, as seen in Figure \ref{Fig:LC_abs}, also argues against nova eruptions being a single ballistic ejection. For a ballistic ejection, after optical peak one would expect to observe a continued deceleration of the absorption components, as the photosphere recedes to inner regions, which are moving at lower velocities.

%Also, the acceleration of the fast component runs counter to expectations for a single ballistic ejection, where velocities should decelerate as the photosphere recedes into the inner, slower-expanding (for a homologous flow) regions  (see e.g., \citealt{Seitter_1990,Cassatella_etal_2004,Friedjung_2011}).

\section{Summary and conclusions}
\label{sec_conc}
Based on the near-peak evolution of the Balmer line profiles of all the novae in our sample and the spectral evolution of novae V906~Car and FM~Cir, we reached the following conclusions: 
%We present medium-/high-resolution rapid spectroscopic follow up for a sample of 12 novae to highlight the early spectral evolution of novae in context of a universal ejection scenario and shocks formed by colliding flows of different velocities. Our results and interpretation of the data led us to the following conclusions:

\begin{itemize}
    \item All 12 novae show the same spectral evolution: before optical peak, the line profiles are dominated by P Cygni profiles characterized by slow velocities ($\lesssim$ 1000\,km\,s$^{-1}$). After optical peak, a broad emission base emerges characterized by faster velocities (more than double that of the slow component), while the pre-existing P Cygni profile is superimposed on top of the broad emission component. 
    
    \item The co-existence of the fast and slow spectral components (slow P Cygni superimposed on top of the fast component), the large difference in velocity between them, and the abrupt transition of the spectral profiles in a matter of $\sim$a day indicate the presence of at least two physically distinct flows. It also indicates that the faster flow originates from inside the slower one. 
    
    \item For novae with multiple observations before and after peak, we notice that the spectral components also show similar velocity evolution. The absorption trough of the slow component decelerates until peak, then accelerates. The fast component accelerates once it appears.
    
    \item The THEA lines in nova V906~Car show the same spectral evolution of the Balmer, \eal{Fe}{II}, and \eal{O}{I} near optical peak, except they do not show the emergence of a fast component. In FM~Cir, THEA lines in the pre-maximum spectra appear to decelerate during the rise to peak. We argue that the THEA lines are associated with the nova slow flow, rather than a pre-existing circumbinary reservoir of gas. 
    %The elements responsible for these absorption possibly reside in the accretion disk or the upper layers of the white dwarf envelop after they were freshly accreted from the companion star. Thereafter, they are carried away with the initial slow ejection during the early days of the nova eruption.  
    
    \item The fast, internal outflow must be launched after the slow outflow, but even so, it is expected to quickly catch up with the slow outflow. However, the fast component persists in the spectrum long after the time when collision is expected, implying that the ejecta are likely aspherical and the fast flow expands relatively unimpeded in some directions.
    
    \item For nova V906~Car we detect the emergence of an intermediate component with slightly bluer velocities compared to the slow component. This intermediate component replaces the slow, pre-maximum one a few days after optical peak. We suggest that this intermediate component originates in a shell formed by the collision of the slow and fast flows. 
    
    \item Since all of the novae in our sample follow the same spectral evolution, we suggest a common scenario to explain the observations. The scenario consists of an initial ejection of the slow flow, which could be expelled preferentially in the orbtial plane during a common envelope phase. This is followed by a fast flow---likely a radiation pressure driven wind---which sweeps through the slow flow and causes it to accelerate. 
    
    \item The shocks formed by the interaction between these two flows could also be responsible for the $\gamma$-ray emission observed in some novae, and may power a significant part of the optical emission of the nova, contributing to its optical peak. This can help explain the coincidence of timing commonly seen in novae: that the fast and intermediate component appear just around optical maximum.

%This scenario is in good agreement with previous studies \citep{Friedjung_1987,Chomiuk_etal_2014,Metzger_etal_2015,Li_etal_2017_nature,Aydi_etal_2020}.
\end{itemize}

%\subsection{The diversity of $\gamma$-ray luminosities in novae}

%The shocks responsible for $\gamma$-ray emission in nova are suggested to occur at the interface between the slow and fast flows (e.g., \citealt{Chomiuk_etal_2014, Metzger_etal_2015,Li_etal_2017_nature,Aydi_etal_2020}).
The $\gamma$-ray luminosities of novae have been shown to span at least two orders of magnitude \citep{Franckowiak_etal_2018}, but the link between shock luminosity and nova properties---and the cause of this diversity in $\gamma$-ray luminosity---remain poorly understood (e.g., 
\citealt{Finzell_etal_2018,Franckowiak_etal_2018}). To first order, we might expect shock luminosity to be determined by the density and velocity of the ejecta \citep{Metzger_etal_2015}. Since we can measure the differential velocity between the fast and slow flows with optical spectroscopy, we can use this to estimate the luminosity of the shock---if we know the distance to the nova and have an estimate of the mass of the ejecta from e.g., radio observations \citep{Chomiuk_etal_2012,Chomiuk_etal_2014,Weston_etal_2016,Aydi_etal_2020}. 
%Consequently, these two parameters, along with a distance estimate, could help us build a parameter space to constrain why only some novae are detected with \textit{Fermi}-LAT.
Future work should be dedicated to making such observations for a large sample of novae, in order to test theories about nova shocks and $\gamma$-ray production, and understand the diversity of shock luminosities observed for $\gamma$-ray detected novae.
%particle acceleration mechanism, and properties of the shocks.

%For novae with extensive spectroscopic follow up, the timing of the emergence of the fast flow and its observed velocity could also tell us when this flow started, and when the slow and fast flows impact one another, producing $\gamma$-ray emitting shocks.

In this paper we proposed a qualitative scenario explaining the spectral evolution of novae near maximum light. The next step would be to construct a quantitative model with radiation hydrodynamical simulations to attempt to describe the observed spectra. Given the modest size of our current sample of early nova spectroscopy and the observed diversity of novae, more pre-maximum spectral observations will also inform our understanding of mass ejection in novae.

\section*{Acknowledgments}

EA, LC, JL-M, and KVS acknowledge NSF award AST-1751874, NASA award 11-Fermi 80NSSC18K1746, and a Cottrell fellowship of the Research Corporation. JS was supported by the Packard Foundation. LI was supported by grants from VILLUM FONDEN (project number 16599 and 25501). DAHB gratefully acknowledges the receipt of research grants from the National Research Foundation (NRF) of South Africa. CSK and BJS are supported by NSF grant AST-1907570/AST-1908952. CSK is also supported by NSF grants AST-1515927 and AST-181440. BJS is also supported by NSF grants AST-1920392 and AST-1911074. PAW kindly acknowledges the National Research Foundation and the University of Cape Town. FMW acknowledges support of the US taxpayers through NSF grant 1611443.

ASAS-SN thanks the Las Cumbres Observatory and its staff for its continuing support of the ASAS-SN project. ASAS-SN is supported by the Gordon and Betty Moore Foundation through grant GBMF5490 to the Ohio State University and NSF grant AST-1515927. Development of ASAS-SN has been supported by NSF grant AST-0908816, the Mt. Cuba Astronomical Foundation, the Center for Cosmology and AstroParticle Physics at the Ohio State University, the Chinese Academy of Sciences South America Center for Astronomy (CASSACA), the Villum Foundation, and George Skestos. 

We thank Robert E. Williams for useful comments and discussion. We thank Kristen Dage and Chelsea Harris for useful comments and support during this work.  
We thank the AAVSO observers from around the world who contributed their magnitude measurements to the AAVSO International Database used in this work.
%We acknowledge with thanks the variable star observations from the AAVSO International Database contributed by observers worldwide and used in this research. 
We acknowledge all the ARAS observers for their optical spectroscopic observations which complement our data. This work is based on observations obtained at the Southern Astrophysical Research (SOAR) telescope, which is a joint project of the Minist\'{e}rio da Ci\^{e}ncia, Tecnologia, Inova\c{c}\~{o}es e Comunica\c{c}\~{o}es (MCTIC) do Brasil, the U.S. National Optical Astronomy Observatory (NOAO), the University of North Carolina at Chapel Hill (UNC), and Michigan State University (MSU). A part of this work is based on observations made with the Southern African Large Telescope (SALT), with the Large Science Programme on transients 2018-2-LSP-001 (PI: DAHB). 
This work is also partly based on observations collected at the European Organisation for Astronomical Research in the Southern Hemisphere under ESO programme(s) PPP.C-NNNN(R). This paper includes data gathered with the 6.5 meter Magellan Telescopes located at Las Campanas Observatory, Chile. Polish participation in SALT is funded by grant no. MNiSW DIR/WK/2016/07.

\bibliographystyle{aasjournal}
\bibliography{biblio}



\appendix
 
\renewcommand\thetable{\thesection.\arabic{table}}    
\renewcommand\thefigure{\thesection.\arabic{figure}}   
\setcounter{figure}{0}

\section{The optical light curves}
\label{appB}
In this Appendix we present the optical light curves of our nova sample. These light curves are produced using $V$, $g$, $CV$ 
(clear filter with $V$ magnitude zero-point)
%calibrated to $V$ magnitude), 
or visual data from ASAS-SN and AAVSO. 
The time ranges were selected to highlight the timing of our spectroscopic observations around light curve peak.

\begin{figure}[h!]
\begin{center}
  \includegraphics[width=0.85\textwidth]{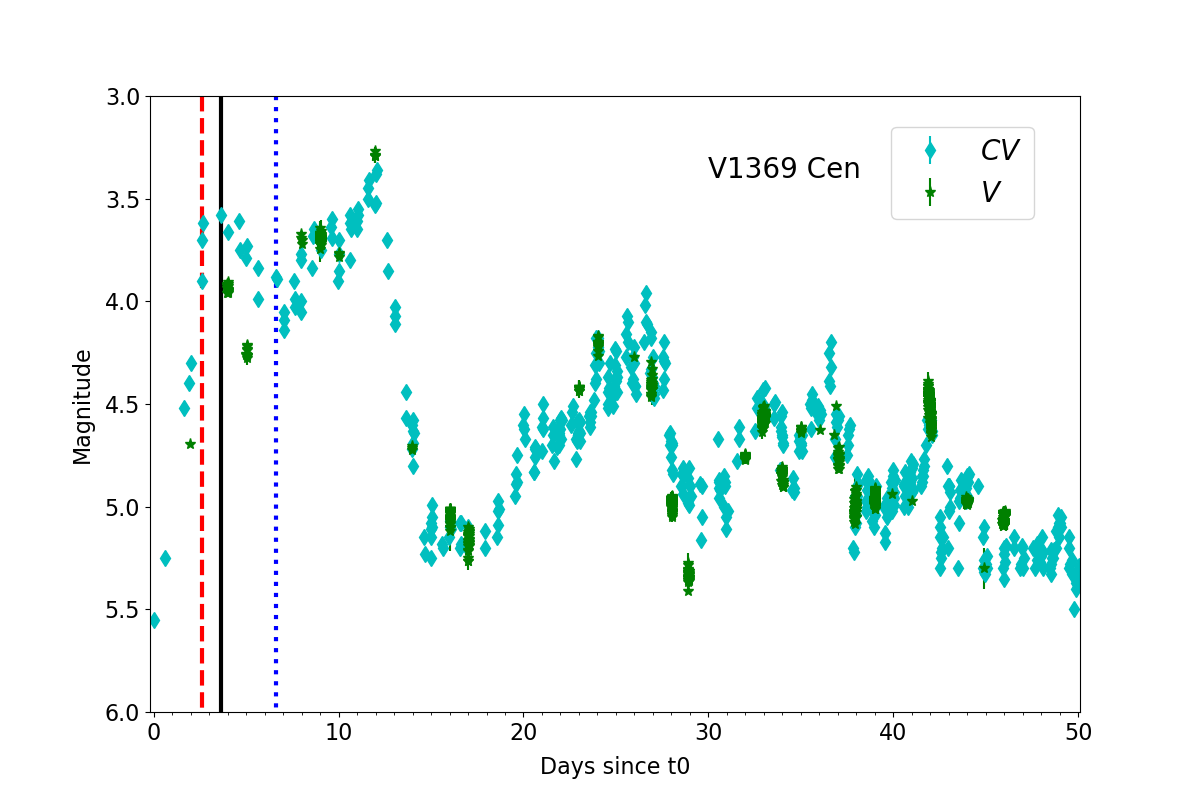}
\caption{The optical light curve of nova V1369~Cen color and symbol coded as indicated in the legend. The black solid line represents the date of the first optical peak (some novae exhibit multiple maxima). The red dashed line represents the date of the first optical spectrum and the blue dotted line represents the data of the second optical spectrum (see Table~\ref{table:sample}).}
\label{Fig:LC_V1369Cen}
\end{center}
\end{figure}

\begin{figure*}
\begin{center}
  \includegraphics[width=0.85\textwidth]{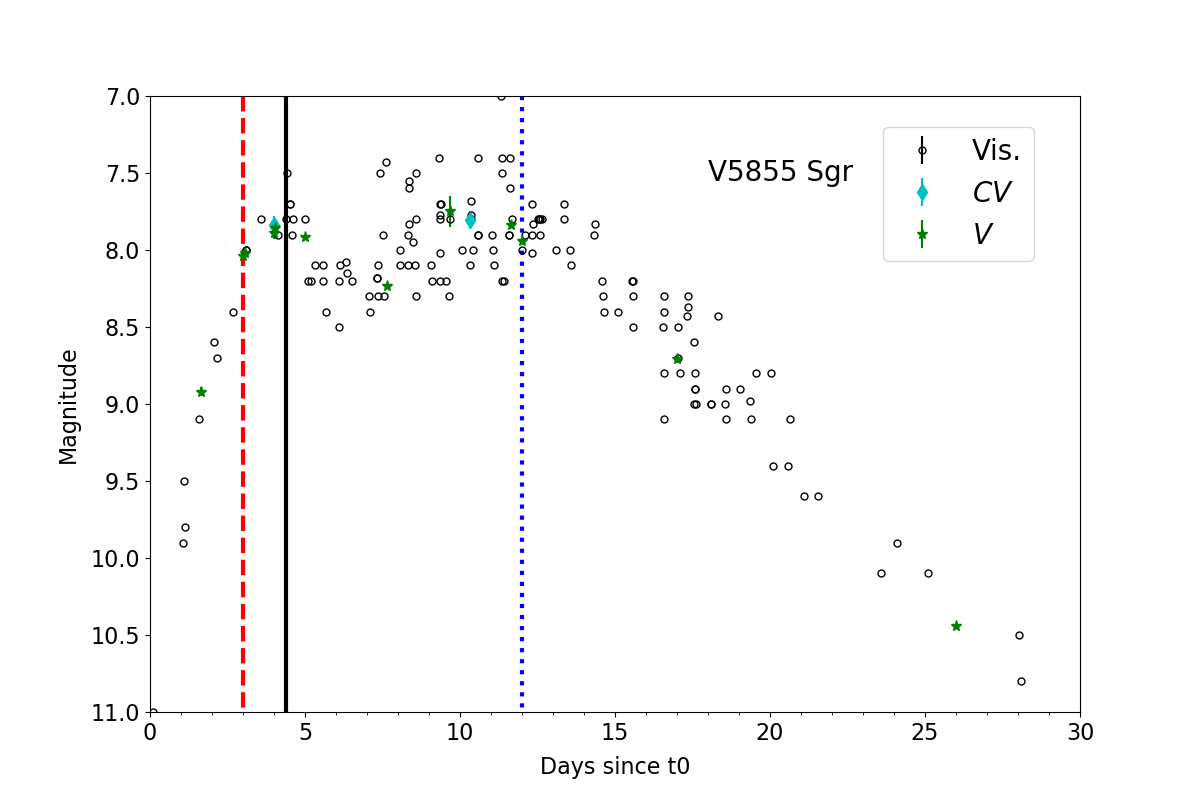}
\caption{Same as Figure~\ref{Fig:LC_V1369Cen} but for nova V5855~Sgr.}
\label{Fig:LC_V5855Sgr}
\end{center}
\end{figure*}

\begin{figure*}
\begin{center}
  \includegraphics[width=0.85\textwidth]{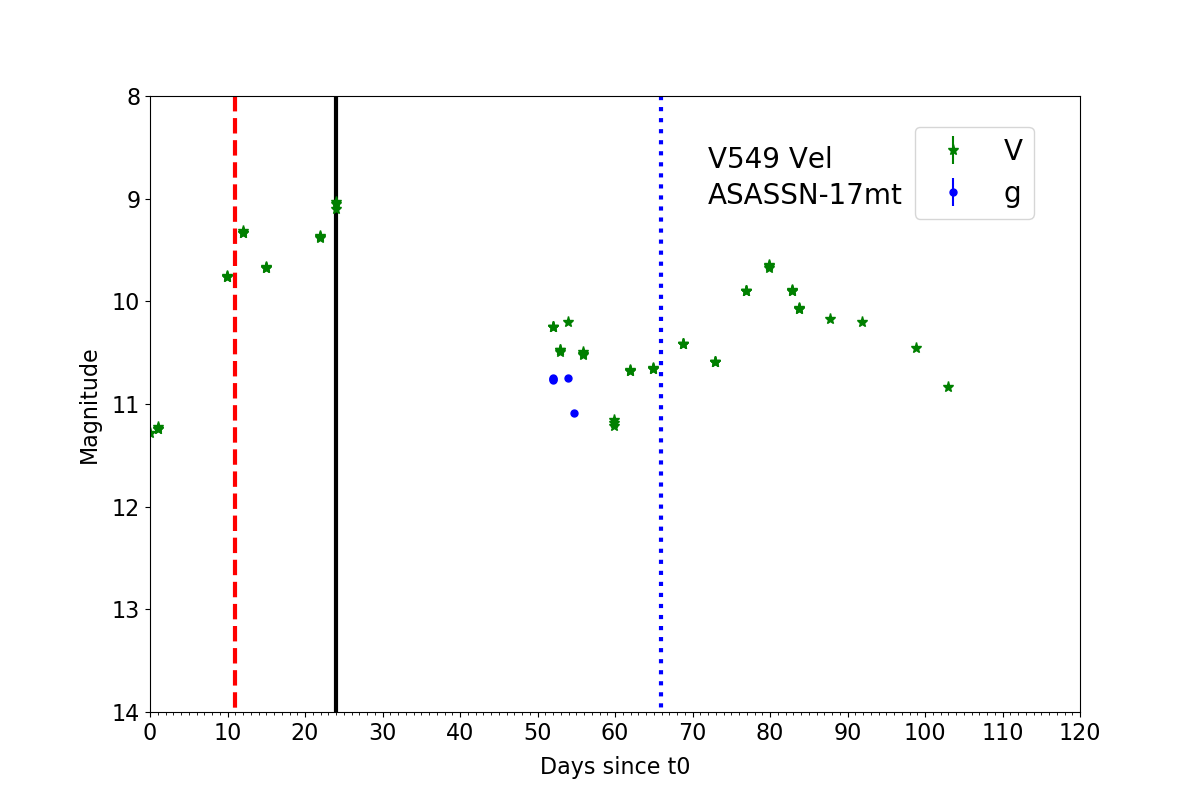}
\caption{Same as Figure~\ref{Fig:LC_V1369Cen} but for nova V549~Vel (ASASSN-17mt).}
\label{Fig:LC_V549Vel}
\end{center}
\end{figure*}

\begin{figure*}
\begin{center}
  \includegraphics[width=0.85\textwidth]{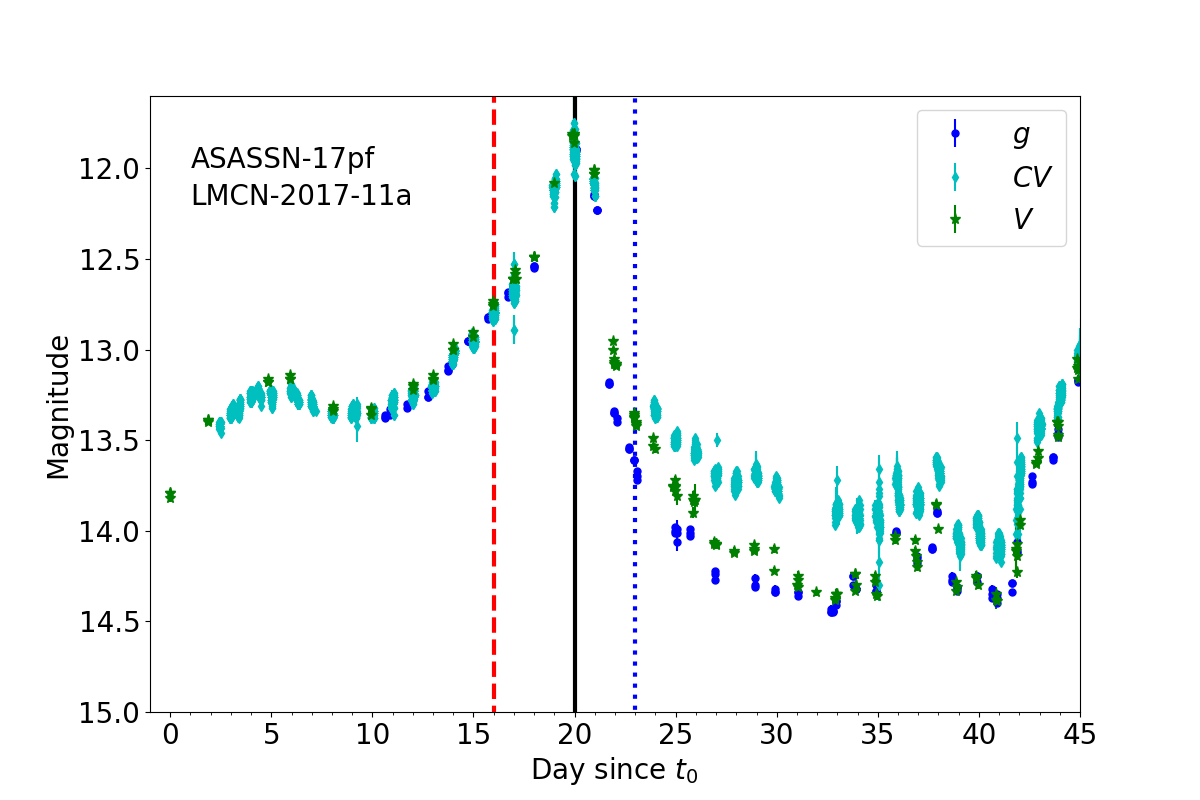}
\caption{Same as Figure~\ref{Fig:LC_V1369Cen} but for nova LMCN-2017-11a (ASASSN-17pf).}
\label{Fig:LC_17pf}
\end{center}
\end{figure*}

\begin{figure*}
\begin{center}
  \includegraphics[width=0.85\textwidth]{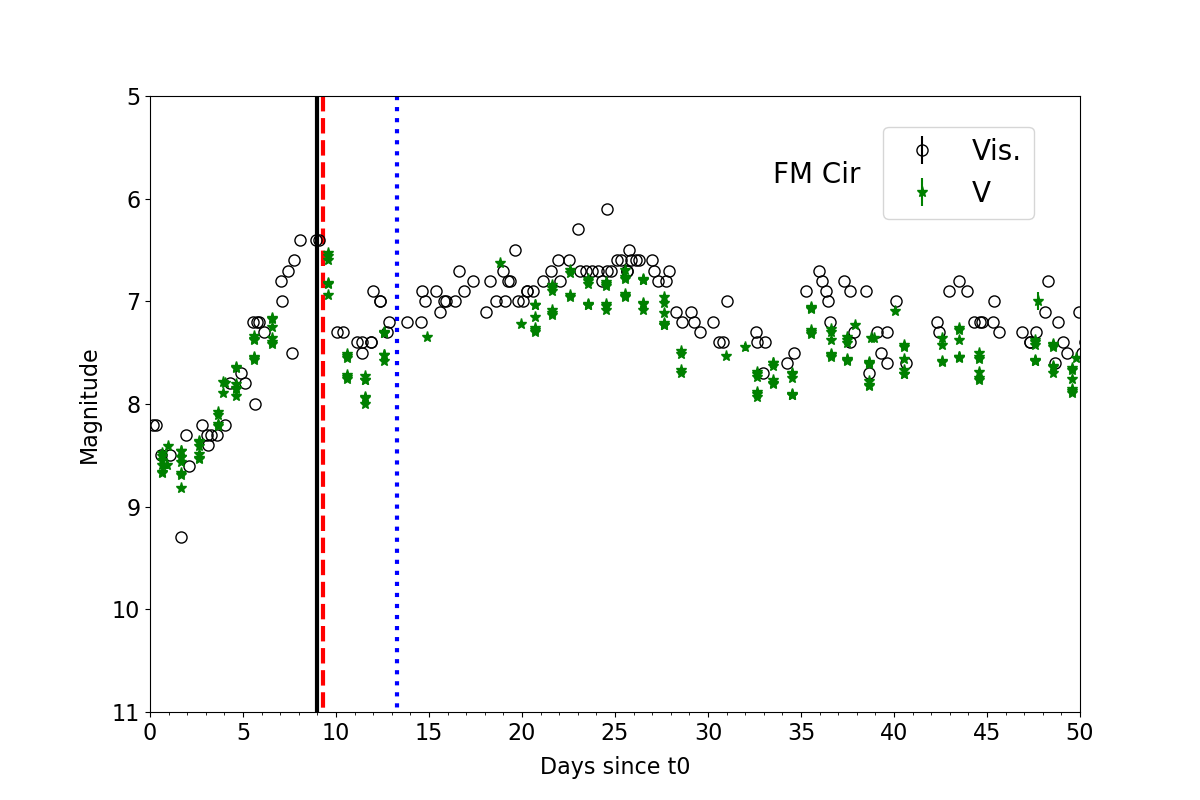}
\caption{Same as Figure~\ref{Fig:LC_V1369Cen} but for nova FM~Cir.}
\label{Fig:LC_FMCir}
\end{center}
\end{figure*}

\begin{figure*}
\begin{center}
  \includegraphics[width=0.85\textwidth]{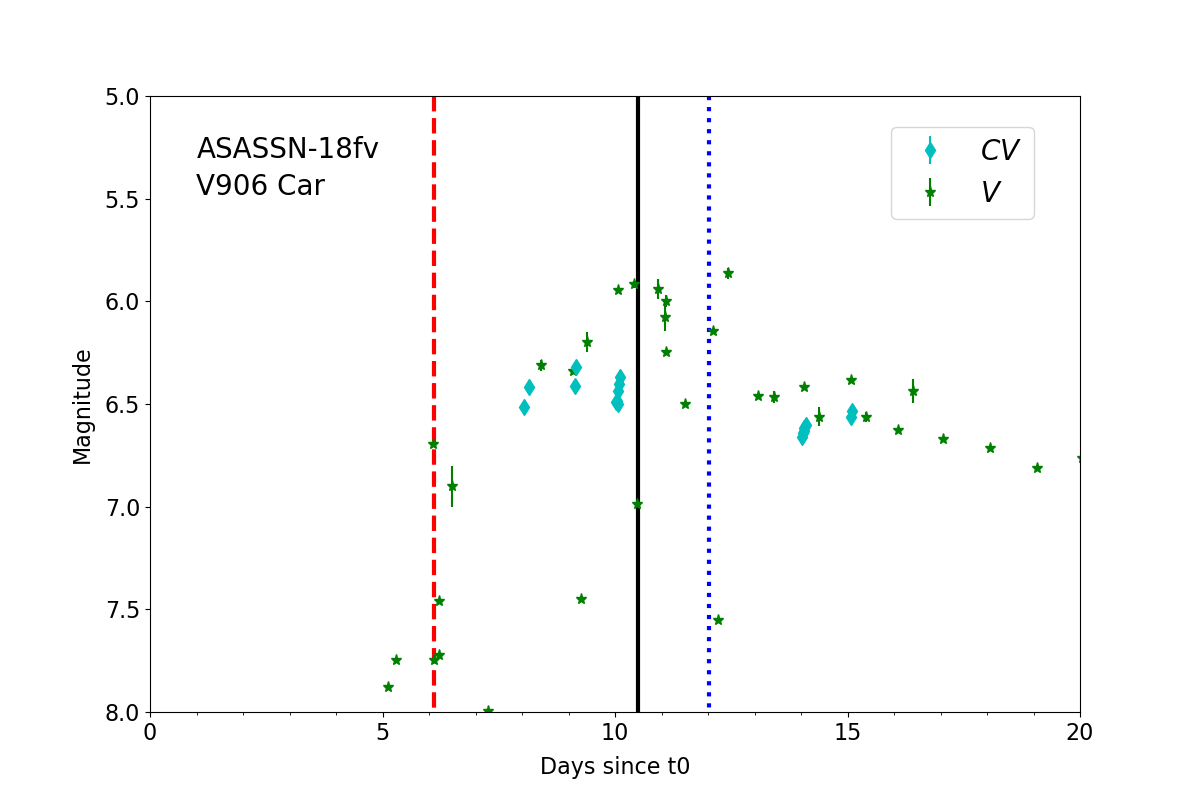}
\caption{Same as Figure~\ref{Fig:LC_V1369Cen} but for nova V906~Car (ASASSN-18fv).}
\label{Fig:LC_18fv}
\end{center}
\end{figure*}

\begin{figure*}
\begin{center}
  \includegraphics[width=0.85\textwidth]{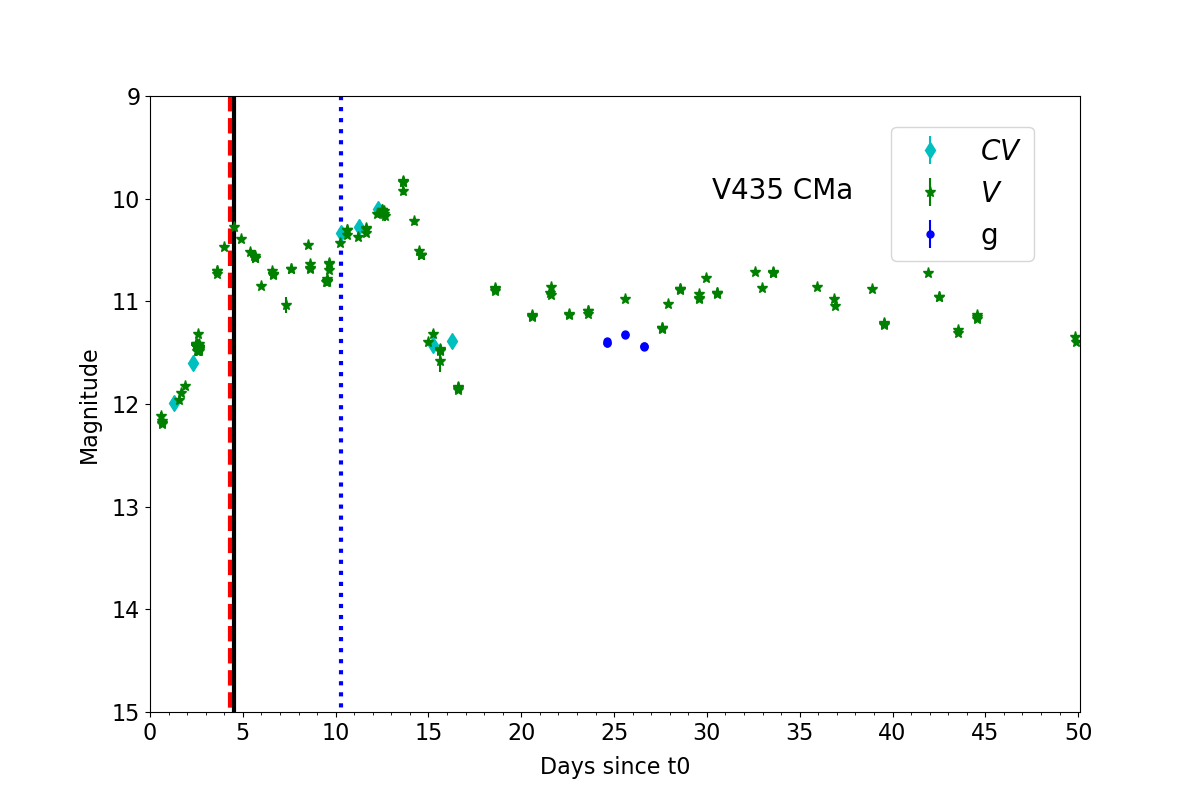}
\caption{Same as Figure~\ref{Fig:LC_V1369Cen} but for nova V435~CMa.}
\label{Fig:LC_V435CMa}
\end{center}
\end{figure*}

\begin{figure*}
\begin{center}
  \includegraphics[width=0.85\textwidth]{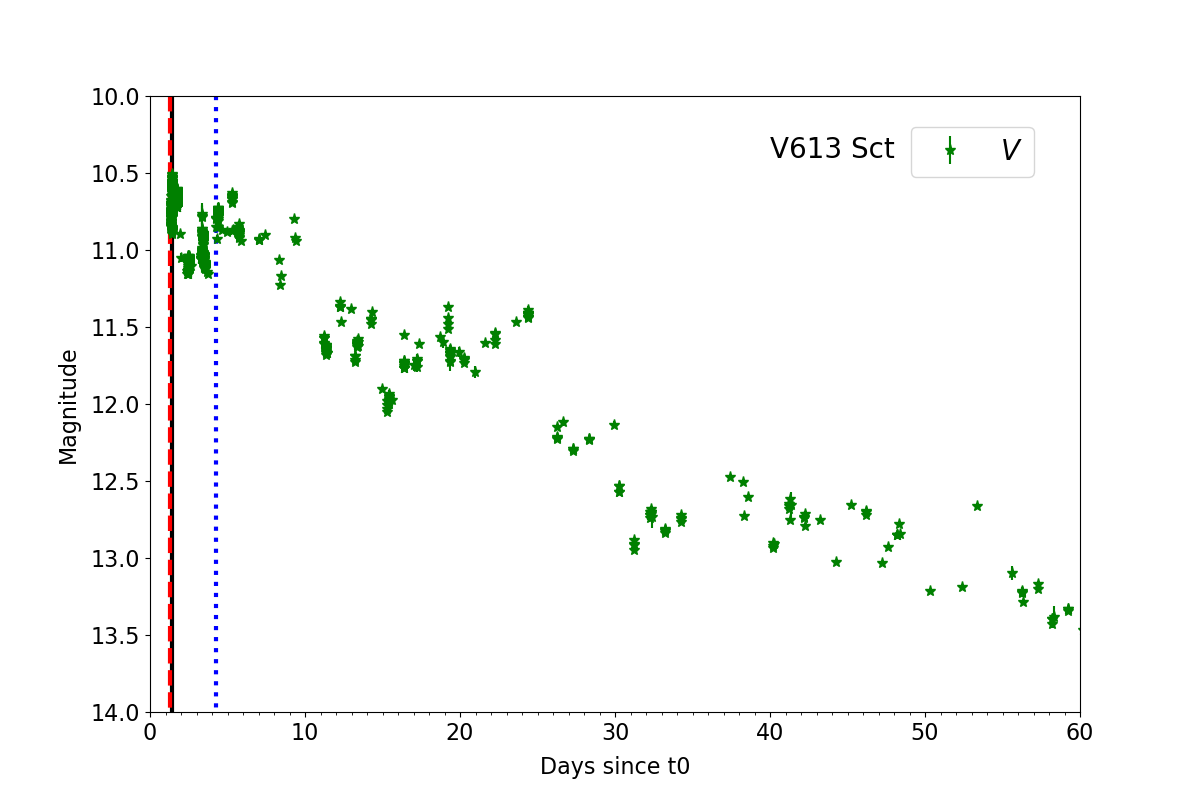}
\caption{Same as Figure~\ref{Fig:LC_V1369Cen} but for nova V613~Sct.}
\label{Fig:LC_V613Sct}
\end{center}
\end{figure*}

\begin{figure*}
\begin{center}
  \includegraphics[width=0.85\textwidth]{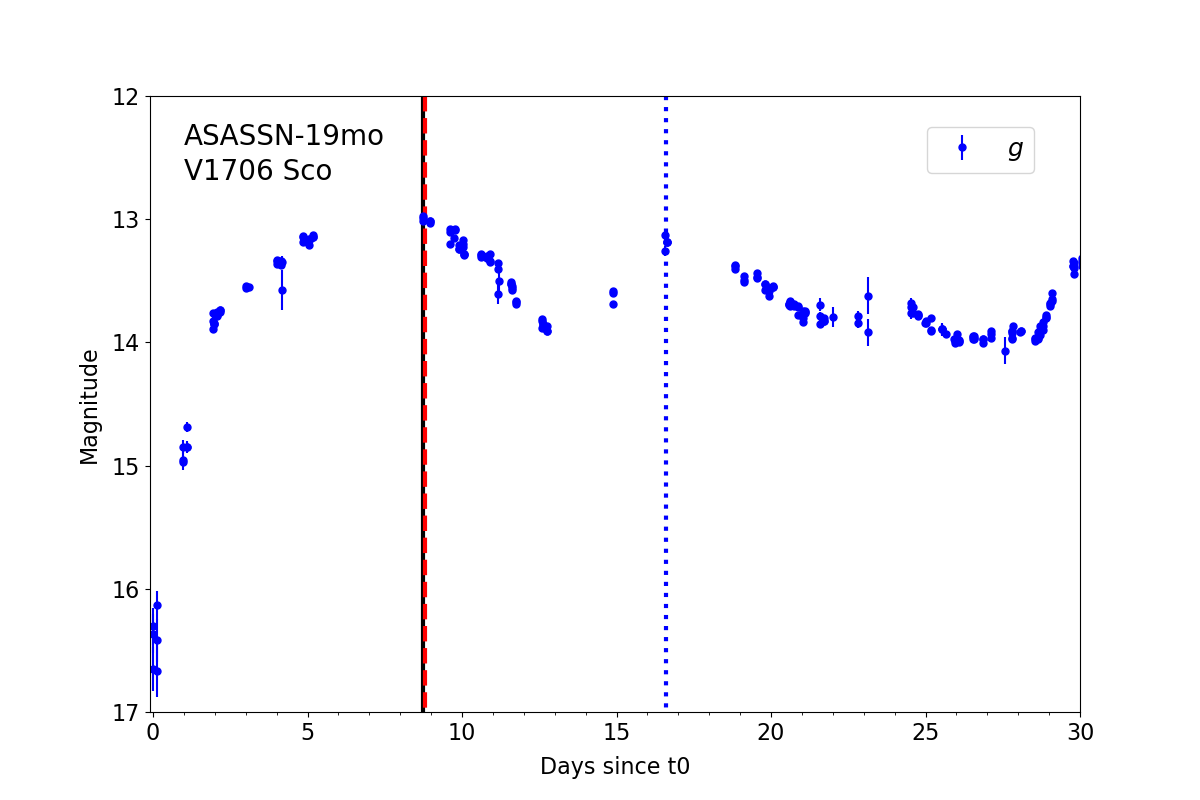}
\caption{Same as Figure~\ref{Fig:LC_V1369Cen} but for nova V1706~Sco (ASASSN-19mo).}
\label{Fig:LC_19mo}
\end{center}
\end{figure*}

\begin{figure*}
\begin{center}
  \includegraphics[width=0.85\textwidth]{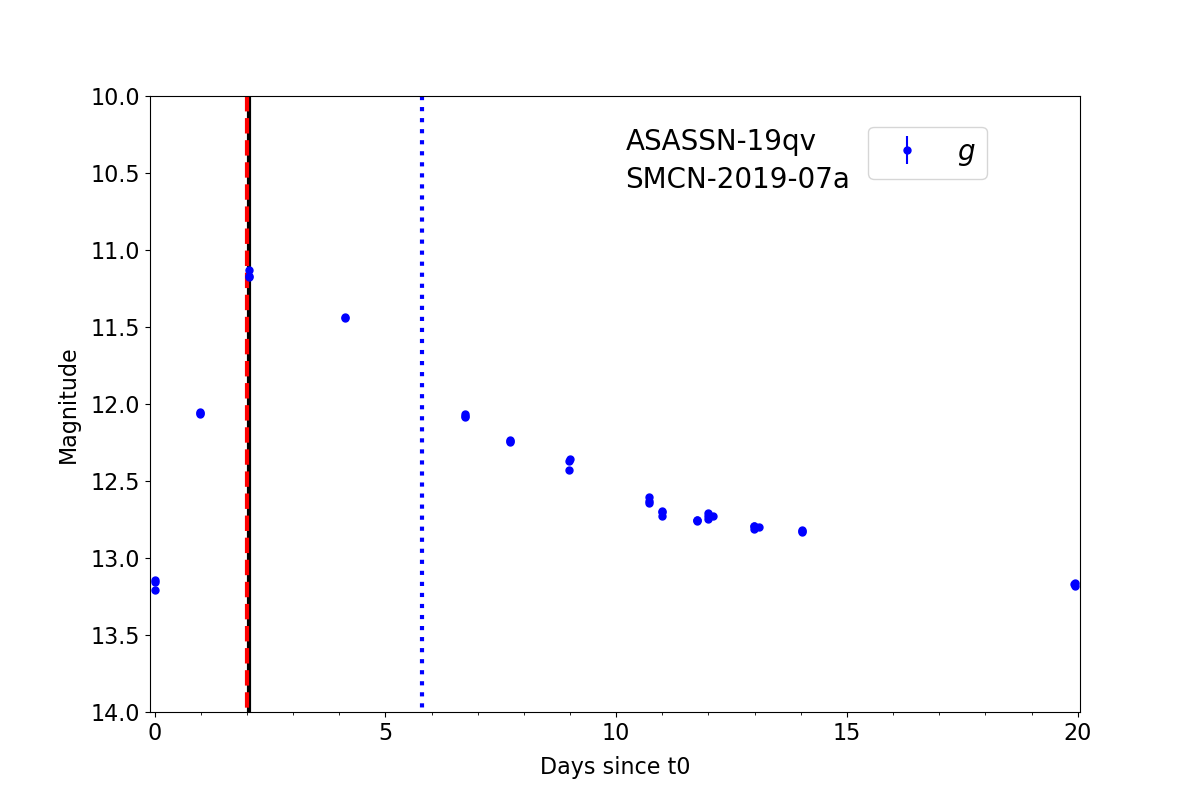}
\caption{Same as Figure~\ref{Fig:LC_V1369Cen} but for nova SMCN-2019-07a (ASASSN-19qv).}
\label{Fig:LC_19qv}
\end{center}
\end{figure*}

\begin{figure*}
\begin{center}
  \includegraphics[width=0.85\textwidth]{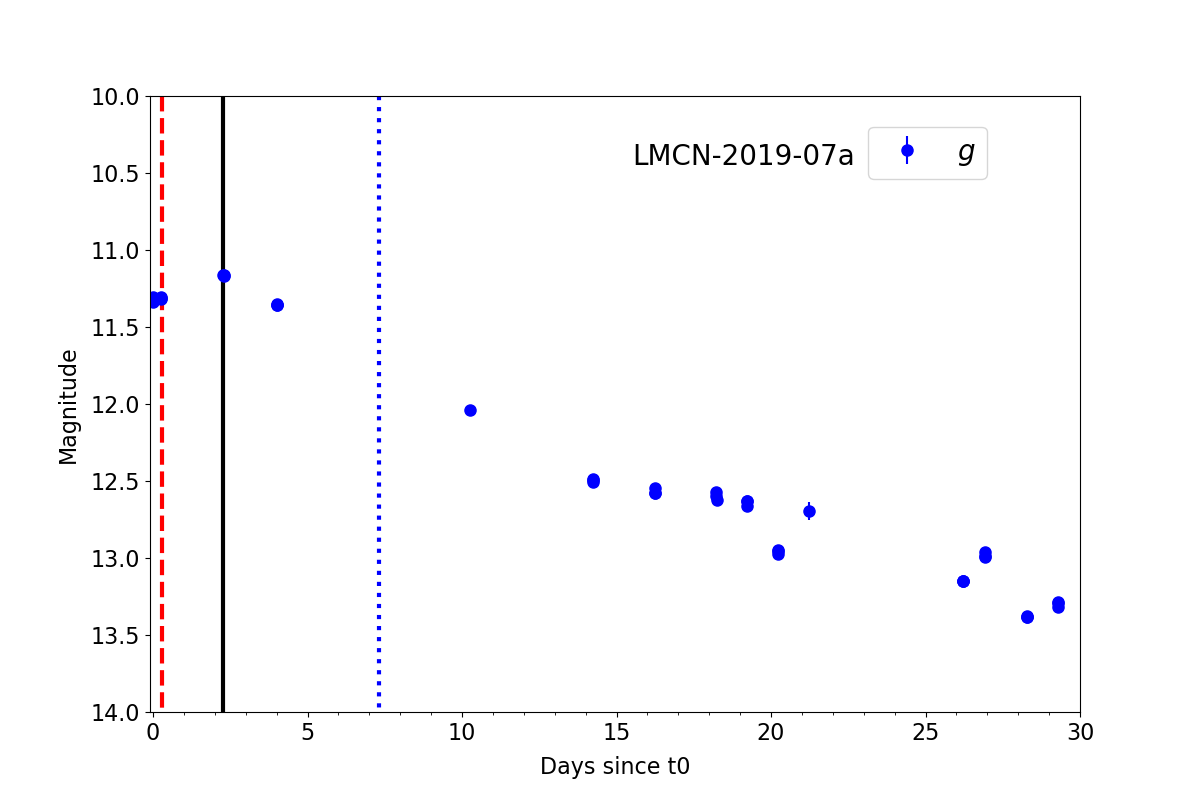}
\caption{Same as Figure~\ref{Fig:LC_V1369Cen} but for nova LMCN-2019-07a.}
\label{Fig:LC_LMCN2019}
\end{center}
\end{figure*}

\begin{figure*}
\begin{center}
  \includegraphics[width=0.85\textwidth]{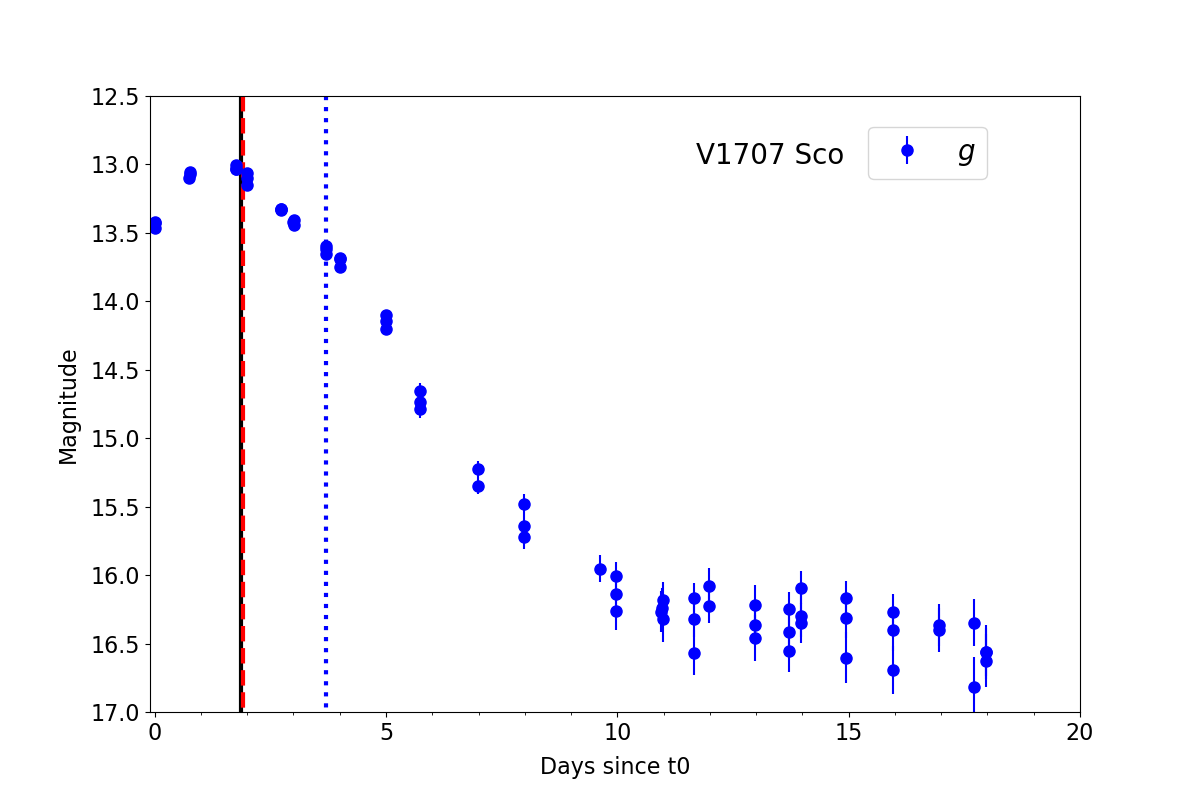}
\caption{Same as Figure~\ref{Fig:LC_V1369Cen} but for nova V1707~Sco.}
\label{Fig:LC_V1707Sco}
\end{center}
\end{figure*}

\renewcommand\thetable{\thesection.\arabic{table}}    
\renewcommand\thefigure{\thesection.\arabic{figure}}   
\setcounter{figure}{0}

\clearpage
\section{The THEA line profiles}
\label{appTHEA}
In this Appendix we present the THEA lines plot in comparison to the Na D, \eal{Fe}{II}(42), and \eal{O}{I} lines.

\begin{figure}[h!]
\begin{center}
  \includegraphics[width=0.75\columnwidth]{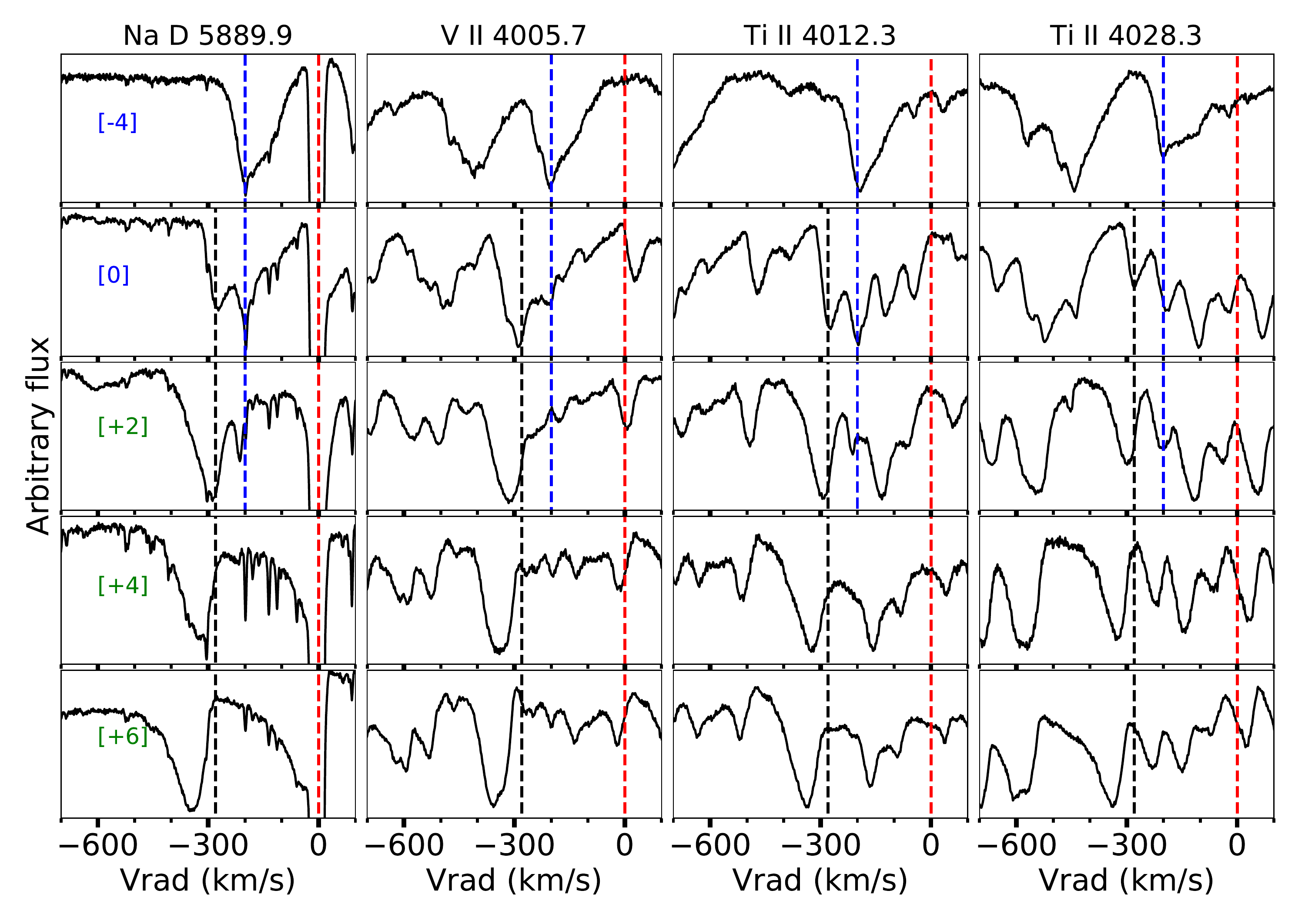}
  \includegraphics[width=0.75\columnwidth]{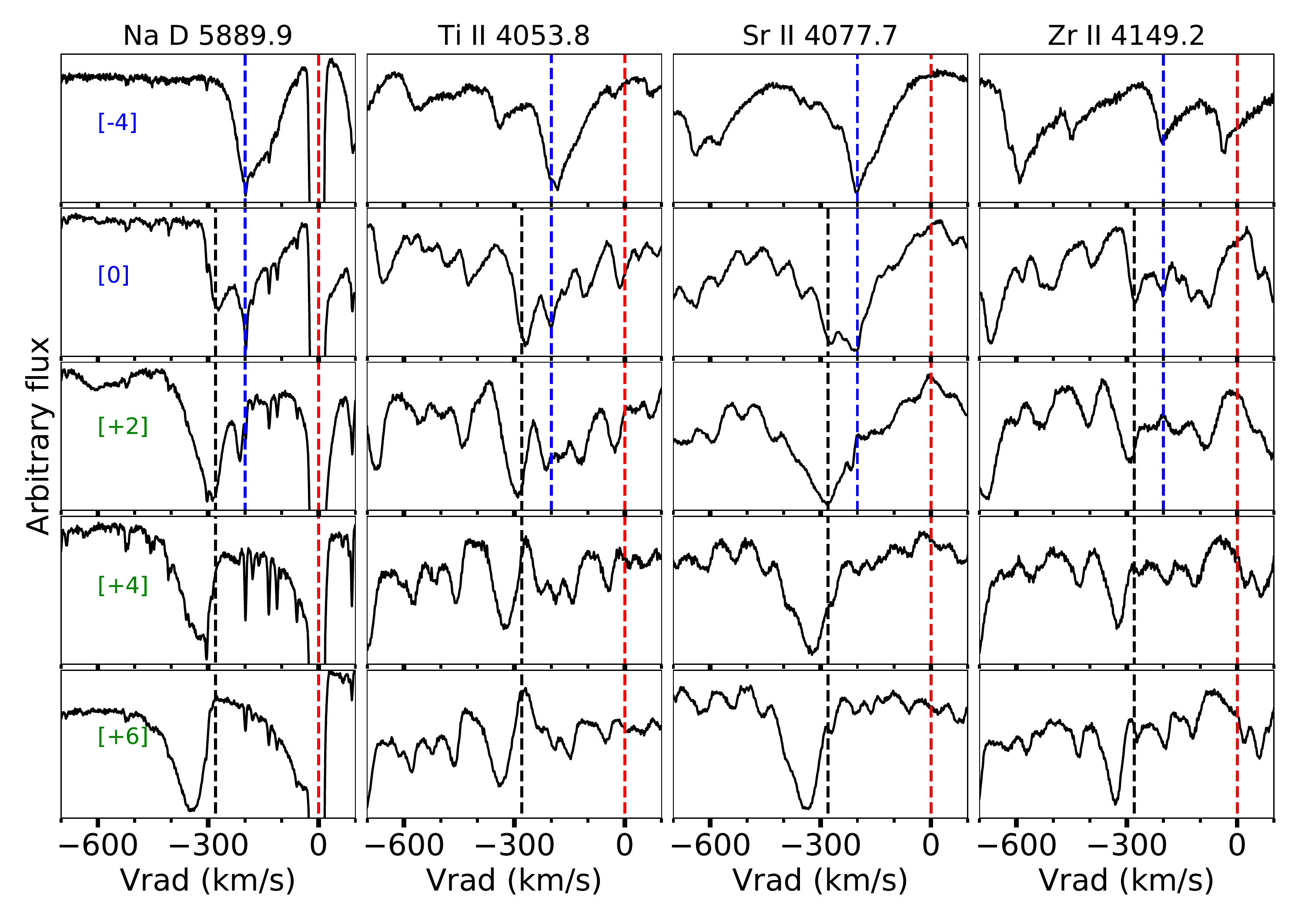}
\caption{Same as Figure~\ref{Fig:THEA_1} but for different THEA lines.}
\label{Fig:THEA_2_3}
\end{center}
\end{figure}

\begin{figure}[h!]
\begin{center}
  \includegraphics[width=0.75\columnwidth]{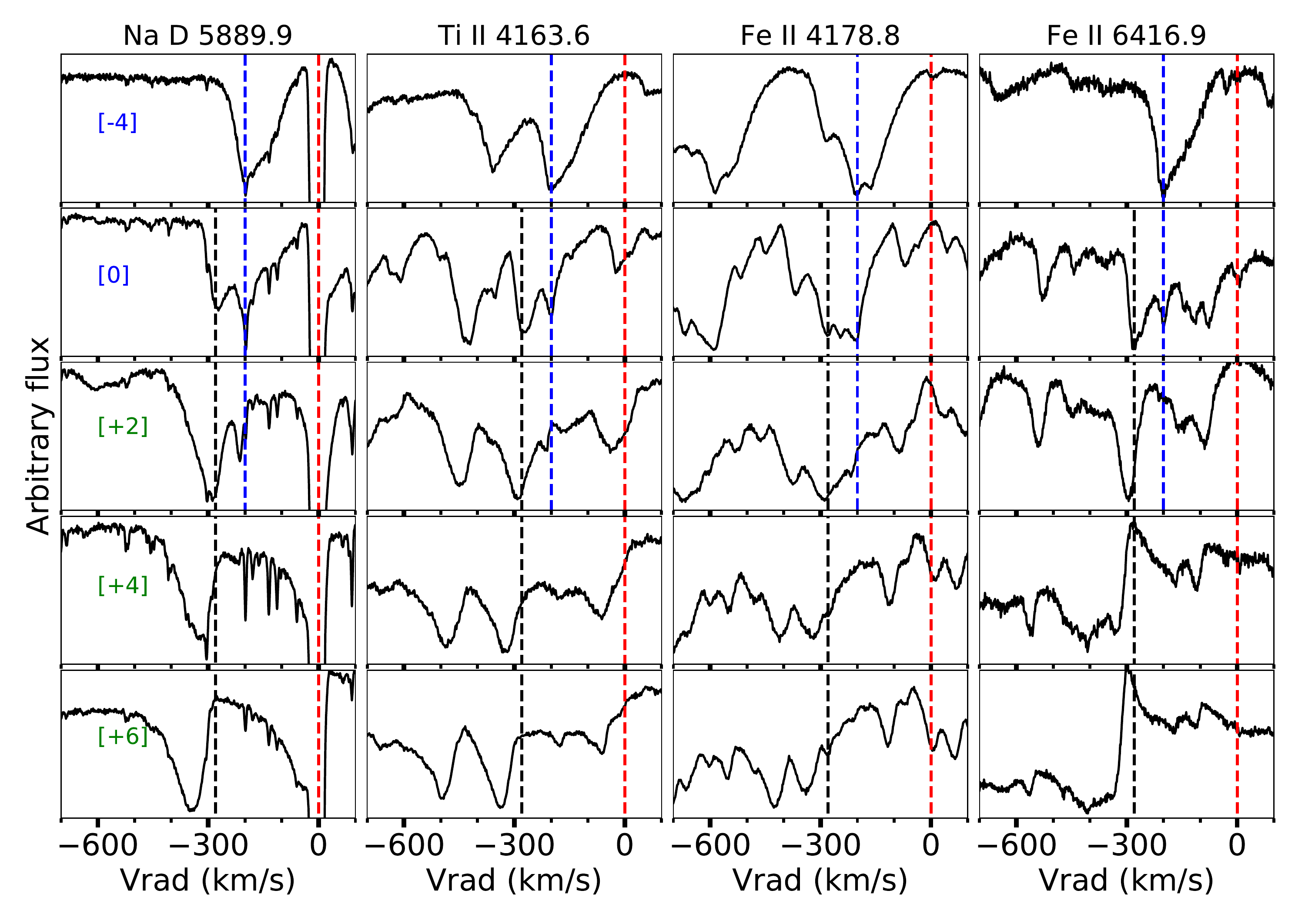}
  \includegraphics[width=0.75\columnwidth]{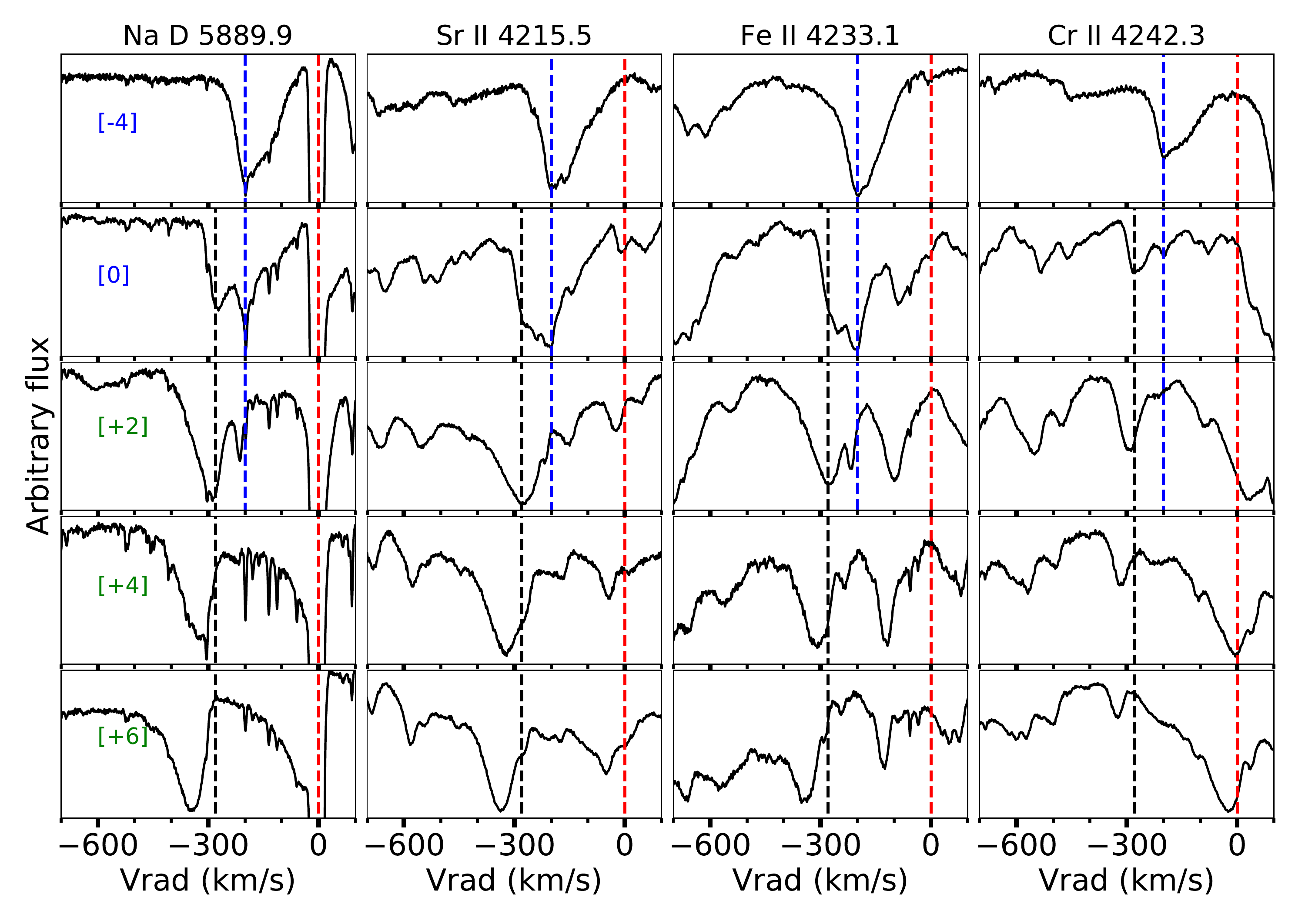}  
\caption{Same as Figure~\ref{Fig:THEA_1} but for different THEA lines.}
\label{Fig:THEA_4_5}
\end{center}
\end{figure}

\begin{figure}[h!]
\begin{center}
  \includegraphics[width=0.75\columnwidth]{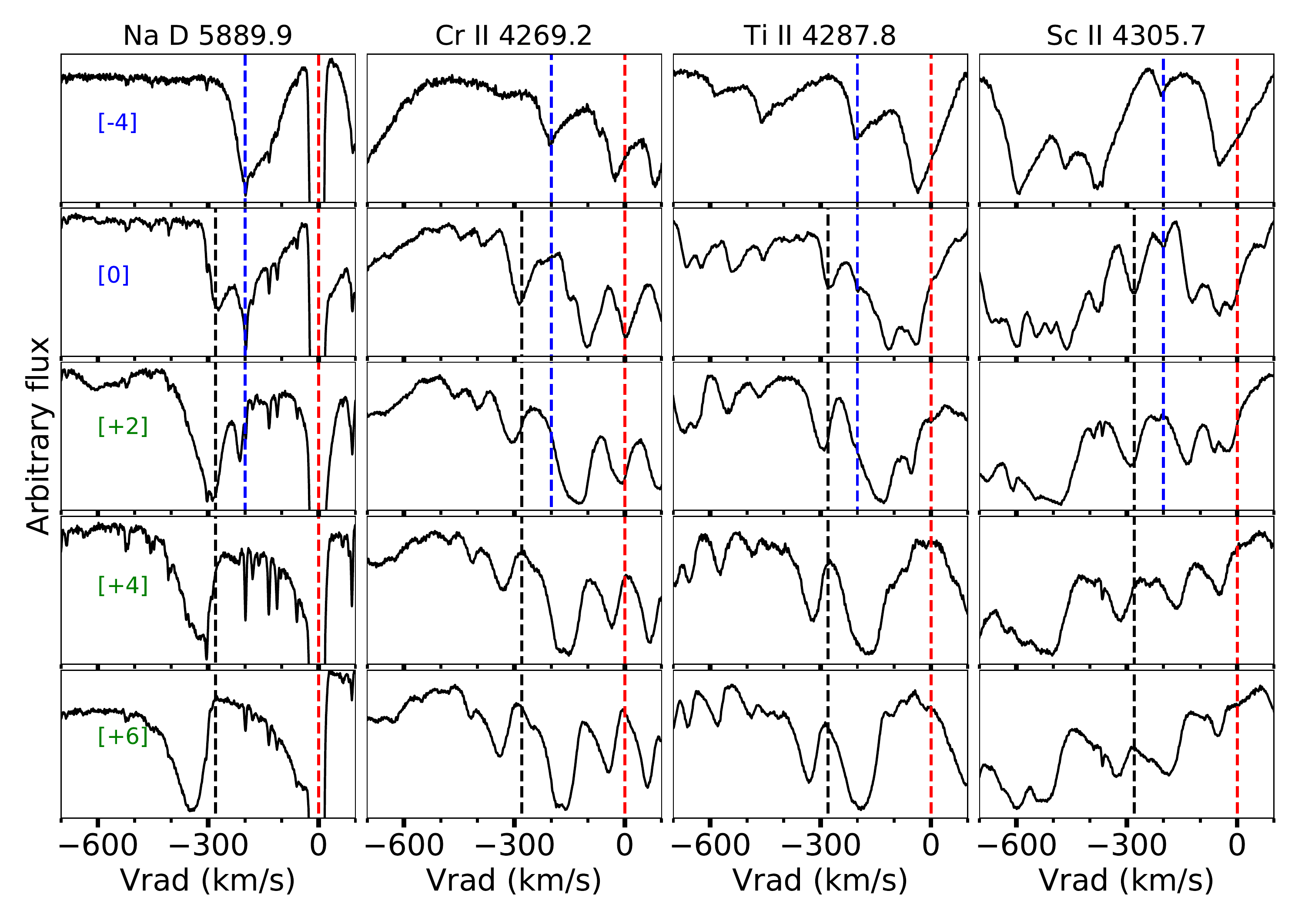}
  \includegraphics[width=0.75\columnwidth]{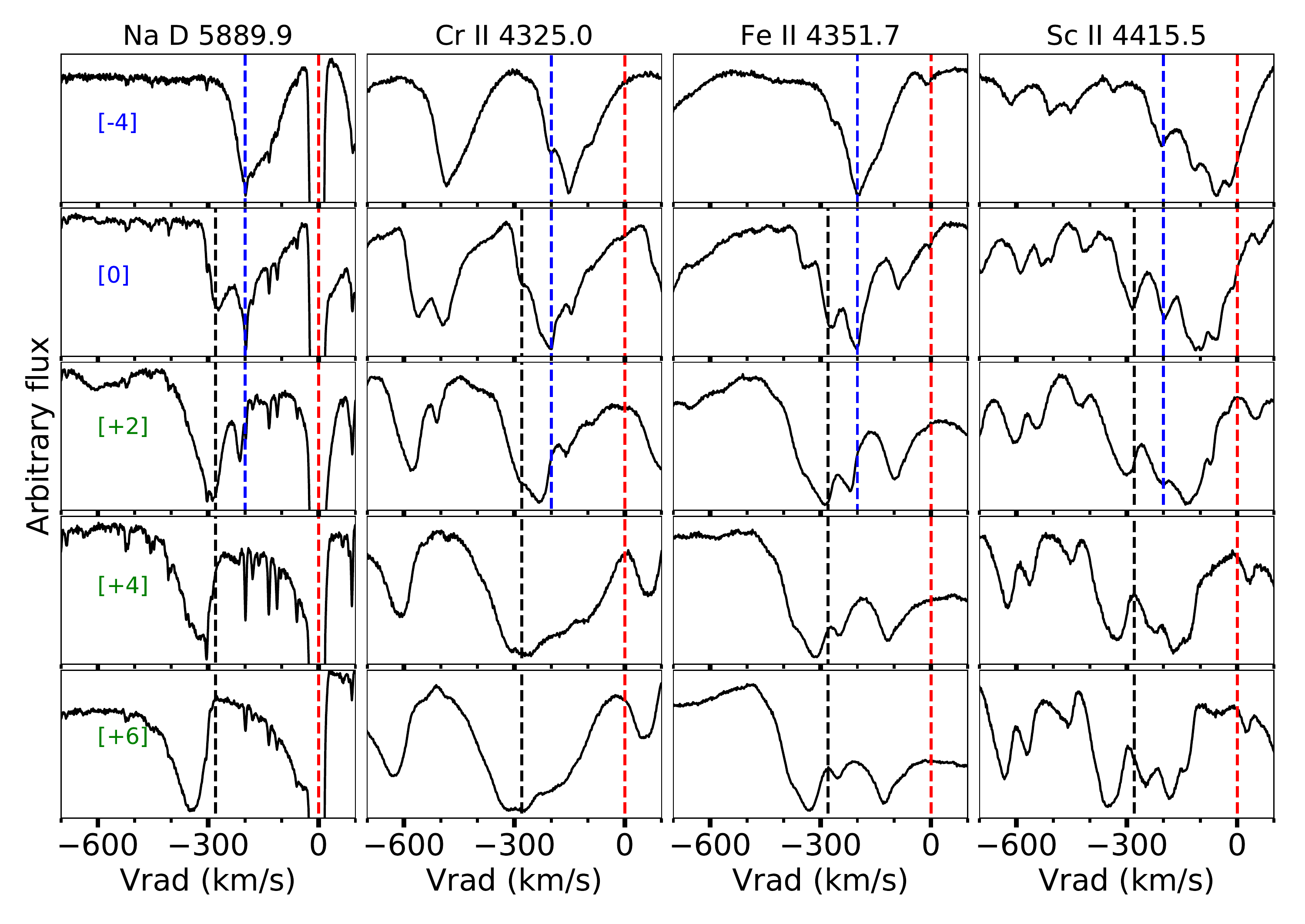}  
\caption{Same as Figure~\ref{Fig:THEA_1} but for different THEA lines.}
\label{Fig:THEA_6_7}
\end{center}
\end{figure}

\begin{figure}[h!]
\begin{center}
  \includegraphics[width=0.75\columnwidth]{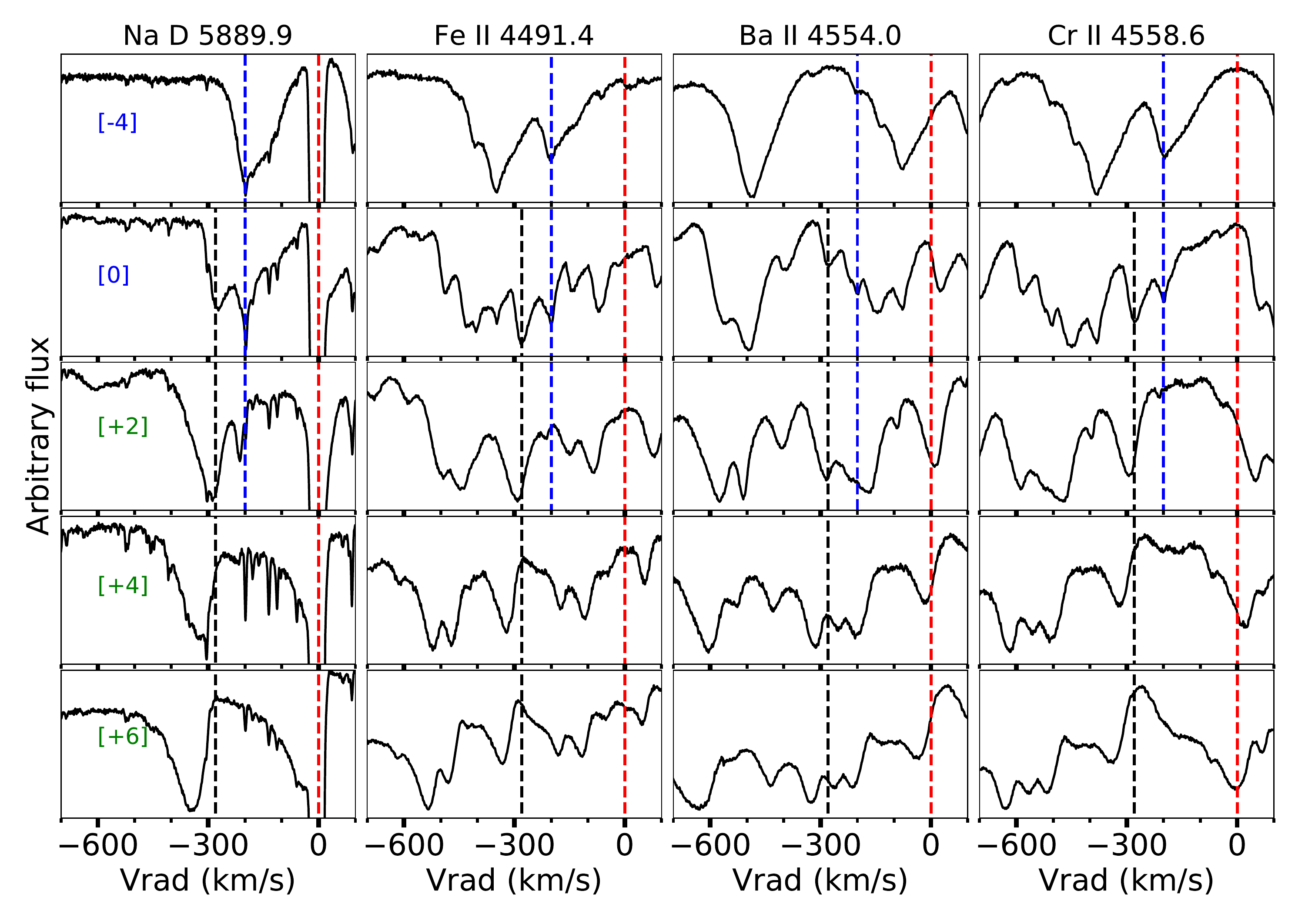}
  \includegraphics[width=0.75\columnwidth]{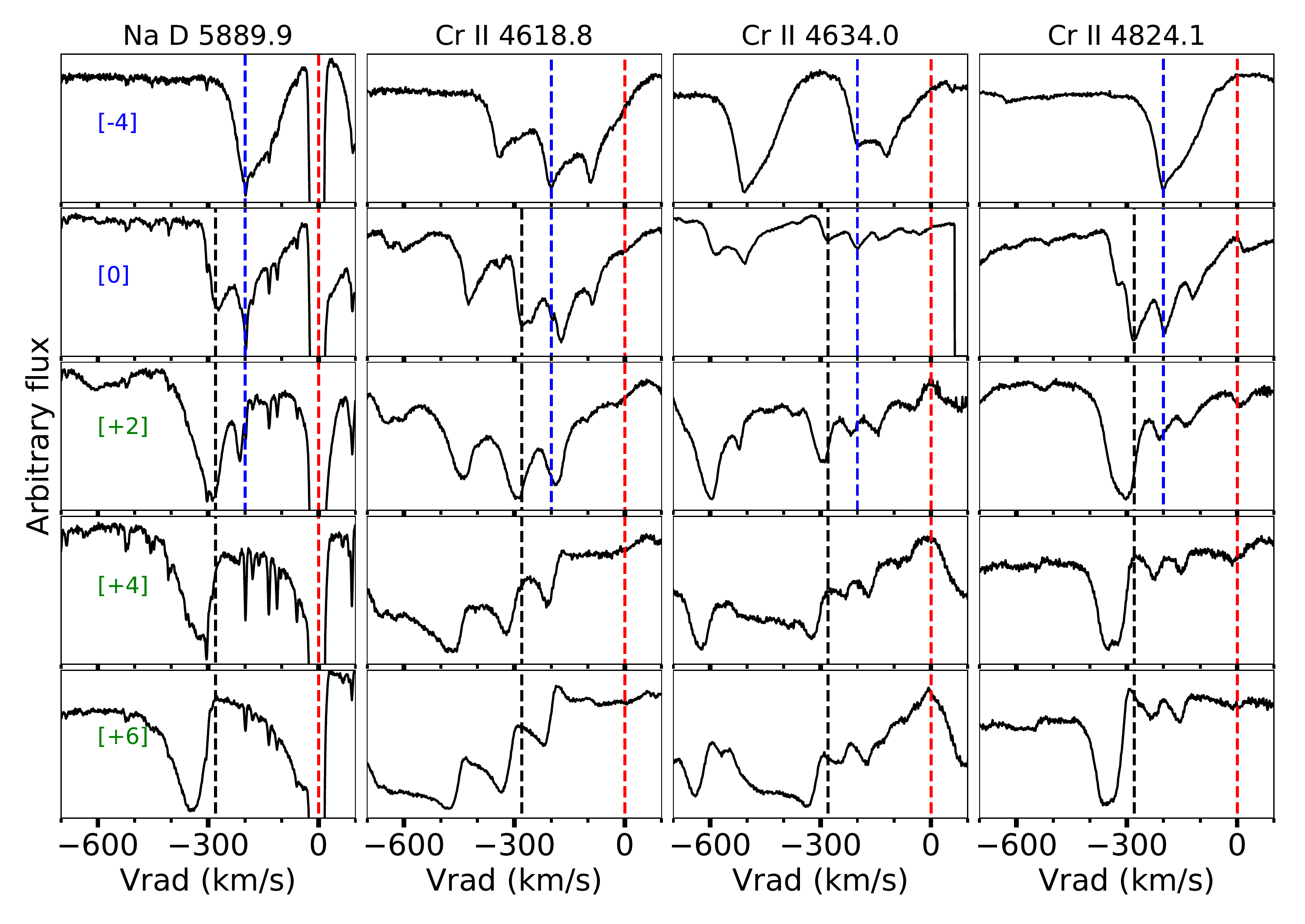}  
\caption{Same as Figure~\ref{Fig:THEA_1} but for different THEA lines.}
\label{Fig:THEA_8_9}
\end{center}
\end{figure}

\begin{figure*}
\begin{center}
  \includegraphics[width=\textwidth]{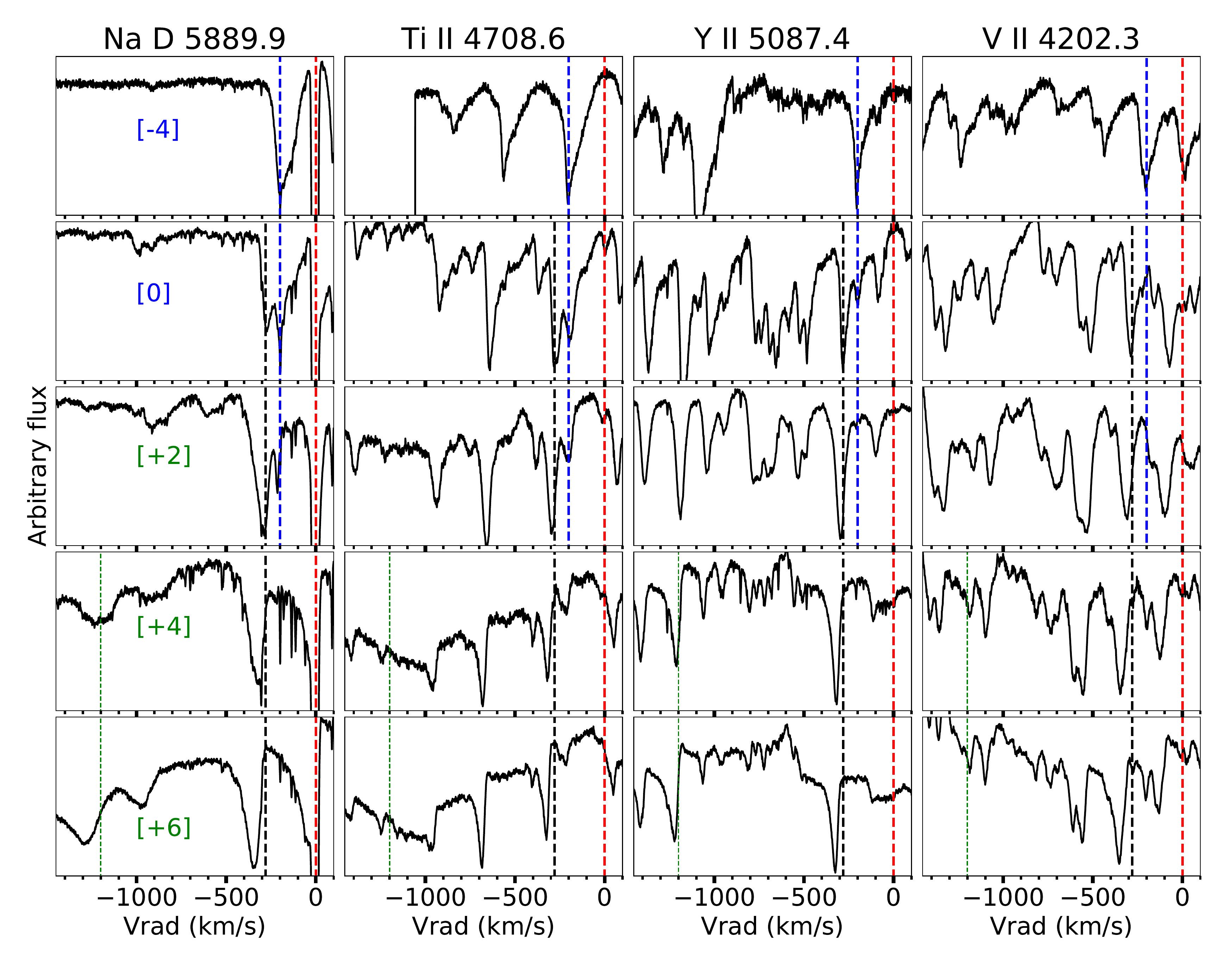}
\caption{The line profile evolution of a sample of THEA lines plotted in comparison to Na D at 5889.9\,$\mathrm{\AA}$. The red vertical dashed line represents $v_{\mathrm{rad}}$ = 0\,km\,s$^{-1}$ (rest wavelength). The blue vertical dashed line marks a velocity of $-200$\,km\,s$^{-1}$ highlighting the slow component. The black dashed line marks a velocity of $-300$\,km\,s$^{-1}$ highlighting the intermediate component. The green dashed line marks a velocity of $-1200$\,km\,s$^{-1}$ highlighting the fast component, which can only be seen in the Na D line.}
\label{Fig:THEA_long_range}
\end{center}
\end{figure*}

\renewcommand\thetable{\thesection.\arabic{table}}    
\renewcommand\thefigure{\thesection.\arabic{figure}}   
\setcounter{figure}{0}

\clearpage
\section{The line profile evolution of some novae}
\label{appA}
In this Appendix we present the evolution of the line profiles of H$\alpha$ for novae V435 CMa, V5855~Sgr, and V549~Vel.

\begin{figure}[h!]
\begin{center}
  \includegraphics[width=1.0\textwidth]{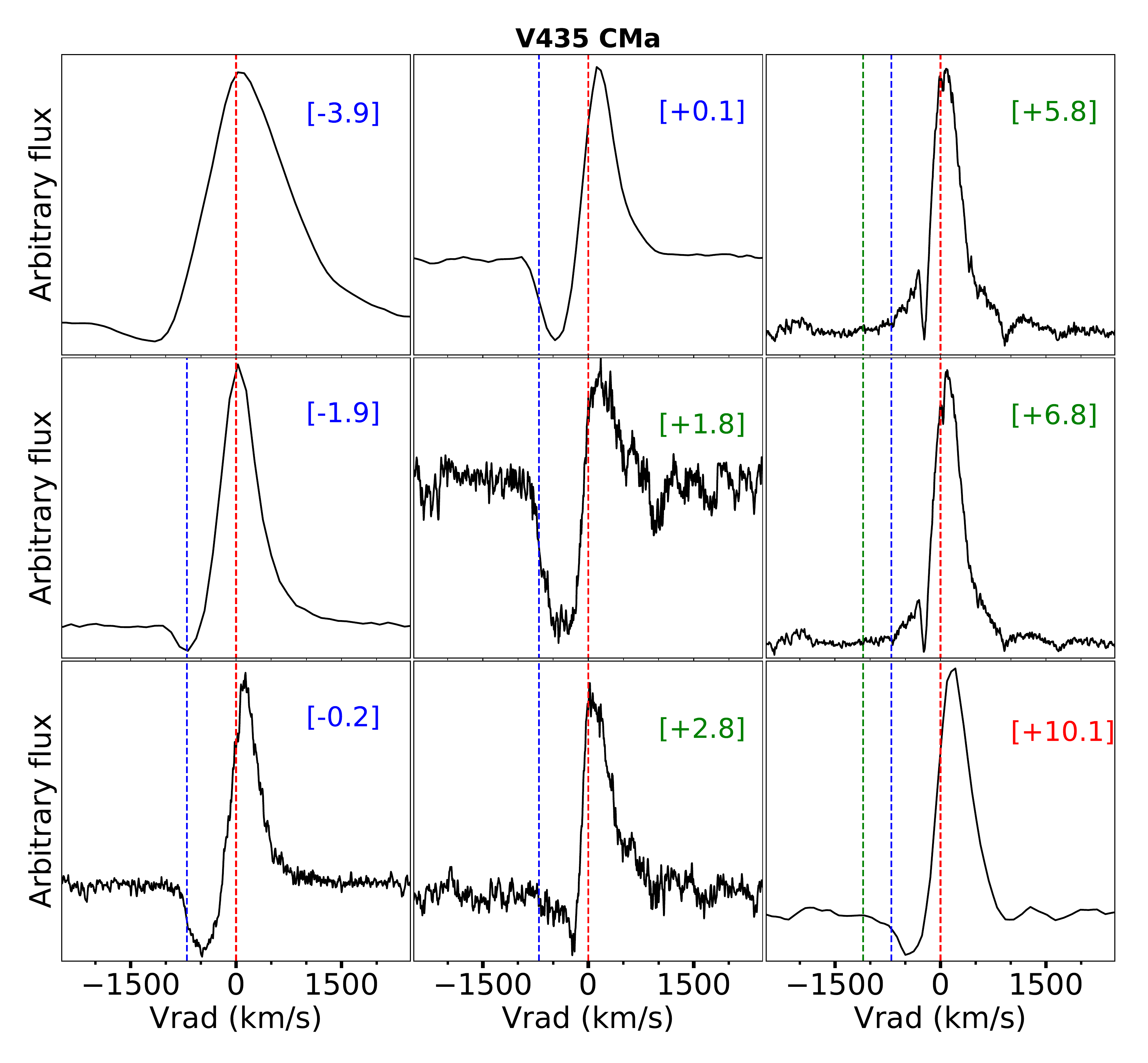}
\caption{The evolution of the H$\alpha$ line profiles for nova \textbf{V435~CMa}. The red, blue, and green dashed lines represent $v_r$ = 0\,km\,s$^{-1}$ (estimate of the line center), $-700$ (the slow component), $-1100$ (the fast component), respectively. The numbers in brackets are days after discovery. The numbers in brackets are highlighted in blue and green are for observations taken before and after the first optical peak, respectively. The one highlighted in red is taken during the second optical peak. Heliocentric corrections are applied to all radial velocities.}
\label{Fig:V435_CMa_profiles}
\end{center}
\end{figure}

\iffalse
\begin{figure}[h!]
\begin{center}
  \includegraphics[width=1.0\textwidth]{line_profiles_V1369Cen.pdf}
\caption{The evolution of the H$\alpha$ line profiles for nova \textbf{V1369~Cen}. The red, blue, and green dashed lines represent $v_r$ = 0\,km\,s$^{-1}$ (rest wavelength), $-900$ (the slow component), $-1400$ (the fast component), respectively. The magenta dashed line highlights the appearance of a new system of absorption features at $-1100$\,km\,s$^{-1}$, coinciding with the second maximum in the optical light curve. The numbers in brackets are days after discovery. The numbers in brackets are highlighted in red for observations taken around a maximum and in blue for these taken around a minimum. Heliocentric corrections are applied to all radial velocities.}
\label{Fig:V1369_Cen_profiles}
\end{center}
\end{figure}
\fi

\begin{figure}[h!]
\begin{center}
  \includegraphics[width=0.7\textwidth]{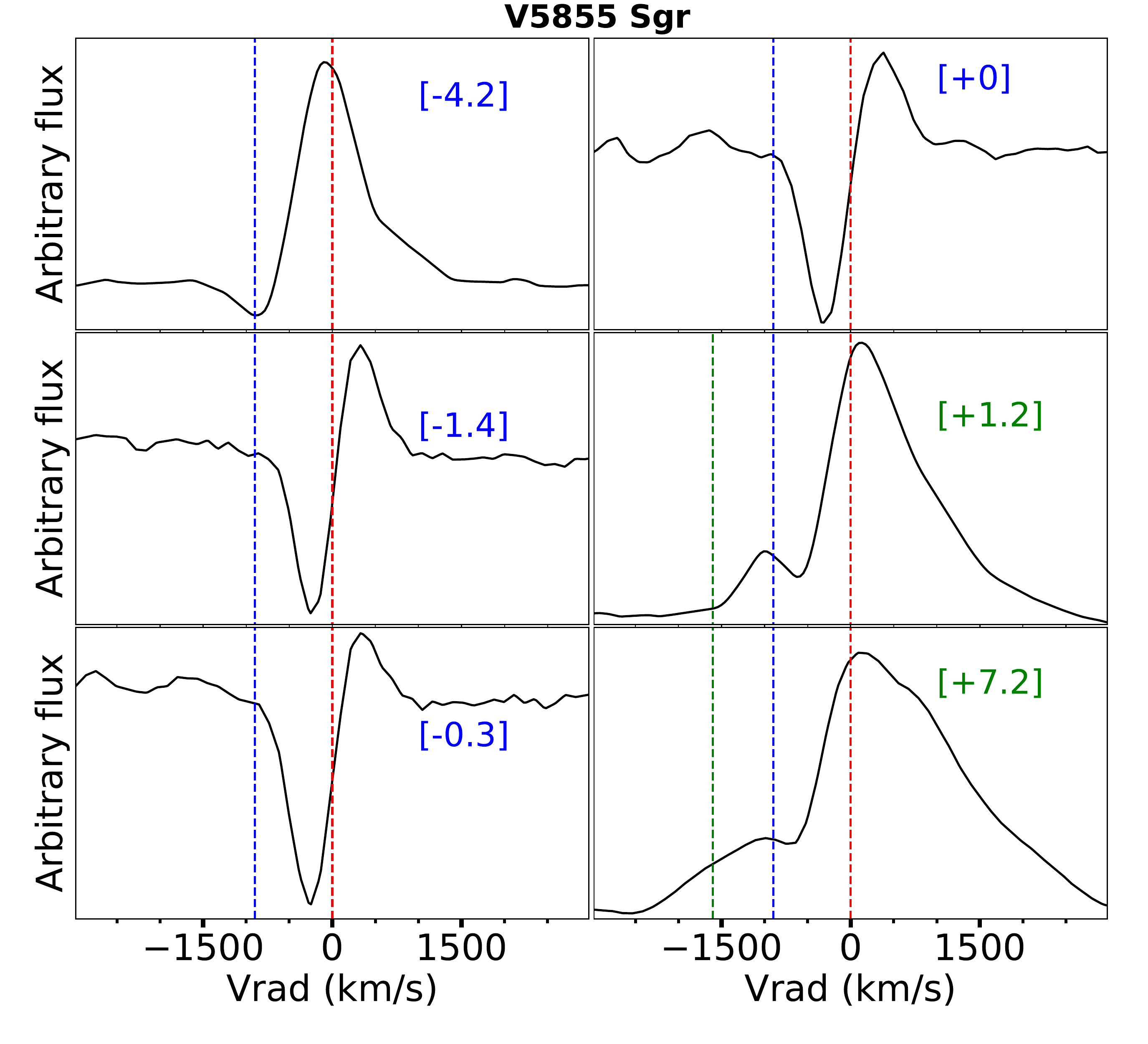}
    \includegraphics[width=0.7\textwidth]{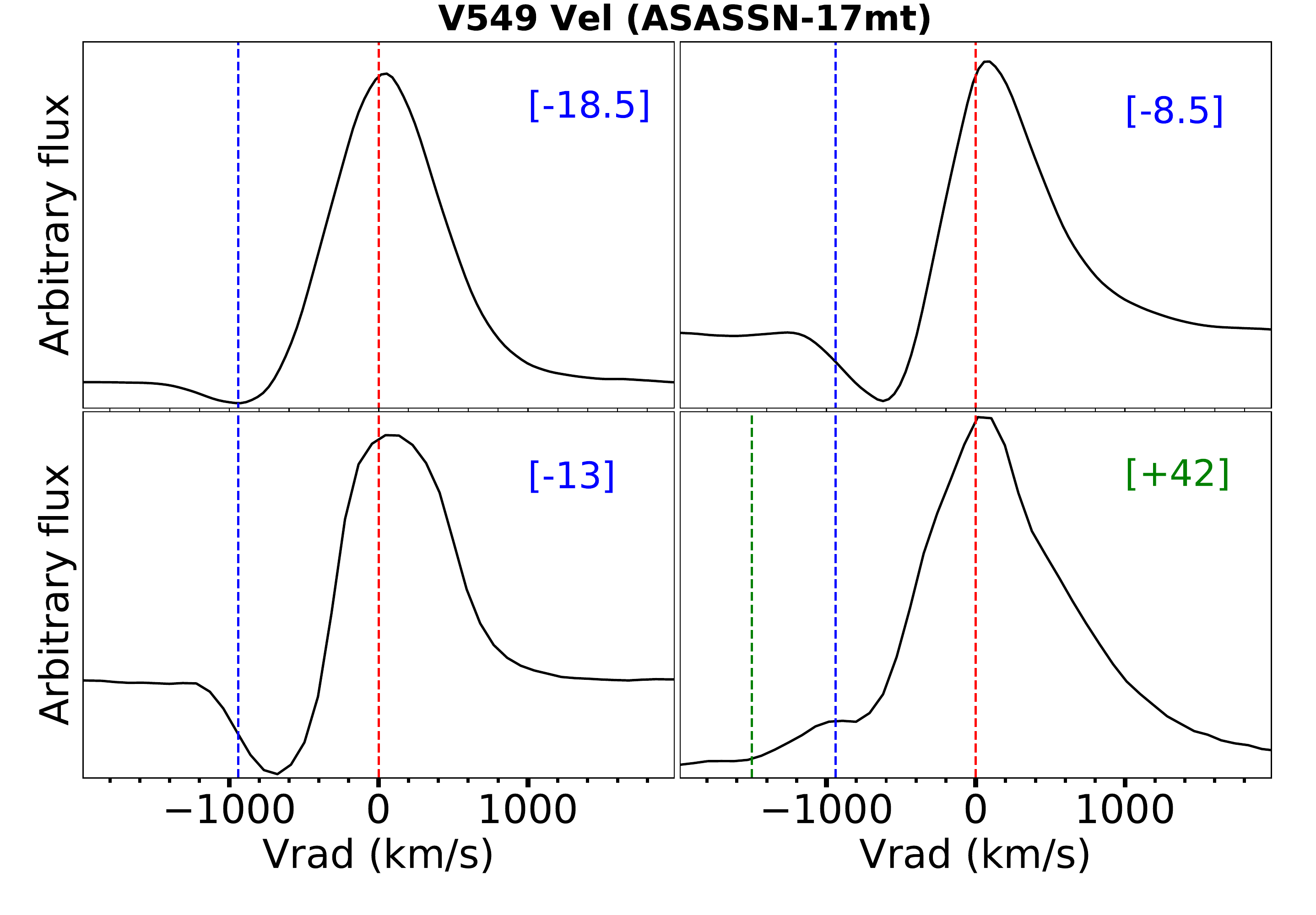}
\caption{The evolution of the H$\alpha$ line profiles for novae \textbf{V5855~Sgr} (\textit{top}) and \textbf{V549~Vel} (\textit{bottom}). In the top panel, the red, blue, and green dashed lines represent $v_r$ = 0\,km\,s$^{-1}$ (rest wavelength), $-900$\,km\,s$^{-1}$ (the slow component), $-1600$\,km\,s$^{-1}$ (the fast component), respectively. In the bottom panel, the red, blue, and green dashed lines represent $v_r$ = 0 (rest wavelength), $-950$\,km\,s$^{-1}$ (the slow component), $-1500$\,km\,s$^{-1}$ (the fast component), respectively. The numbers in brackets are days after discovery. The numbers in brackets are highlighted in blue and green are for observations taken before and after the optical peak, respectively. Heliocentric corrections are applied to all radial velocities.}
\label{Fig:V5855_Sgr_profiles}
\end{center}
\end{figure}

\iffalse
\begin{figure}
\begin{center}
  \includegraphics[width=\columnwidth]{CaII.pdf}
\caption{The line profile evolution of a sample of \eal{Fe}{II} 5018\,$\mathrm{\AA}$ and H$\beta$ in comparison \eal{Ca}{II} 8498 and 8542\,$\mathrm{\AA}$. The red vertical dashed line represents $v_{\mathrm{rad}}$ = 0\,km\,s$^{-1}$ (rest wavelength). The blue vertical dashed line marks a velocity of $-200$\,km\,s$^{-1}$ highlighting the slow component. The black dashed line marks a velocity of $-300$\,km\,s$^{-1}$ highlighting the intermediate component. The top three panels extend between $-$1000\,km\,s$^{-1}$ and 1000\,km\,s$^{-1}$ while the bottom two panels extend between $-$2000\,km\,s$^{-1}$ and 2000\,km\,s$^{-1}$ to show the fast component.}
\label{Fig:CaII}
\end{center}
\end{figure}
\fi

\renewcommand\thetable{\thesection.\arabic{table}}    
\renewcommand\thefigure{\thesection.\arabic{figure}}   
\setcounter{figure}{0}

\clearpage

\section{Complementary Tables}
\label{appTables}
In this Appendix we present complementary tables including log of observations for nova V906~Car and FM~Cir.

\begin{table*}
\centering
\caption{Log of the spectroscopic observations of nova V906~Car adopted from \citet{Aydi_etal_2020}.}
\begin{tabular}{lccc}
\hline
($t_{\mathrm{s}} - t_{\mathrm{max}}$) & Instrument & $R$ & Range\\
(days) & & & ($\mathrm{\AA}$)\\
\hline
\hline
$-4$ & VLT-UVES & 59,000 & 3050\,--\,9000\\
$-2.5$ & VLT-UVES & 59,000 & 3050\,--\,9000\\
$-2$ & ARAS & 11,000 & 6400\,--\,6720\\
$-1$ & ARAS & 11,000 & 6400\,--\,6720\\
$-0.5$ & VLT-UVES & 59,000 & 3050\,--\,9000\\
$0$ & ARAS & 11,000 & 6400\,--\,6720\\
$+1$ & ARAS & 11,000 & 6400\,--\,6720\\
$+1.5$ & VLT-UVES & 59,000 & 3050\,--\,9000\\
$+2$ & ARAS & 11,000 & 6400\,--\,6720\\
$+4$ & VLT-UVES & 59,000 & 3050\,--\,9000\\
$+5$ & HERCULES & 41,000 & 4000\,--\,10000\\
$+6$ & ARAS & 11,000 & 6400\,--\,6720\\
$+7$ & HERCULES & 41,000 & 4000\,--\,10000\\
$+8$ & ARAS & 11,000 & 6400\,--\,6720\\
$+9$ & HERCULES & 41,000 & 4000\,--\,10000\\
$+10$ & HERCULES & 41,000 & 4000\,--\,10000\\
$+11.5$ & VLT-UVES & 59,000 & 3050\,--\,9000\\
$+13.5$ & VLT-UVES & 59,000 & 3050\,--\,9000\\
$+14$ & ARAS & 11,000 & 6400\,--\,6720\\
$+16$ & ARAS & 11,000 & 6400\,--\,6720\\
$+16.5$ & VLT-UVES & 59,000 & 3050\,--\,9000\\
$+20.5$ & SALT & 67,000 & 3900\,--\,8800\\
$+21.5$ & SALT & 67,000 & 3900\,--\,8800\\
$+24.5$ & VLT-UVES & 59,000 & 3050\,--\,9000\\
\hline
\end{tabular}
\label{table:spec_V906_Car}
\end{table*}

\begin{table*}
\centering
\caption{Log of the CHIRON spectroscopic observations of nova FM~Cir.}
\begin{tabular}{lccc}
\hline
($t_{\mathrm{s}} - t_{\mathrm{max}}$) & Instrument & $R$ & Range\\
(days) & & & ($\mathrm{\AA}$)\\
\hline
\hline
$-7$ & CHIRON-SMARTS & 27,000 & 4100\,--\,8900\\
$-6$ & CHIRON-SMARTS & 27,000 & 4100\,--\,8900\\
$-5$ & CHIRON-SMARTS & 78,000 & 4100\,--\,8900\\
$-4$ & CHIRON-SMARTS & 78,000 & 4100\,--\,8900\\
$-2$ & CHIRON-SMARTS & 78,000 & 4100\,--\,8900\\
$-0.5$ & CHIRON-SMARTS & 27,000 & 4100\,--\,8900\\
\hline
\end{tabular}
\label{table:spec_FM_Cir}
\end{table*}

\end{document}